\documentclass[a4paper,12pt]{article}

\usepackage[top=3.5cm, bottom=3.5cm, left=2.5cm, right=2.5cm]{geometry} 

\usepackage{wrapfig}
\usepackage[sc,margin=1cm,textfont=small]{caption}
\usepackage{array}
\usepackage{arydshln}
\usepackage{graphicx}
\usepackage[lofdepth,lotdepth]{subfig}
\usepackage{float}
\usepackage{amsthm}
\usepackage{amssymb}
\usepackage{epstopdf}
\usepackage[utf8]{inputenc}
\DeclareUnicodeCharacter{221E}{$\infty$}
\DeclareUnicodeCharacter{00B4}{\'}
\DeclareUnicodeCharacter{03BB}{$\lambda$}
\DeclareUnicodeCharacter{0393}{$\Gamma$}
\DeclareUnicodeCharacter{211D}{$\mathbb{R}$}
\DeclareUnicodeCharacter{2212}{-} 

\usepackage{csquotes}
\usepackage[english]{babel} 
\usepackage{amsmath}
\usepackage{mathrsfs}
\usepackage{enumerate}
\usepackage{color}
\usepackage[colorlinks=false,linkcolor=blue]{hyperref}
\usepackage[nottoc, notlof, notlot, numbib]{tocbibind}
\usepackage{titling}
\usepackage{mathtools}
\usepackage[]{pdfpages}
 \usepackage[numbers]{natbib}
\usepackage{tikz}
\usetikzlibrary{arrows}
\usetikzlibrary[patterns]
\usepackage[T1]{fontenc}
\usepackage[final]{microtype}
\usepackage{comment}
\usepackage[boxed]{algorithm2e}
\usepackage{setspace}
\usepackage{etoolbox}
\AtBeginEnvironment{algorithm}{\setstretch{1.2}}

\usepackage{array}
\usepackage{tabularx}
\usepackage{ltablex} 
\usepackage{caption}

\usepackage{authblk}
\usepackage{mathrsfs}
\usepackage{thmtools}
\usepackage{bbm}

\usepackage{stmaryrd}

\makeatletter
\newsavebox{\@brx}
\newcommand{\llangle}[1][]{\savebox{\@brx}{\(\m@th{#1\langle}\)}%
  \mathopen{\copy\@brx\kern-0.5\wd\@brx\usebox{\@brx}}}
\newcommand{\rrangle}[1][]{\savebox{\@brx}{\(\m@th{#1\rangle}\)}%
  \mathclose{\copy\@brx\kern-0.5\wd\@brx\usebox{\@brx}}}
\makeatother

\newtheorem{theorem}{Theorem}[section]

\newtheorem{lemma}[theorem]{Lemma}

\theoremstyle{definition}

\theoremstyle{remark}

\newtheorem{video}{Video}

\title{\bf \LARGE Bulk topological states in a new collective dynamics model}
\author[1]{Pierre \textsc{Degond}}
\author[2]{Antoine \textsc{Diez}}
\author[3]{Mingye \textsc{Na}}

\affil[1]{\small
Institut de Math\'ematiques de Toulouse; UMR5219; 

Universit\'e de Toulouse; CNRS; 

UPS; F-31062 Toulouse Cedex 9, France

\url{pierre.degond@math.univ-toulouse.fr}
\bigskip
}
\affil[2]{\small
Department of Mathematics, 

Imperial College London, South Kensington Campus,

London, SW7 2AZ, UK

\url{antoine.diez18@imperial.ac.uk}
\bigskip
}
\affil[2]{\small
Department of Mathematics, 

Southern University of Science and Technology,

Shenzhen, 518055, China

\url{12121231@mail.sustech.edu.cn}
\bigskip
}

\date{}

\begin{document}

\maketitle

\abstract{In this paper, we demonstrate the existence of topological states in a new collective dynamics model. This individual-based model (IBM) describes self-propelled rigid bodies moving with constant speed and adjusting their rigid-body attitude to that of their neighbors. In previous works, a macroscopic model has been derived from this IBM in a suitable scaling limit. In the present work, we exhibit explicit solutions of the macroscopic model characterized by a non-trivial topology. We show that these solutions are well approximated by the IBM during a certain time but then the IBM transitions towards topologically trivial states. Using a set of appropriately defined topological indicators, we reveal that the breakage of the non-trivial topology requires the system to go through a phase of maximal disorder. We also show that similar but topologically trivial initial conditions result in markedly different dynamics, suggesting that topology plays a key role in the dynamics of this system.}



\medskip
\noindent
\textit{Acknowledgements:} Part of this research was done when PD and MN were affiliated to Department of Mathematics, Imperial College London, London, SW7 2AZ, United Kingdom. PD acknowledges support by the Engineering and Physical Sciences Research Council (EPSRC) under grants no. EP/M006883/1 and EP/P013651/1, by the Royal  Society and the Wolfson Foundation through a Royal Society Wolfson Research Merit Award no. WM130048. The work of AD is supported by an EPSRC-Roth scholarship cofounded by the Engineering and Physical Sciences Research Council and the Department of Mathematics at Imperial College London.

\medskip
\noindent
\textit{Data statement:} no new data were collected in the course of this research.

\tableofcontents

\section{Introduction}

Systems of particles (or agents) which exhibit self-organized collective behavior are ubiquitous in the living world at all scales, from bird flocks \cite{lukeman2010inferring} to sperm \cite{creppy2016symmetry} or bacterial colonies \cite{czirok1996formation}. Examples are also found in social sciences \cite{castellano2009statistical, degond2017continuum} or for inert matter~\cite{bricard2015emergent}. In such systems, the agents interact locally with a limited number of neighbors through rather simple rules such as attraction, repulsion or alignment \cite{aoki1982simulation, couzin2002collective, gautrais2012deciphering} without any leader or centralized control. When the number of agents becomes large, vast structures encompassing many agents appear, such as clusters \cite{martin2018collective, vicsek2012collective}, traveling bands \cite{chate2008collective}, vortices \cite{costanzo2018spontaneous, czirok1996formation}, lanes \cite{couzin2003self}, etc. As there is no direct or apparent relation between these structures and the nature of the agents interactions, such a phenomenon is named ``emergence''. Its study has stimulated a vast literature (see e.g. \cite{vicsek2012collective} for a review). 

There are mainly two levels of description of particle systems: the most detailed one consists of individual based models (IBM) where the agents dynamics are described by coupled ordinary or stochastic differential equations. When the number of agents becomes large, a macroscopic description in terms of average quantities such as the agents mean density or velocity is preferred. The rigorous link between these two levels of description involves two successive limits by which the number of agents is first sent to infinity (mean-field limit) and then, the system size relative to the typical interaction distance between the agents is also sent to infinity (hydrodynamic limit), see e.g. \cite{cercignani2013mathematical, degond2004macroscopic}. In collective dynamics, particles are capable of self-propulsion by transforming an internal source of chemical energy into motion \cite{vicsek2012collective}. There are two main classes of IBM of self-propelled particles. The first class is based on the Cucker-Smale model \cite{barbaro2014phase, cucker2007emergent, ha2009simple, ha2008from} where self-propulsion is treated as an external force. The second class is based on the Vicsek model \cite{aldana2009emergence, caussin2014emergent, chate2008collective, czirok1996formation, degond2008continuum, dimarcomotsch16, martin2018collective, vicsek1995novel} where self-propulsion is modeled by imposing the norm of the particle velocity to be a constant. At the mean-field or hydrodynamic levels, the two frameworks give rise to corresponding models (see e.g. \cite{aceves2019hydrodynamic, barbaro2016phase} for Cucker-Smale type models and \cite{bolley2012mean, degondfrouvelleliu15, degond2008continuum, dimarcomotsch16, peruani2008mean, toner1998flocks} for Vicsek type models). The two categories are linked by an asymptotic limit \cite{bostan2013asymptotic, bostan2017reduced}. Of course, there are many variants of these models and we refer to \cite{bertin2006boltzmann, bertin2009hydrodynamic, carrillo2009double, cavagna2015flocking, degond2011macroscopic, dorsogna2006self, motsch2011new} for a non-exhaustive set of examples. 

Recently, a series of studies has investigated the existence of topological states in collective dynamics. Topological states have appeared with the quantum Hall effect \cite{Klitzing_1980, Laughlin_1981, NB1985, Thouless_1983} which relies on so-called conducting chiral edge states: when a sample of a 2-dimensional insulator is placed in a magnetic field, its bulk conductance is nil but a current can flow around its edges in only one direction (hence the 'chiral' terminology). Then, materials that exhibit chiral edge states without a magnetic field have been discovered, the so-called ``topological insulators'' \cite{hasan2010colloquium, NB2016, qi2011topological}. Chiral edge states are robust against perturbations because of their non trivial topology which can be characterized by a integer, the winding number. Any destruction of the chiral edge state would require a finite jump of this integer, which consumes a finite amount of energy. Hence lower energy perturbations will fail to destroy the chiral edge state. This property is of strategic interest for various applications such as quantum computers. Recently a series of works have explored the occurrence of topological states in collective dynamics (see e.g. \cite{shankar2017topological, sone2019anomalous, souslov2017topological}). They are based on numerical simulations of the Toner and Tu model \cite{toner1998flocks}, which is a continuum analog of the Vicsek model \cite{vicsek1995novel}. Investigating appropriate geometrical configurations (a sphere in \cite{shankar2017topological}, a network of rings in \cite{sone2019anomalous, souslov2017topological}), they show that linearized perturbations of the stationary state (i.e. sound waves) generate chiral edge states which propagate uni-directionally, revealing an underpinning non-trivial topology. However, the question of whether this effect could be realized with a finite (even large) number of discrete particles and whether the topological states would survive the noise induced by this finite particle number long enough is not investigated. 

In this paper, we demonstrate the existence of non-trivial bulk topological states in a new collective dynamics model. Bulk states propagate in the whole domain, by opposition to edge states which are localized at the boundary. The collective dynamics model studied here has first been proposed in \cite{degondfrouvellemerino17} and later analyzed and expanded in \cite{degond2019phase, degondfrouvellemerinotrescases18, degond2018alignment}. Referred to below as the ``Body-Alignment Individual-Based Model'' (BA-IBM or IBM for short), it describes self-propelled rigid bodies moving with constant speed and trying to adjust their rigid body attitude to that of their neighbors. In \cite{degondfrouvellemerinotrescases18, degondfrouvellemerino17} the BA-IBM was based on Stochastic Differential Equations (SDE) and a macroscopic model named the ``Self-Organized Hydrodynamics for Body-orientation (SOHB)'' was derived. In \cite{degond2018alignment, degond2019phase}, SDE were replaced by Piecewise Deterministic Markov Processes (PDMP) in the IBM but the macroscopic model remained the SOHB model (with possibly different coefficients). In \cite{degond2019phase}, a variant of the  BA-IBM was shown to exhibit phase transitions which were rigorously studied. In the present work, we derive explicit solutions of the SOHB model which exhibit striking non-trivial topologies revealed by non-zero winding numbers. We explore how these non-trivial topologies are maintained at the level of the IBM by solving the PDMP of \cite{degond2018alignment}. In particular, we observe that, due to noise induced by the finite particle number, topological phase transitions from states with non-trivial topology to states with trivial one may occur and we study these phase transitions in detail. Using a set of appropriately defined topological indicators, we reveal that the breakage of the non-trivial topology requires the system to go through a phase of maximal disorder. We also show that similar but topologically trivial initial conditions result in markedly different dynamics, suggesting that topology plays a key role in the dynamics of this system. We are led to question the possible existence of topological protection against perturbations as mentioned above for topological insulators. Compared to previous works on topological states in collective dynamics, we deal with bulk states instead of edge states and we explore them at the level of the IBM and not just at the continuum level, which is closer to realistic particle systems. The present work adds a new item to the list of collective dynamics models exhibiting topological states. The topological protection concept could bring new perspectives to poorly understood questions such as the robustness of morphogenesis or the emergence of symmetries in growing organisms.

\begin{figure}[ht!]
\centering
\includegraphics[trim={4cm 19.5cm 4cm 4.5cm},clip,width=8.cm]{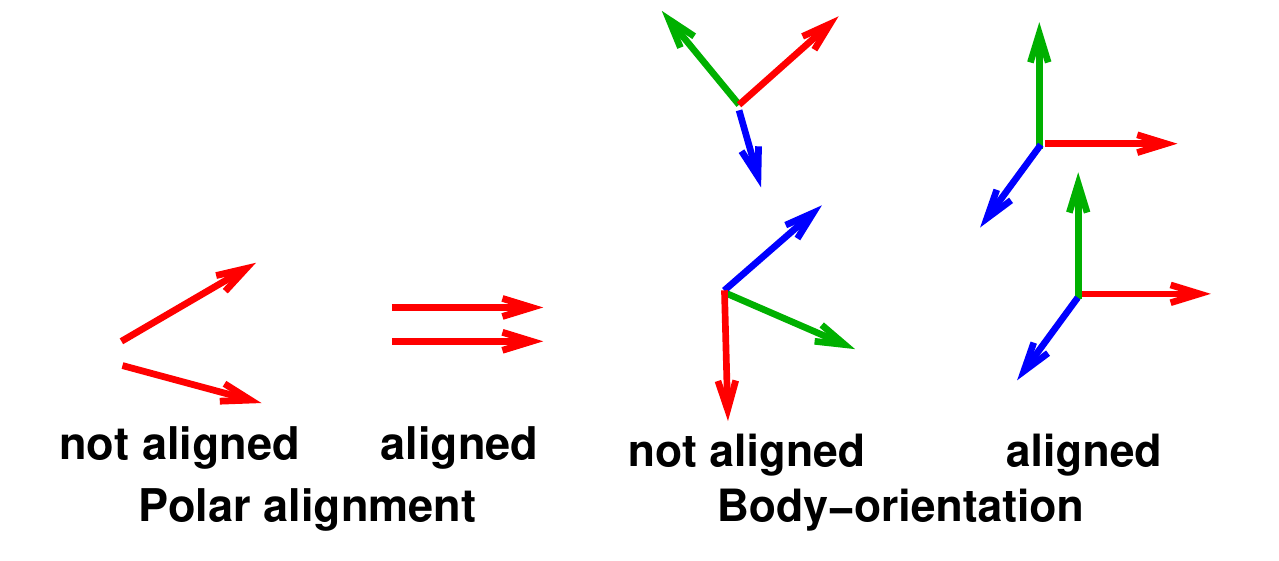} 
\caption{Vicsek model versus body-alignment model. Left:  polar alignment of velocity orientations (red vectors) of two agents. Right: alignment of body-orientations: in addition to its velocity orientation (red), each agent has two other axes (green and blue), the three vectors forming a direct orthogonal frame.}
\label{fig:vicsekvsbo}
\end{figure}

The present model belongs to the category of Vicsek-like models in the sense that it introduces a geometrical constraint within the degrees of freedom of the particles. In the Vicsek model, the particle velocities were constrained to belong to the unit sphere (after convenient normalization). In the present IBM,  the particles carry an orthonormal frame, or equivalently, a rotation matrix, that describes their body attitude. Thus their degrees of freedom are constrained to belong to the manifold SO$_3({\mathbb R})$ of $3 \times 3$ rotation matrices. Fig. \ref{fig:vicsekvsbo} highlights the difference between the Vicsek and body orientation models. The left picture shows alignment of two agents in the Vicsek sense, while the right picture shows alignment in the body-alignment sense. We mention that models involving full body attitudes have already been considered in \cite{cavagna2015flocking, hemelrijk2012schools, hemelrijk2010emergence, hildenbrandt2010self} in the context of flocking, but the alignment rules were different and essentially based on a velocity orientation (and not full body attitude) alignment.

We complete this introduction by a review of the mathematical literature on the Vicsek model and the BA-IBM. The mean-field limit of the IBM has been proven in \cite{bolley2012mean} for the Vicsek model and in \cite{diez2019propagation} for the body orientation model. Existence theory for the mean-field Vicsek model is available in \cite{briant2020cauchy, figalli2018global, gamba2016global} but the corresponding theory for the mean-field body orientation model is still open. The mean-field kinetic models exhibit phase transitions which have been studied in \cite{degond2013macroscopic, degondfrouvelleliu15, frouvelle2012dynamics} and \cite{degond2019phase} for the Vicsek and body orientation models respectively. The numerical approximation of the mean-field kinetic model has been undertaken for the Vicsek model only in \cite{gamba2015spectral, griette2019kinetic}. The derivation of macroscopic equations from the mean-field Vicsek kinetic equations has first been formally achieved in \cite{degond2008continuum} and later rigorously proved in \cite{jiang2016hydrodynamic}. Corresponding works for the body alignment model are only formal \cite{degondfrouvellemerino17, degondfrouvellemerinotrescases18, degond2018alignment}. Existence theory for the hydrodynamic models derived from the Vicsek model can be found in \cite{degond20133hydrodynamic, zhang2017local} and numerical methods in \cite{dimarcomotsch16, gamba2015spectral, motsch2011numerical}. Both questions are still open for the body orientation model.

The organization of this paper is as follows. Section \ref{sectionmodels} is devoted to the exposition of the IBM and macroscopic models. Then explicit solutions of the macroscopic model are derived in Section \ref{sec:compar} and are shown to exhibit non-trivial topology. They also serve as benchmarks to show that the macroscopic model is an accurate approximation of the IBM. But after a some time, the IBM departs from the special solutions of the macroscopic model and undergoes a topological phase transition. The study of these phase transitions require appropriate topological indicators which are developed in Section \ref{sec:tools}. Then, the topological phase transitions are analyzed in Section \ref{sectionnumericalexperiments}. A discussion and some open questions raised by these observations can be found in Section \ref{sectiondiscussion}. The supplementary material (SM) collects additional information: a list of supplementary videos (Section \ref{appendix:listvideos}), a summary of the quaternion framework (Section~\ref{sectionquaternion}), a description of the numerical methods (Section \ref{sectionsimulation}), a summary of the derivation of the macroscopic models (Section \ref{appendix:derivationmacro}) and finally a derivation of the explicit solutions presented in Section  \ref{sec:compar} (Section \ref{appendixgeneralsolutions}).


\section{Models}
\label{sectionmodels}

\subsection{The Individual-Based body-alignment Model}

\subsubsection{Description of the model}

In this section, we present the Individual-Based body-alignment Model (IBM). This model was first proposed in \cite{degond2018alignment}. We consider $N$ particles (or individuals, or agents) indexed by $k \in \{1, \ldots, N\}$ whose spatial locations are denoted by $\mathbf{X}_k(t) \in {\mathbb R}^3$ where $t \in [0,\infty)$ is the time. A direct orthonormal frame $\{\Omega_k(t),\mathbf{u}_k(t),\mathbf{v}_k(t)\}$ is attached to each particle (i.e. $\Omega_k, \, \mathbf{u}_k, \,\mathbf{v}_k \in {\mathbb S}^2$, $\Omega_k \cdot \mathbf{u}_k = 0$ and $\mathbf{v}_k = \Omega_k \times \mathbf{u}_k$). Likewise, if $(\mathbf{e}_1,\mathbf{e}_2,\mathbf{e}_3)$ is a fixed direct orthonormal reference frame, we define $A_k(t)$ to be the unique element of the special orthonormal group SO$_3({\mathbb R})$ which maps $(\mathbf{e}_1,\mathbf{e}_2,\mathbf{e}_3)$ onto $(\Omega_k(t),\mathbf{u}_k(t),\mathbf{v}_k(t))$. We will choose $(\mathbf{e}_1,\mathbf{e}_2,\mathbf{e}_3)$ once for all and write $A_k(t) = [\Omega_k(t),\mathbf{u}_k(t),\mathbf{v}_k(t)]$. This will be referred to as the local particle frame or as the particle's body orientation. $\Omega_k(t)$ is the self-propulsion direction: Particle $k$ moves in straight line in the direction of $\Omega_k$ with unchanged local frame $A_k$ except at exponentially distributed times at which the local frame jumps and  adjusts itself to the average neighbors' local frame up to some noise. The motion of the particles is thus described by the functions $[0,\infty) \ni t \mapsto (\mathbf{X}_k(t), A_k(t)) \in {\mathbb R}^3 \times \mbox{SO}_3({\mathbb R})$ for $k \in \{1, \ldots, N\}$.  

We first describe how the average neighbors' local frame is defined. We introduce a fixed observation (or sensing) kernel $K$: ${\mathbb R}^3 \ni \mathbf{x} \mapsto K(\mathbf{x}) \in [0,\infty)$. We assume that $K$ is a radial function (i.e. there exists $\tilde K$: $[0,\infty) \ni r \mapsto \tilde K(r) \in [0,\infty)$ such that $K(\mathbf{x}) = \tilde K(|\mathbf{x}|)$, where $|\mathbf{x}|$ is the euclidean norm of $\mathbf{x}$).  For a collection of~$N$ particles $\{(\mathbf{X}_k,A_k)\}_{k \in \{1, \ldots, N \}} \in ({\mathbb R}^3 \times \mbox{SO}_3({\mathbb R}))^N$, we define the local flux as the following $3 \times 3$ matrix:
$$ J_k = \frac{1}{N} \sum_{j=1}^N K(\mathbf{X}_k-\mathbf{X}_j) \, A_j.  $$
Typically, we can think of $K(\mathbf{x})$ as the indicator function of the ball centered at zero with radius $R$. In this case, $J_k$ is just the sum of the matrices $A_j$ of all particles~$j$ located within a distance $R$ to Particle $k$, divided by the total number of particles~$N$. However, more sophisticated sensing functions can be used to account for the fact that e.g. distant particles will contribute to $J_k$ less than neighboring particles. In general, $J_k$ is not a rotation matrix. To recover a rotation matrix, we need to map $J_k$ back onto the manifold SO$_3({\mathbb R})$. To do so, the space $\mathcal{M}_3({\mathbb R})$ of $3 \times 3$ matrices, is equipped with the inner product: 
\begin{equation}\label{eq:innerproduct}
A\cdot B:=\frac{1}{2} \mbox{Tr} (A^\mathrm{T}B),
\end{equation}
where Tr denotes the trace operator and $A^\mathrm{T}$ is the transpose of the matrix $A$. Now, we define the average neighbors' local frame ${\mathbb A}_k$ of Particle $k$ as follows:
\begin{equation}
\label{projection}
{\mathbb A}_k:= \mbox{arg\,max}_{A \in \mbox{\scriptsize{SO}}_3({\mathbb R})} A\cdot J_k . 
\end{equation}
This expression stands for the element ${\mathbb A}_k \in \mbox{SO}_3({\mathbb R})$ that maximizes the function $\mbox{SO}_3({\mathbb R}) \ni A \mapsto A\cdot J_k \in {\mathbb R}$. 
The maximization procedure \eqref{projection} has a unique solution as soon as $J_k$ is not singular, i.e. $\det J_k \not = 0$ where $\det$ stands for the determinant. Since the singular matrices form a zero-measure set in $\mathcal{M}_3({\mathbb R})$ it is legitimate to assume that, except for a zero-measure set of initial data, this situation will not occur. Furthermore, when $\det J_k > 0$, ${\mathbb A}_k$ is nothing but the unique rotation matrix involved in the polar decomposition of $J_k$.  

We let the particles evolve according to the following Piecewise Deterministic Markov Process (PDMP). 
\begin{itemize}
\item To each agent $k \in \{1,\ldots,N\}$ is attached an increasing sequence of random times (jump times) $T_k^1, \, T_k^2, \ldots$ such that the intervals between two successive times are independent and follow an exponential law with constant parameter $\nu>0$ (Poisson process). At each jump time $T_k^n$, the function $\mathbf{X}_k$ is continuous and the function $A_k$ has a discontinuity between its left and right states respectively denoted by $A_k(T_k^n-0)$ and $A_k(T_k^n+0)$.
\item Between two jump times $(T_k^n, T_k^{n+1})$, the evolution is deterministic: the orientation of Agent $k$ does not change and it moves in straight line at speed $c_0 > 0$ in the direction $A_k(T_k^n+0) \, \mathbf{e}_1$, i.e. for all $t\in[T_k^n,T_k^{n+1})$, we have 
\begin{equation}
\label{deterministicpart}
\mathbf{X}_k(t) = \mathbf{X}_k(T_k^n) + c_0 \, (t-T_k^n) \, A_k(t) \, \mathbf{e}_1,\,\,\,A_k(t)=A_k(T_k^n+0).
\end{equation}
\item To compute $A_k(T_k^n+0)$ from $A_k(T_k^n-0)$,  
we compute the local flux defined at time $T_k^n-0$ given by:
\begin{equation}\label{eq:flux}
J_k^{n-} := \frac{1}{N} \sum_{j=1}^N K \big( \mathbf{X}_k(T_k^n)- \mathbf{X}_j(T_k^n) \big) A_j(T_k^n-0),
\end{equation}
having in mind that $A_j(T_k^n-0) = A_j(T_k^n)$ for $j \not = k$. From $J_k^{n-}$, which we assume is a non-singular matrix, we compute ${\mathbb A}_k^n$ as the unique solution of the maximization problem \eqref{projection} (with $J_k$ replaced by $J_k^{n-}$). 
Then, $A_k(T_k^n+0)$ is drawn from a von Mises distribution: 
\begin{equation}
\label{jumppart}
A_k(T_k^n+0) \sim M_{{\mathbb A}_k^n}.
\end{equation}
The von Mises distribution on SO$_3({\mathbb R})$ with parameter ${\mathbb A} \in$ SO$_3({\mathbb R})$ is defined to be the probability density function: 
\begin{equation}
\label{eq:vonmises}
M_{{\mathbb A}}(A):=\frac{\mathrm{e}^{\kappa {\mathbb A}\cdot A}}{\int_{\mbox{{\scriptsize SO}}_3({\mathbb R})} \mathrm{e}^{\kappa {\mathbb A}\cdot A'} \mathrm{d} A'},
\end{equation} 
where $\kappa>0$ is a supposed given parameter named concentration parameter, or inverse of the noise intensity. The von Mises distribution, also known in the literature as the matrix Fisher distribution \cite{kent2018new,lee2018bayesian}, is an analog (in the case of SO$_3({\mathbb R})$) of the Gaussian distribution in a flat space. The new orientation of Agent $k$ at time $T_n$ can therefore be interpreted as a small random perturbation of the average local orientation given by ${\mathbb A}_k^n$, where the perturbation size is measured by $1/\sqrt{\kappa}$. 
\end{itemize}

\begin{figure}[ht!]
  \begin{center}
    \includegraphics[trim={4cm 11cm 1cm 4cm},clip,width=10cm]{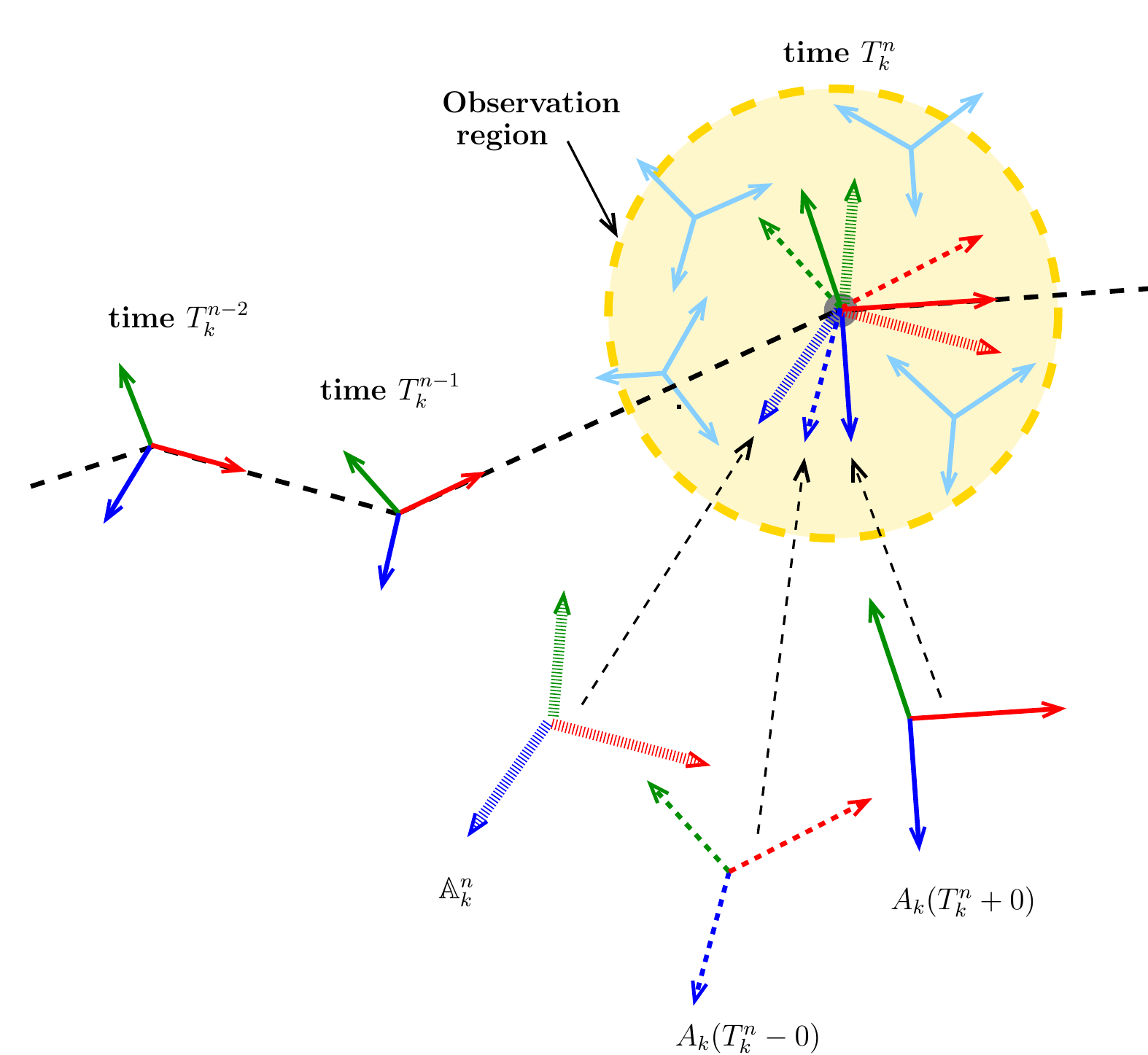} 
    \caption{Schematic representation of the PDMP described in the text: the motion of Particle~$k$ is represented in physical space as the black broken dotted line. The body frame $A_k$ is represented with $\Omega_k$ in red, $\mathbf{u}_k$ in green and $\mathbf{v}_k$ in blue. Each angular point of the trajectory corresponds to one of the jump times $T_k^n$. Between two jump times, the trajectory is the straight line spanned by $\Omega_k$ and the body frame stays constant. The jump dynamics is depicted at time $T_k^n$. At this time, the observation region is colored in yellow and body frames of the other particles present in this region are depicted in light blue. The averaged body frame ${\mathbb A}_k^n$ is depicted with thick lightly colored arrows. The body frame before the jump $A_k(T_k^n-0)$ is drawn in broken lines whereas that after the jump $A_k(T_k^n+0)$ is drawn in plain lines. $A_k(T_k^n+0)$ is close, but not equal to ${\mathbb A}_k^n$ because of the noise intensity proportional to $1/\kappa$. For clarity, the frames involved in the description of the jump are magnified.}
    \label{fig:PDMP3}
  \end{center}
\end{figure}

In Formula \eqref{eq:vonmises} and in the remainder of this paper, the manifold SO$_3({\mathbb R})$ is endowed with its unique normalized Haar measure defined for any test function $\varphi$ by: 
\begin{equation}
\label{eq:haar}
\int_{\mbox{{\scriptsize SO}}_3({\mathbb R})} \varphi(A) \, \mathrm{d} A := \frac{2}{\pi} \int_0^\pi \int_{\mathbb{S}^2} \varphi({\mathcal A}(\theta,\mathbf{n})) \, \sin^2(\theta/2)\, \mathrm{d} \theta \, \mathrm{d} \mathbf{n},
\end{equation}
where $\mathrm{d} \mathbf{n}$ is the uniform probability measure on the sphere $\mathbb{S}^2$. Here, a rotation matrix $A \equiv {\mathcal A}(\theta,\mathbf{n})$ is parametrized by its rotation angle $\theta\in[0,\pi]$ and its axis $\mathbf{n} \in \mathbb{S}^2$ through Rodrigues' formula: 
\begin{equation}
\label{eq:rodrigues}
{\mathcal A}(\theta,\mathbf{n}) := I_3 + \sin\theta \, [\mathbf{n}]_\times + (1 - \cos\theta) \, [\mathbf{n}]_\times^2 = \exp (\theta[\mathbf{n}]_\times) 
\end{equation}
with $\mathbf{n}=(n_1,n_2,n_3)^\mathrm{T}$ and $I_3$ is the $3 \times 3$ identity matrix. For any vector $\mathbf{w} = (w_1,w_2,w_3)^\mathrm{T} \in {\mathbb R}$, $[\mathbf{w}]_\times$ is the antisymmetric matrix of the linear map ${\mathbb R}^3 \ni \mathbf{u} \mapsto \mathbf{w} \times \mathbf{u}$ (where $\times$ denotes the cross product) which has the following expression:
\begin{equation} 
[\mathbf{w}]_\times :=\left(\begin{array}{ccc} 0 & -w_3 & w_2 \\ w_3 & 0 & -w_1 \\ -w_2 & w_1 & 0\end{array}\right).
\label{eq:def[]x}
\end{equation}
Additional details on the structure of $SO_3({\mathbb R})$ can be found for instance in \cite{huynh2009metrics}. The IBM \eqref{deterministicpart}, \eqref{jumppart} is schematically represented in Fig. \ref{fig:PDMP3}. 

\subsubsection{Numerical simulations of the IBM}
\label{subsec:ibm_num_sim}

Unless otherwise specified, throughout this paper, a square box of side length $L$ with periodic boundary conditions is used. As sensing kernel $K$, we use the indicator function of the ball centered at $0$ and of radius $R$. Thus, an agent interacts with all its neighbors at a distance less than $R$ (radius of interaction). Table~\ref{tab_parameters_IBM} summarizes the model parameters.

\begin{table}[ht!]
\begin{center}
\begin{tabular}{|l|c|}
\hline
\hline
Parameter                                  & Symbol        \\
\hline
\hline
Number of particles                        &$N$            \\
\hline
Computational box side length              &$L$            \\
\hline
Interaction radius                         &$R$            \\
\hline
Particle speed                             &$c_0$          \\
\hline
Concentration parameter                    &$\kappa$            \\
\hline
Alignment frequency                        &$\nu$          \\
\hline
\hline
\end{tabular}
\caption{Parameters of the IBM \eqref{deterministicpart}, \eqref{jumppart}.}
\label{tab_parameters_IBM}
\end{center}
\end{table}

For the numerical simulations presented in this paper, we have used the convenient framework offered by quaternions. Indeed, there is a group isomorphism between $\mathrm{SO}_3({\mathbb R})$ and ${\mathbb H}/\{\pm 1\}$ where ${\mathbb H}$ is the group of unit quaternions. We can express the IBM \eqref{deterministicpart}, \eqref{jumppart} using this representation (see \cite{degond2018alignment} and Section \ref{sectionquaternion}). Roughly speaking, body-alignment as described here is equivalent to nematic alignment of the corresponding quaternions (nematic alignment of a unit quaternions $\mathbf{q}$ to the mean direction $\mathbf{Q}$ is unchanged if $\mathbf{q}$ is replaced by $-\mathbf{q}$, as opposed to polar alignment where the result depends on the sign of $\mathbf{q}$). This is because a given rotation can be represented by two opposite quaternions and thus, the outcome of the alignment process should not depend of the choice of this representative. The numerical algorithm is described in Section \ref{sectionsimulation}. Additionally, the quaternion framework also suggests to use order parameters derived from nematic alignment dynamics (such as in liquid crystal polymers). We shall use this analogy to define appropriate order parameters in Section \ref{sec:order}. 

All the simulations were written in Python using the SiSyPHE library \cite{diez2021sisyphe} specifically developed for the simulation of large-scale mean-field particle systems by the second author. The implementation is based on the PyTorch \cite{NEURIPS2019_9015} library and more specifically on the GPU routines introduced by the KeOps \cite{charlier2020kernel} library. The computational details as well as the source code are freely available on the documentation website \url{https://sisyphe.readthedocs.io/}. The outcomes of the simulations were analyzed and plotted using the NumPy \cite{harris2020array} and Matplotlib~\cite{Hunter:2007} libraries. The 3D particle plots were produced using VPython \cite{vpython}. All the particle simulations have been run on a GPU cluster at Imperial College London using an Nvidia GTX 2080 Ti GPU chip.  

\begin{figure}[ht!]
\centering
\subfloat[Graphical representation of particles]{\includegraphics[trim={3.5cm 13.5cm 1.5cm 3.5cm},clip,width=3.8cm]{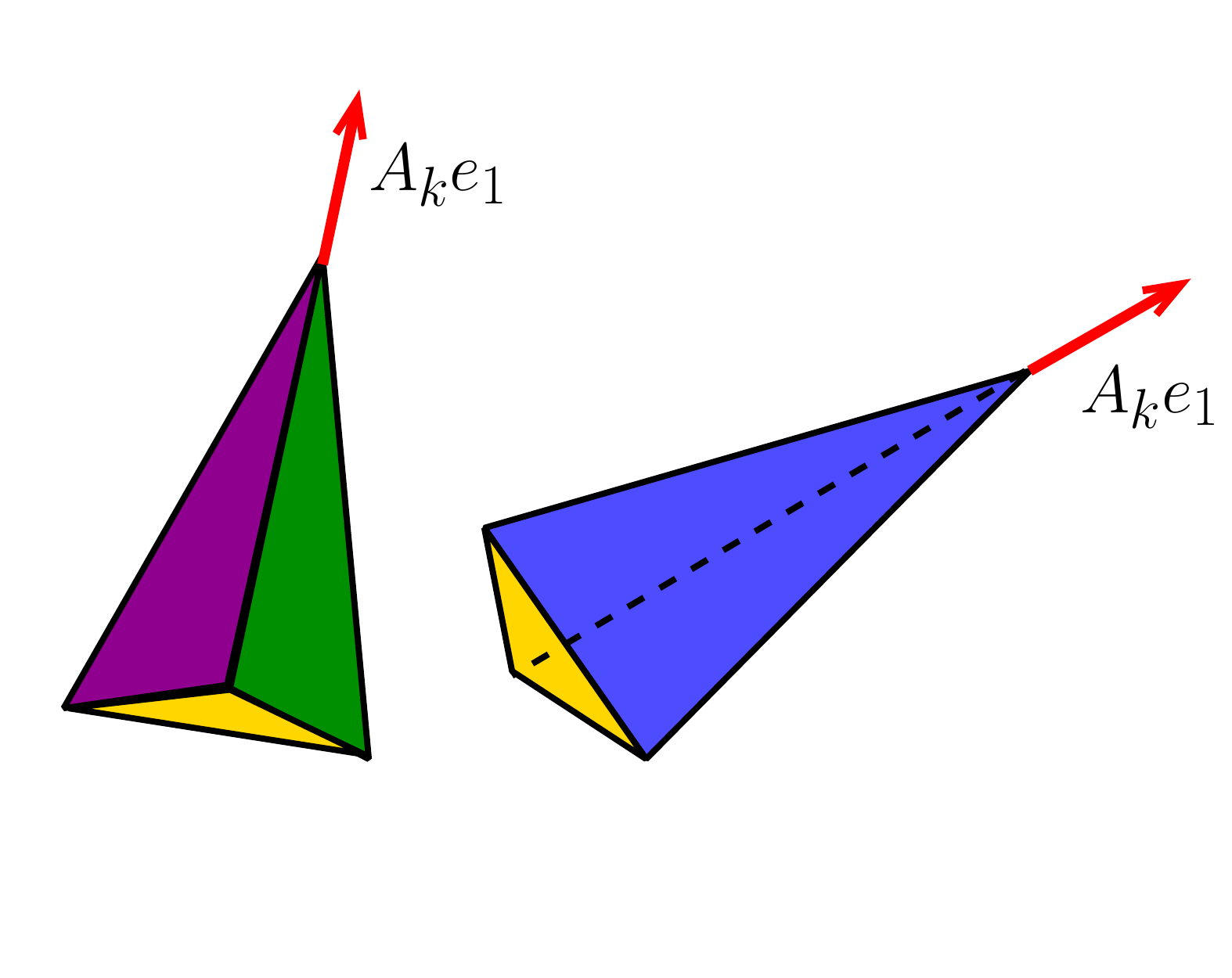}\label{fig:tetra}}
\subfloat[Time=0]{\includegraphics[width= 3.8cm]{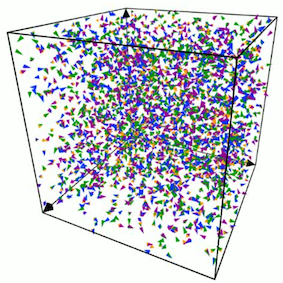}\label{fig:example0}}
\subfloat[Time=4]{\includegraphics[width= 3.8cm]{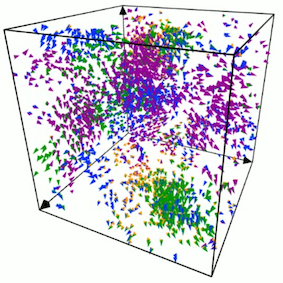}\label{fig:example4}}
\subfloat[Time=40]{\includegraphics[width= 3.8cm]{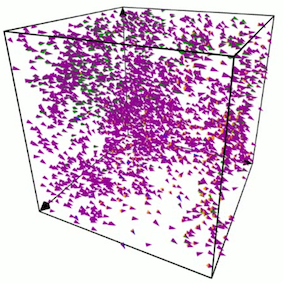}\label{fig:example40}}
\caption{(a) Graphical representation of particles and their body orientations as elongated tetrahedra pointing towards the self-propulsion direction with blue, magenta and green large faces and gold bases. (b,c,d) Snapshots of a typical output of the simulation at three different times (b) Time=0, (c) Time=4 and (d) Time=40. Parameters: $N=3000$, $L=1$, $R=0.075$, $\kappa = 20$, $\nu=5$, $c_0=0.2$. see also Section \ref{appendix:listvideos}, Video \ref{vid:example}.}
\label{figureexample}
\end{figure}

A typical outcome of the IBM is shown in Figure \ref{figureexample} (see also Section \ref{appendix:listvideos}, Video \ref{vid:example}) for a moderate number of particles ($N=3000$). Throughout this paper, in the plots, we will represent each agent graphically by an elongated tetrahedron pointing in the direction of motion. The three large faces around the height will be painted in blue, green and magenta and the base will be in gold, as described in Fig. \ref{fig:tetra}. We notice that, starting from a uniformly random initial state (Fig. \ref{fig:example0}), the system self-organizes in small clusters (Fig. \ref{fig:example4}) and finally reaches a flocking equilibrium where all the agents have roughly the same body-orientation (Fig. \ref{fig:example40}). We will see below that flocking is not necessarily the ultimate fate of the system, because it may be trapped in a so-called topologically protected state. To better understand these aspects, we first need to develop the continuum (or macroscopic) description of the system. This is done in the next section. 

\subsubsection{Relation with other collective dynamics models}
\label{subsec:ibm_relation}

We finally make a comparison with previous models. First, there is a version of the IBM where particles follow a stochastic differential equation (SDE) instead of a jump process \cite{degondfrouvellemerino17, degondfrouvellemerinotrescases18}. Both the current and previous models have the same hydrodynamic model as macroscopic limit (see forthcoming section). There are two reasons for us to prefer the jump process. First, its simulation is slightly easier and second, the coefficients of the macroscopic model are explicit, which is not so in the SDE case where they require the resolution of an auxiliary elliptic problem \cite{degondfrouvellemerino17, degondfrouvellemerinotrescases18}. 

Beyond the present body-orientation model, numerous models of self-propelled particles have been proposed in the literature (see the review \cite{vicsek2012collective}). The most closely related one is the celebrated Vicsek model \cite{vicsek1995novel}. There are several versions of this model: time-discrete ones \cite{chate2008collective, vicsek1995novel}, time-continuous ones relying on an SDE description of the particle trajectories \cite{degond2008continuum} and time-continuous ones using a jump process instead \cite{dimarcomotsch16}. The latter version is the most closely related to the present work. In \cite{dimarcomotsch16}, the difference is that particles carry a single direction vector $\Omega_k$ instead of a whole body frame. This vector gives the direction of self-propulsion. The particles follow a similar PDMP, namely 
\begin{itemize}
\item The random jump times are defined in the same way: they follow an exponential law with constant parameter $\nu>0$. At jump times, the position is continuous and the direction vector $\Omega_k$ is discontinuous with left and right states respectively denoted by $\Omega_k(T_k^n-0)$ and $\Omega_k(T_k^n+0)$.
\item Between two jump times $T_k^n$, $T_k^{n+1}$, the direction vector $\Omega_k$ does not change and the particle moves in straight line at speed $c_0 > 0$ in the direction given by $\Omega_k(T_k^n+0)$.  
\item To pass from $\Omega_k(T_k^n-0)$ to $\Omega_k(T_k^n+0)$,  we compute the local flux given by $\mathbf{J}_k^{n-} =$ $\frac{1}{N} \sum_{j=1}^N K \big( \mathbf{X}_k(T_k^n)- \mathbf{X}_j(T_k^n) \big) \, \Omega_j(T_k^n-0) \in {\mathbb R}^3$ and, assuming that it is non-zero, the mean direction $\bar \Omega_k^n = \mathbf{J}_k^{n-}/|\mathbf{J}_k^{n-}| \in {\mathbb S}^2$ at time $T_k^n-0$. Then, $\Omega_k(T_k^n+0)$ is drawn from a von Mises distribution on ${\mathbb S}^2$: $\Omega_k(T_k^n+0) \sim \tilde M_{\bar \Omega_k^n}$, with $\tilde M_{\bar \Omega}(\Omega) = e^{\kappa (\bar \Omega \cdot \Omega)} / \int_{{\mathbb S}^2} e^{\kappa (\bar \Omega \cdot \Omega)} \, \mathrm{d} \Omega$, for $\Omega$ and $\bar \Omega$ in ${\mathbb S}^2$. 
\end{itemize}

So, the current model is an elaboration of \cite{dimarcomotsch16} replacing self-propulsion directions by whole body frames and polar alignment of unit vectors (as expressed by the von Mises distribution on the sphere) by alignment of rotations matrices. Outcomes of numerical simulations of the Vicsek model do not show striking differences whether one uses any of the above mentioned versions (time-discrete, time-continuous with SDE or time-continuous with jump process). Results given in \cite{chate2008collective, vicsek1995novel} for the time-discrete version display the emergence of a global alignment together with the formation of clusters when the noise intensity $1/\kappa$ is not too big. The outcome strongly resembles what is shown in Fig. \ref{figureexample} for the body-orientation model, but for the depiction of the body orientation itself which is not provided by the Vicsek model. So, it is legitimate to wonder whether the inclusion of the full body orientation instead of the mere self-propulsion direction makes any change in the dynamics of the particle positions and direction vectors. In particular, do the particle positions and directions follow the same dynamics in the Vicsek and body orientation model? We will see below that this is not the case and that in certain circumstances, striking differences between the two models are obtained. To show this, the use of the macroscopic limit of the IBM, as developed in the forthcoming section, will be of crucial importance.


\subsection{The macroscopic body-alignment model}
\label{sectionmacro}

\subsubsection{Description of the model}

As soon as $N$ is not very small, the IBM \eqref{deterministicpart}, \eqref{jumppart} involves a large number of unknowns which makes its mathematical analysis virtually impossible. A reduced description, more amenable to mathematical analysis, is obtained through the macroscopic limit of the IBM, and consists of a system of partial differential equations. This reduced description gives a valid approximation of the IBM in an appropriate range of parameters, namely 
\begin{equation}
N \gg 1, \qquad \frac{R}{L} \sim \frac{c_0}{\nu \, L} \ll 1.
\label{macroscalingnumeric}
\end{equation} 
Throughout the remainder of this paper, we will focus on this regime. The macroscopic limit of the IBM \eqref{deterministicpart}, \eqref{jumppart} has first been proposed in \cite{degond2018alignment} and leads to a model called ``Self-Organized Hydrodynamics for Body orientation (SOHB)''. The derivation relies on earlier work \cite{degondfrouvellemerino17, degondfrouvellemerinotrescases18}. This derivation is ``formally rigorous'' in the sense that, if appropriate smoothness assumptions are made on the involved mathematical objects, the limit model can be identified rigorously as being the SOHB. For the reader's convenience, we summarize the main steps of this mathematical result in Section \ref{appendix:derivationmacro}.

The unknowns in the SOHB are the particle density $\rho(t,\mathbf{x})$ and mean body-orientation ${\mathbb A}(t,\mathbf{x})\in$ SO$_3({\mathbb R})$ at time $t$ and position $\mathbf{x}=(x,y,z) \in {\mathbb R}^3$. They satisfy the following set of equations:
\begin{subequations}\label{equationA}
\begin{align}
& \partial_t \rho + c_1 \, \nabla_\mathbf{x} \cdot (\rho \, {\mathbb A} \mathbf{e}_1)=0,\label{equationrhoA}\\
& \big(\partial_t + c_2 ({\mathbb A} \mathbf{e}_1) \cdot \nabla_\mathbf{x} \big) {\mathbb A} + \big[ ({\mathbb A} \mathbf{e}_1) \times ( c_3 \nabla_\mathbf{x} \log \rho + c_4 \, \mathbf{r}) + c_4 \, \, \delta \, {\mathbb A} \mathbf{e}_1 \big]_\times {\mathbb A} = 0.  \label{equationbigA}
\end{align}
\end{subequations}
The quantities $\mathbf{r}$ and $\delta$ have intrinsic expressions in terms of ${\mathbb A}$ \cite{degondfrouvellemerino17}. However, it is more convenient to write the rotation field ${\mathbb A}$ in terms of the basis vectors 
$$ \Omega = {\mathbb A} \mathbf{e}_1, \quad \mathbf{u}= {\mathbb A} \mathbf{e}_2, \quad \mathbf{v} = {\mathbb A} \mathbf{e}_3.$$
With these notations, the vector $\mathbf{r}(t,\mathbf{x}) \in {\mathbb R}^3$ and scalar $\delta(t,\mathbf{x}) \in {\mathbb R}$ fields are defined by
\begin{eqnarray}
\mathbf{r} &:=& (\nabla_\mathbf{x} \cdot \Omega) \, \Omega + (\nabla_\mathbf{x}\cdot \mathbf{u}) \, \mathbf{u} + (\nabla_\mathbf{x} \cdot \mathbf{v}) \, \mathbf{v},  \label{eq:def_r} \\
\delta &:=& [(\Omega \cdot \nabla_\mathbf{x})\, \mathbf{u}] \cdot \mathbf{v} + [(\mathbf{u} \cdot \nabla_\mathbf{x}) \mathbf{v}] \cdot\Omega + [(\mathbf{v} \cdot \nabla_\mathbf{x}) \Omega] \cdot \mathbf{u}. \label{eq:def_delta}
\end{eqnarray}
Here, for a vector field $\mathbf{B}(\mathbf{x}) \in {\mathbb R}^3$ and a scalar field $\lambda(\mathbf{x}) \in {\mathbb R}$ we denote by $\nabla_\mathbf{x} \cdot \mathbf{B}$, and $\nabla_\mathbf{x} \times \mathbf{B}$ the divergence and curl of $\mathbf{B}$ respectively, by $\nabla_\mathbf{x} \lambda$, the gradient of $\lambda$ and we set $(\mathbf{B} \cdot \nabla_\mathbf{x}) \lambda = \mathbf{B} \cdot \nabla_\mathbf{x} \lambda$ with $\cdot$ the inner product of vectors in ${\mathbb R}^3$. We remind that $\times$ denotes the cross product and we refer to formula \eqref{eq:def[]x} for the definition of $[\mathbf{w}]_\times$ when $\mathbf{w}$ is a vector in ${\mathbb R}^3$. Alternate expressions of $\delta$ can be found in Section \ref{section_delta_alternate} of the Supplementary Material. 

\begin{figure}[ht!]
\centering
\includegraphics[width= 5cm]{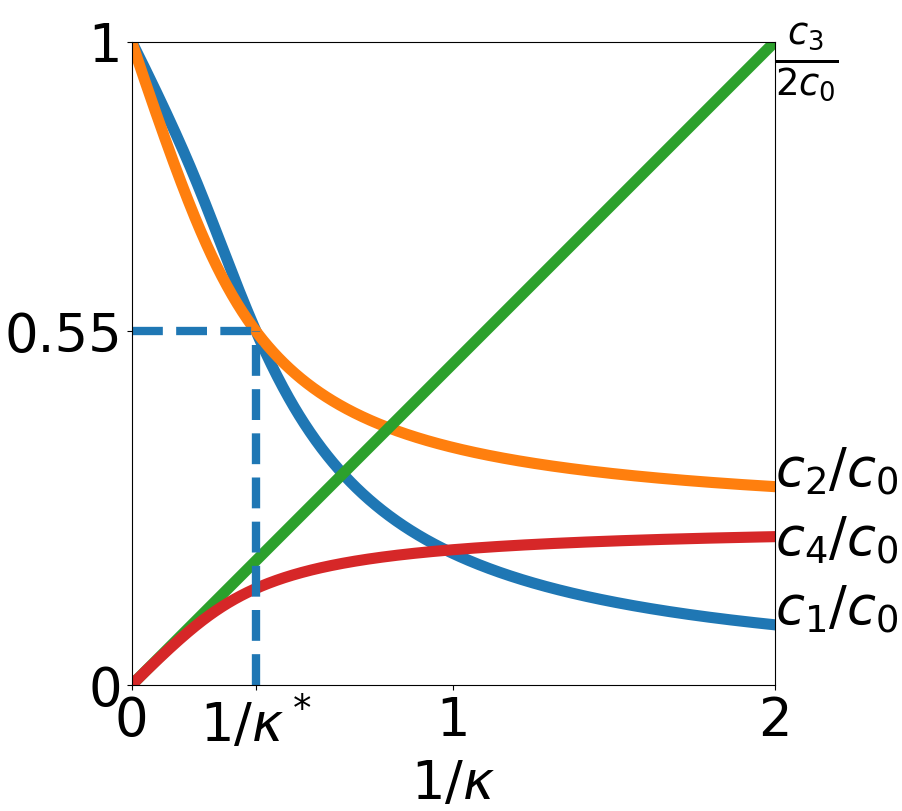}
\caption{Dimensionless coefficients $c_i/c_0$ as functions of the inverse of concentration parameter $1/\kappa$. Blue curve $c_1/c_0$, orange curve $c_2/c_0$, green curve $c_3/2c_0$ and red curve $c_4/c_0$. At the crossover value $\kappa^*\simeq 2.58$, the sign of $c_2 - c_1$ changes (see Section \ref{sec:properties_solutions}).}
\label{fig:ci}
\end{figure}

The quantities $c_1$, $c_2$, $c_3$, $c_4$ are functions of $\kappa$ and $c_0$ given as follows: 
\begin{align}
\frac{c_1}{c_0} &= \frac{2}{3} \, \big\langle \frac{1}{2} + \cos \theta \big\rangle_{\exp \left( \kappa \left(\frac{1}{2} + \cos \theta \right) \right) \, \sin^2 \left(\frac{\theta}{2} \right)}, \label{c1}\\
\frac{c_2}{c_0} &= \frac{1}{5} \, \left\langle 2 + 3 \cos \theta \right\rangle_{\exp \left( \kappa \left( \frac{1}{2} + \cos \theta \right) \right) \, \sin^4 \left(\frac{\theta}{2} \right) \, \cos^2 \left(\frac{\theta}{2} \right)}, \label{c2}\\
\frac{c_3}{c_0}&= \frac{1}{\kappa}, 
\\
\frac{c_4}{c_0} &= \frac{1}{5} \, \left\langle 1 - \cos \theta \right\rangle_{\exp \left( \kappa \left( \frac{1}{2} + \cos \theta \right) \right) \, \sin^4 \left(\frac{\theta}{2} \right) \, \cos^2 \left(\frac{\theta}{2} \right)}, \label{c4}
\end{align}
where, for two functions $f$ and $g$: $[0,\pi] \to {\mathbb R}$, we write 
$$ \langle f \rangle_g = \frac{ \int_0^\pi f(\theta) \, g(\theta) \, \mathrm{d} \theta}{ \int_0^\pi g(\theta) \, \mathrm{d} \theta}. $$
Fig. \ref{fig:ci} provides a graphical representation of these functions. 

\subsubsection{Interpretation of the model}

To better understand what the SOHB system \eqref{equationA} does, we re-write it as follows:
\begin{subequations}
\label{equationA'}
\begin{align}
& \partial_t \rho + c_1 \, \nabla_\mathbf{x} \cdot (\rho \, \Omega)=0, \label{equationrho}\\
& D_t {\mathbb A} +  [\mathbf{w}]_\times \, {\mathbb A} = 0, \label{equationbigA'} 
\end{align}
\end{subequations}
where the convective derivative $D_t$ and the vector $\mathbf{w}$ are given by:
\begin{align}
& D_t = \partial_t + c_2\Omega\cdot\nabla_\mathbf{x}, \label{eq:Dt} \\
& \mathbf{w} = -\Omega \times \mathbf{F} + c_4 \, \delta \, \Omega, \quad \mbox{ with } \quad \mathbf{F} = -c_3 \, \nabla_\mathbf{x} \, \log \rho - c_4 \, \mathbf{r},   
\label{eq:defw_F}
\end{align}

Eq. \eqref{equationrho} is the mass conservation equation of the fluid. The vector $\Omega$ gives the direction of the fluid motion. The fluid velocity deduced from \eqref{equationrho} is $c_1\Omega$. Since  $c_1/c_0 \in [0,1]$ as can be seen from Fig. \ref{fig:ci} (see also \cite{degondfrouvellemerino17} for a rigorous proof), the fluid motion is oriented positively along $\Omega$ and its magnitude is smaller than the particles self-propulsion velocity $c_0$. This is because the average of vectors of identical norms has smaller norm. The quantity $c_1/c_0$ can be seen as an order parameter \cite{degond2019phase} but we will not dwell on this issue here. 

Eq. \eqref{equationbigA'} provides the rate of change of ${\mathbb A}$ with time along the integral curves of the vector field $c_2\Omega$ as expressed by the convective derivative $D_t$. Note that this vector field is not the fluid velocity $c_1\Omega$ since $c_2 \not = c_1$. It can be interpreted as the propagation velocity of ${\mathbb A}$ when $\mathbf{w}$ is zero. Since $D_t {\mathbb A}$ is the derivative of an element of SO$_3({\mathbb R})$, it must lie in the tangent space to SO$_3({\mathbb R})$ at ${\mathbb A}$ which consists of all matrices of the form ${\mathbb W} \, {\mathbb A}$ with ${\mathbb W}$ antisymmetric. This structure is indeed satisfied by Eq. \eqref{equationbigA'} since, from the definition \eqref{eq:def[]x}, the matrix $[\mathbf{w}]_\times$ is antisymmetric. It can be shown that the SOHB system is hyperbolic \cite{degond2021hyperbolicity}. 

\begin{figure}[ht!]
\centering
\subfloat[Action of $\Omega \times {\mathbf F}$]{\includegraphics[trim={3.5cm 16cm 5.5cm 4.5cm},clip,width=7.cm]{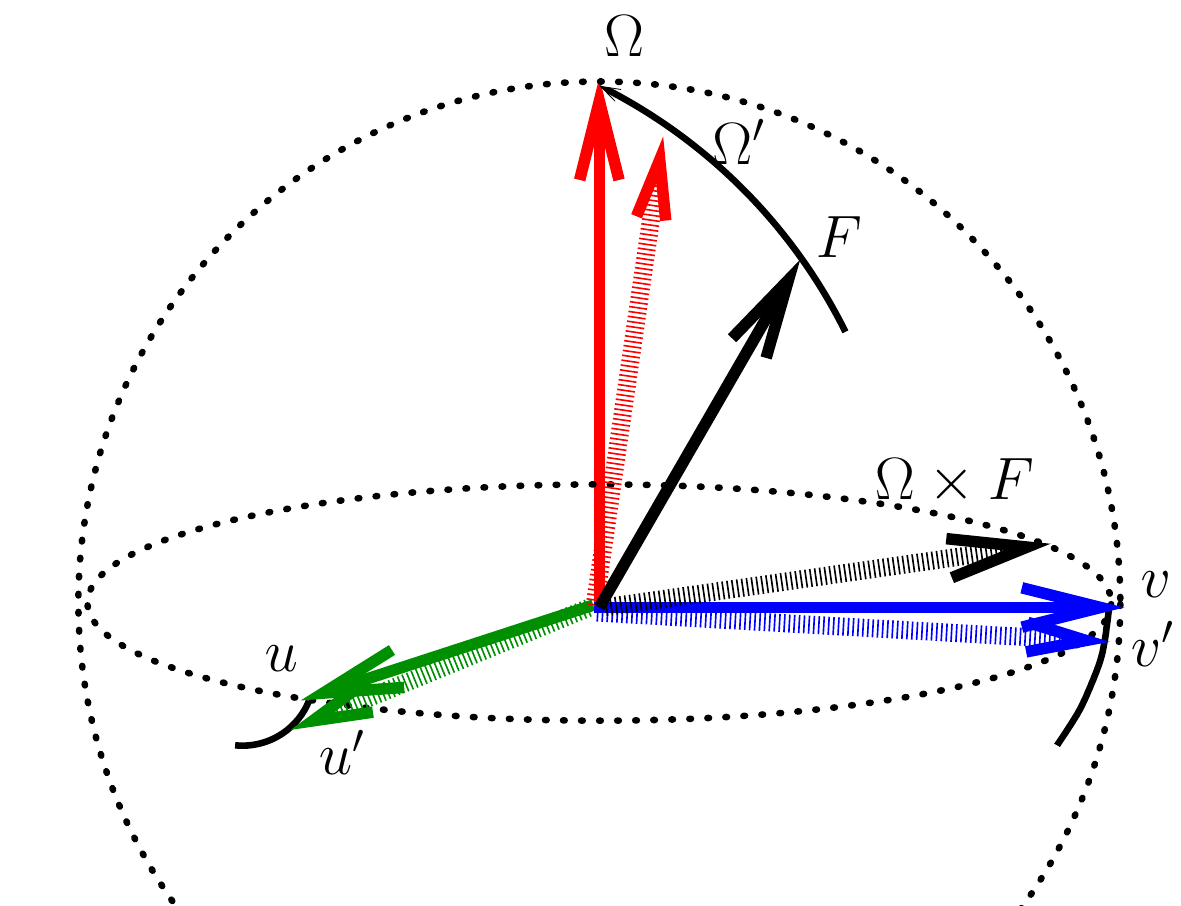} 
}
\subfloat[Action of $- c_4 \delta \Omega$]{\includegraphics[trim={3.5cm 16.5cm 5.5cm 4.5cm},clip,width=7.cm]{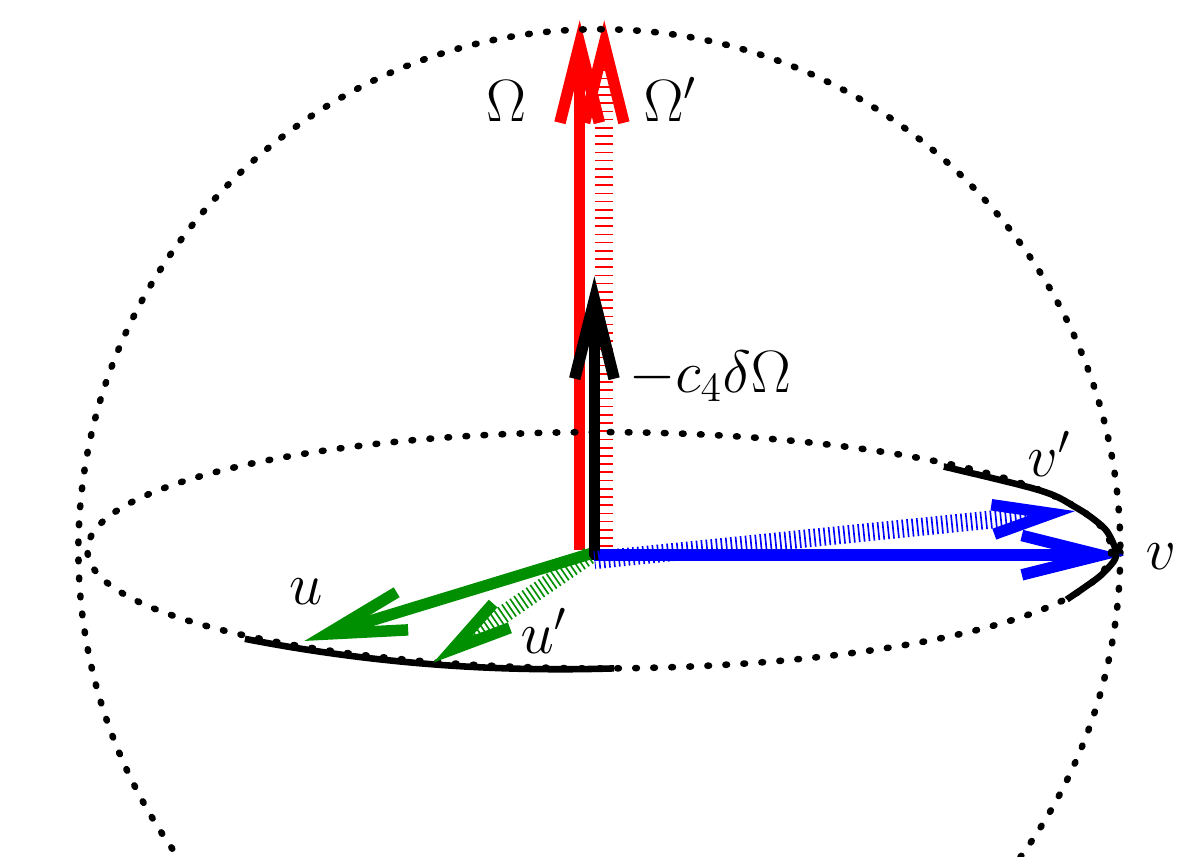} 
}
\caption{Graphical representations of the two components of the infinitesimal rotation. 
$(\Omega, {\mathbf u}, {\mathbf v})$ denotes the position of the frame at time $t$ while $(\Omega', {\mathbf u}', {\mathbf v}')$ is its position at time $t + dt$ with $dt \ll 1$. The frame at time $t$ is denoted in plain colors (red for $\Omega$, green for $\mathbf{u}$ and blue for $\mathbf{v}$) while that at time $t + dt$ is in light colors. The motion of the vectors is indicated by a segment of circle in black color. 
(a) Action of $\Omega \times {\mathbf F}$: the vectors ${\mathbf F}$ and $\Omega \times {\mathbf F}$ are in plain and light black respectively. The vector ${\mathbf F}$ is shown with unit norm for the ease of the representation but could be of any norm in reality. The passage from $(\Omega, {\mathbf u}, {\mathbf v})$ to $(\Omega', {\mathbf u}', {\mathbf v}')$ is via an infinitesimal rotation of axis $\Omega \times {\mathbf F}$. (b) Action of $\delta$: the vector $- c_4 \delta \Omega$ is shown in black. The vectors $\Omega$ and $\Omega'$ are identical and collinear to $- c_4 \delta \Omega$. The passage from $(\Omega, {\mathbf u}, {\mathbf v})$ to $(\Omega', {\mathbf u}', {\mathbf v}')$ is via an infinitesimal rotation of axis $\Omega$.}
\label{fig:action_F_del}
\end{figure}

In fact, Eq. \eqref{equationbigA'} shows that the vector $\mathbf{w}$ is the instantaneous rotation vector of the frame ${\mathbb A}(t, \mathbf{X}(t))$, where $t \mapsto \mathbf{X}(t)$ is any solution of $\frac{\mathrm{d} \mathbf{X}}{\mathrm{d} t} = c_2 \, \Omega (t, \mathbf{X}(t))$. Indeed, Eq. \eqref{equationbigA'} can be equivalently written as a system of equations for $(\Omega, \mathbf{u}, \mathbf{v})$ of the form $D_t \mathbf{Z} = \mathbf{w} \times \mathbf{Z}$, with ${\mathbf Z} = \Omega, \, {\mathbf u}, \, {\mathbf v}$. This describes a rigid body rotation of the frame $\{\Omega, \mathbf{u}, \mathbf{v}\}$  with angular velocity $\mathbf{w}$. The rotation vector $\mathbf{w}$ has two components. The first one is $\Omega \times {\mathbf F}$ and tends to relax $\Omega$ towards ${\mathbf F}$. Due to its expression \eqref{eq:defw_F}, the force ${\mathbf F}$ includes two contributions: that of the pressure gradient $-c_3 \, \nabla_\mathbf{x} \, \log \rho $ and that of gradients of the body orientation through the vector~$- c_4 \, \mathbf{r}$. The second component of the rotation vector is $- c_4 \delta \Omega$ and corresponds to a rotation of the body frame about the self propulsion direction $\Omega$ driven by gradients of the body orientation through the scalar $- c_4 \, \delta$. The contributions of gradients of body orientation in the two components of the rotation vector are under the control of the single coefficient $c_4$. Fig. \ref{fig:action_F_del} gives a graphical representation of the actions of these two infinitesimal rotations. 

\subsubsection{Relation with other models}

To better understand how the SOHB model \eqref{equationA} relates to other models, we re-write the equation for $\Omega$ as follows: 
\begin{equation}
D_t\Omega = \mathrm{P}_{\Omega^\perp} \mathbf{F} , \label{equationOmega} 
\end{equation}
where $\mathrm{P}_{\Omega^\perp}$ is the $3 \times 3$ projection matrix on the orthogonal plane to the vector $\Omega$ and is written $\mathrm{P}_{\Omega^\perp} = \mbox{I}_3 - \Omega \otimes \Omega$ with $\otimes$ standing for the tensor (or outer) product. Eq. \eqref{equationOmega} bears similarities and differences with the momentum equation of isothermal compressible fluids. The latter is exactly recovered if the following three modifications are made: 
\begin{enumerate}
\item the projection matrix $\mathrm{P}_{\Omega^\perp}$ is removed from \eqref{equationOmega} (i.e. it is replaced by I$_3$); 
\item  $c_2=c_1$ in the convective derivative $D_t$ (see \eqref{eq:Dt}); 
\item $c_4=0$ in the expression of $\mathbf{F}$ (see \eqref{eq:defw_F}). 
\end{enumerate}
Indeed, under these three modifications, we get the following system for $(\rho,\mathbf{U})$ where~$\mathbf{U}= c_1 \Omega$ is the fluid velocity: 
\[\partial_t \rho + \nabla_\mathbf{x} \cdot (\rho \mathbf{U})=0, \quad (\partial_t + \mathbf{U} \cdot\nabla_\mathbf{x}) \mathbf{U} = - \Theta \, \nabla_\mathbf{x} \, \log \rho.\]
This is the isothermal compressible Euler equations with the fluid temperature $\Theta = c_1 \, c_3$. 

We now investigate what consequences follow from undoing the above three modifications, one by one. 

\begin{enumerate}
\item Introducing the projection $\mathrm{P}_{\Omega^\perp}$ in \eqref{equationOmega} guarantees that the constraint $|\Omega|=1$ is preserved in the course of time, if it is satisfied at time $0$. Indeed, dotting Eq. \eqref{equationOmega} with $\Omega$ (and assuming that all functions are smooth) leads to $D_t |\Omega|^2 = 0$, which guarantees that $|\Omega|$ is constant along the integral curves of the vector field $c_2 \Omega$. Thus, if $|\Omega|=1$ at time $t=0$, it will stay so at any time. 

\item Having $c_2 \not = c_1$ is a signature of a loss of Galilean invariance. This is consistent with the fact that the microscopic system is not Galilean invariant as well, Indeed, there is a distinguished reference frame where the particle speed is $c_0$. Of course, this speed does not remain equal to $c_0$ in frames that translate at constant speed with respect to this frame. 

So far, with the introduction of $\mathrm{P}_{\Omega^\perp}$ and different constants $c_2 \not = c_1$ but still with $c_4 =0$, the system for $(\rho, \Omega)$ is decoupled from the equations for $u$ and $v$ and is written (see Eqs. \eqref{equationrho},  \eqref{equationOmega} with $\mathbf{F}$ given by \eqref{eq:defw_F} in which $c_4=0$): 
\begin{subequations}
\label{equation_SOH}
\begin{align}
& \partial_t \rho + c_1 \, \nabla_\mathbf{x} \cdot (\rho \, \Omega)=0, 
\\
& D_t\Omega = -c_3 \, \mathrm{P}_{\Omega^\perp} \nabla_\mathbf{x} \, \log \rho . 
\end{align}
\end{subequations}
This is nothing but the hydrodynamic limit of the Vicsek particle model (known as ``Self-Organized Hydrodynamics (SOH)'') as established in \cite{degond2008continuum, dimarcomotsch16}. This system has been shown to be hyperbolic \cite{degond2008continuum} and to have local-in-time smooth solutions \cite{degond20133hydrodynamic}. 

\item When $c_4 \not = 0$, in addition to the pressure gradient, a second component of the force $\mathbf{F}$ appears. This component depends on the full rotation matrix ${\mathbb A}$ through $\Omega$, $\mathbf{u}$, $\mathbf{v}$ and their gradients (see Eq. \ref{eq:def_r}). It is thus truly specific of the body orientation model. 
\end{enumerate}

We are now going to compare the IBM and the SOHB models on a set of explicit stationary solutions of the SOHB model described in the next section.

\section{Special solutions of the macroscopic model}
\label{sec:compar}

\subsection{Three classes of explicit solutions}
\label{sectionthreesolutions}

In this section, we exhibit three different classes of global-in-time solutions of the SOHB model~\eqref{equationA'}. They are special classes of a larger family of solutions which will also be introduced. All these solutions are characterized by uniform (i.e. independent of the spatial coordinate) fields $\rho$, $\mathbf{r}$ and $\delta$. From now on we fix a wave-number (inverse of the length) $\xi \in {\mathbb R} \setminus \{0\}$ and define 
\begin{equation}
\omega = \xi \, c_4, \qquad \lambda = c_2 + c_4.  
\label{eq:defxi}
\end{equation}
We denote by $\mathbf{x} = (x,y,z)^\mathrm{T}$ the coordinates of $\mathbf{x}$ in the basis $(\mathbf{e}_1, \mathbf{e}_2, \mathbf{e}_3)$.


\subsubsection{Flocking state}

The flocking state (FS) is a trivial but important special solution of the SOHB model~\eqref{equationA'} where both the density and rotation fields are constant (i.e. independent of time) and uniform: 
$$ 
\rho(t,\mathbf{x}) \equiv \rho_0 = \text{constant}, \quad {\mathbb A}(t,\mathbf{x})\equiv {\mathbb A}_0 = \text{constant}, \quad \forall (t, \mathbf{x}) \in [0,\infty) \times {\mathbb R}^3. 
$$


\subsubsection{Milling orbits}
\label{subsubsec:milling}

We have the following

\begin{lemma}
The pair $(\rho, {\mathbb A})$ consisting of a constant and uniform density $\rho(t,\mathbf{x}) = \rho_0 =$ constant and the following rotation field:
\begin{eqnarray}
\label{millingsolution}
{\mathbb A}(t,\mathbf{x}) &=& \tilde {\mathbb A}_{\mbox{\scriptsize mill}}(t,z) \nonumber \\ 
&=& \left( \begin{array}{lll}
\cos(\omega t) & \sin(\omega t) \, \cos(\xi z) & -\sin(\omega t) \, \sin(\xi z) \\ 
-\sin(\omega t) &  \cos(\omega t) \, \cos(\xi z) & -\cos(\omega t) \, \sin(\xi z)\\ 
0 & \sin(\xi z) &\cos(\xi z)
\end{array} \right) \\
\label{millingsolution2}
&=& {\mathcal A}(-\omega t, \mathbf{e}_3) \, {\mathcal A}(\xi z, \mathbf{e}_1),
\end{eqnarray}
is a solution of the SOHB system~\eqref{equationA'}, where $\omega$ and $\xi$ are given by \eqref{eq:defxi}. We recall that ${\mathcal A}(\theta, \mathbf{n})$ is the rotation of axis $\mathbf{n} \in {\mathbb S}^2$ and angle $\theta \in {\mathbb R}$ defined by \eqref{eq:rodrigues}. This solution will be referred to as a milling orbit (MO). 
\label{lem:MO}
\end{lemma}

\begin{figure}[ht!]
\centering
\subfloat[$t=0$]{\includegraphics[trim={3.5cm 12.cm 2.5cm 4.5cm},clip,width=7cm]{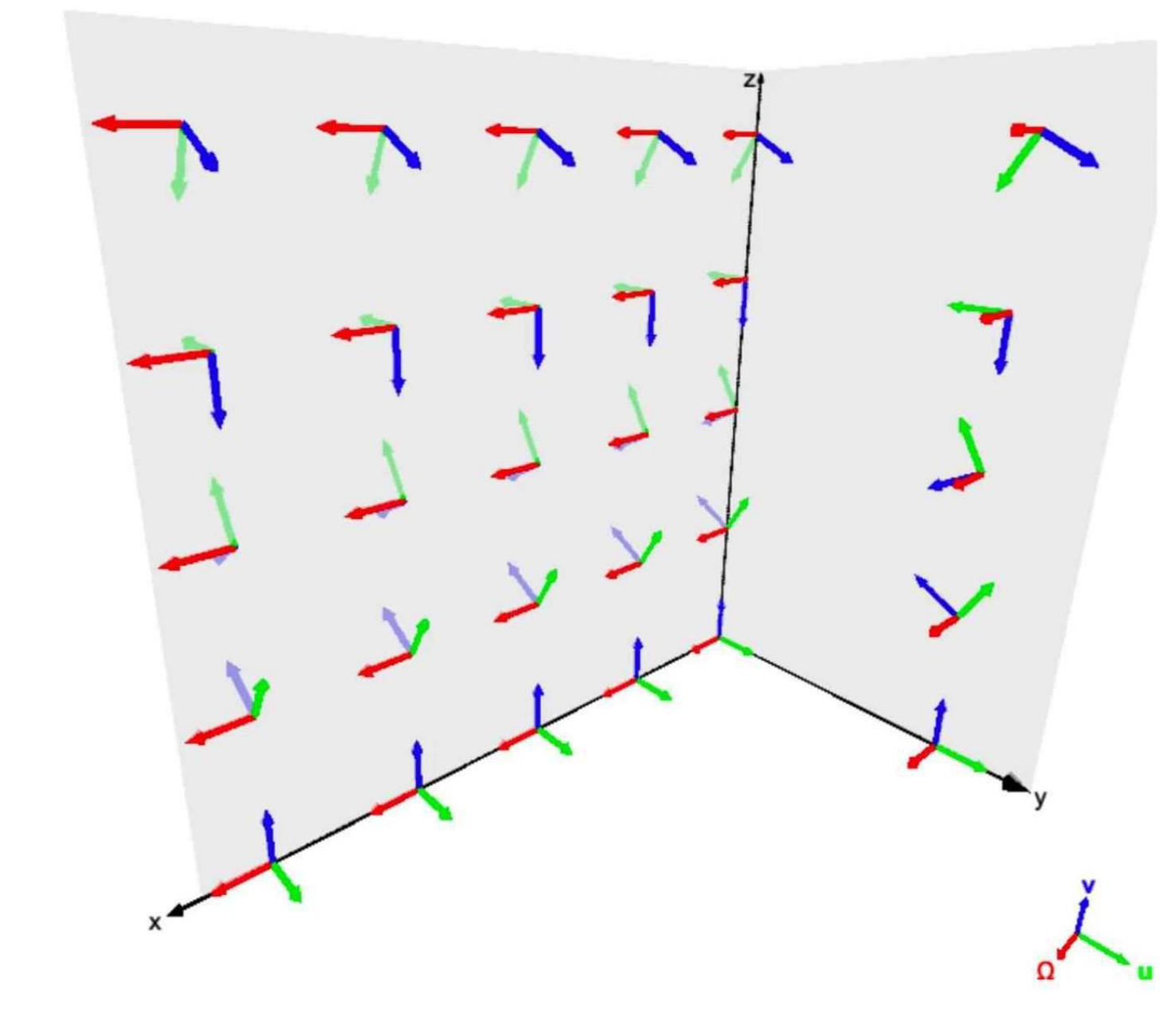}\label{fig:milling0}}
\subfloat[$t>0$]{\includegraphics[trim={3.5cm 12.cm 2.5cm 4.5cm},clip,width=7cm]{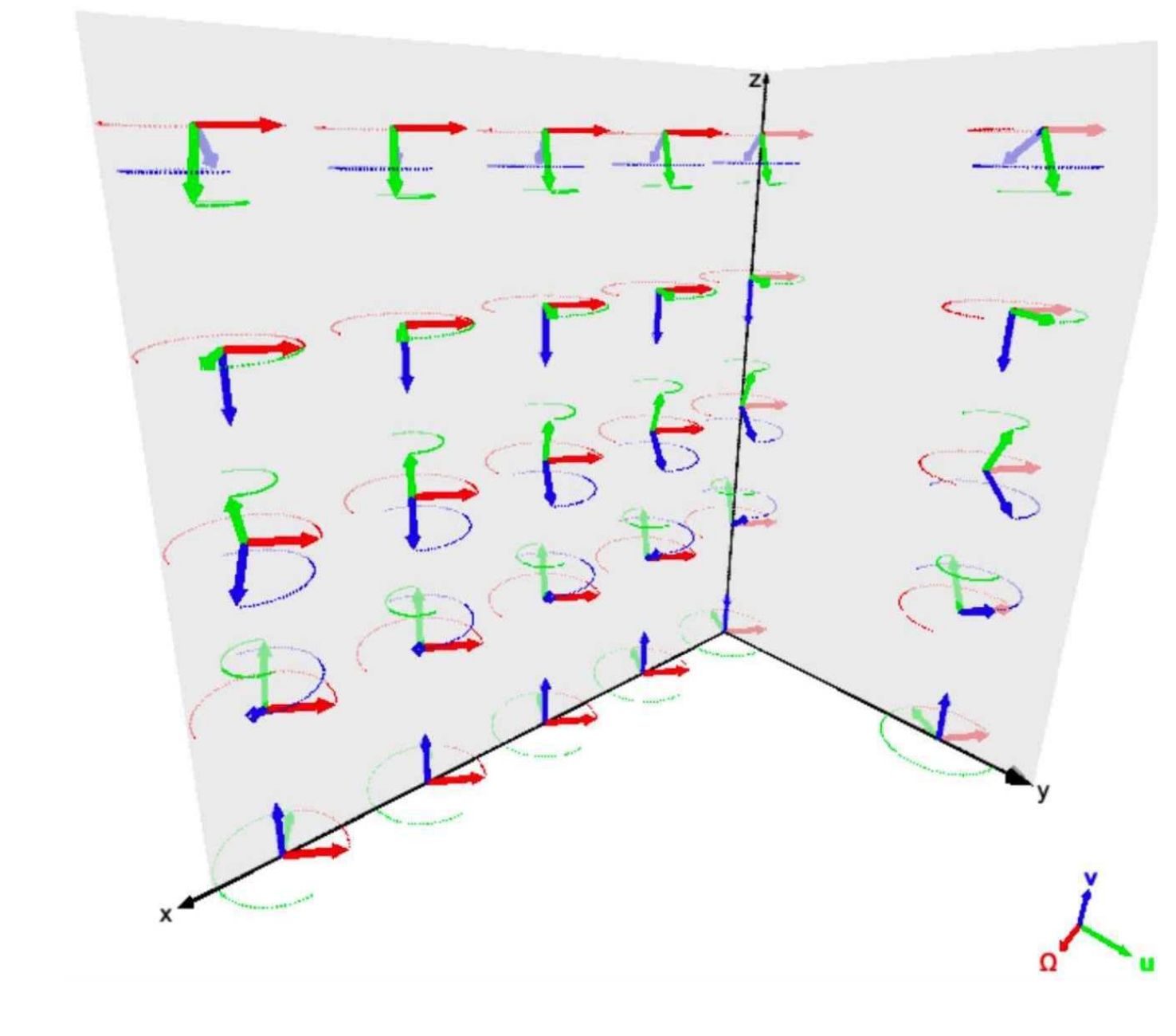}\label{fig:millingt}}
\caption{Graphical representation of the milling orbit (MO) at (a): initial time, and (b): time $t>0$. The frame vectors $\Omega$, $\mathbf{u}$ and $\mathbf{v}$ are represented at a certain number of points of the $(O,x,y)$ and $(O,y,z)$ planes. In (b), the rotation motion of the frame vectors is depicted by dotted circles of the color of the corresponding frame vector. The red dotted circle can be seen as a depiction of the fluid streamlines. See also Section \ref{appendix:listvideos}, Video \ref{vid:milling_solution}.}
\label{fig:milling}
\end{figure}

The proof of this lemma is deferred to Section \ref{appendixgeneralsolutions}. The MO is independent of $x$ and $y$. Its initial condition is 
\begin{equation}
\label{perptwist}
{\mathbb A}_{\mbox{\scriptsize mill}} (0,z) = {\mathcal A}(\xi z, \mathbf{e}_1) = \left( \begin{array}{ccc}
1 & 0 & 0\\
0 &\cos(\xi z) & -\sin(\xi z)\\
0 & \sin(\xi z) & \cos(\xi z)
\end{array}\right).
\end{equation}
The initial direction of motion (the first column of ${\mathbb A}_{\mbox{\scriptsize mill}} (0,z)$) is independent of $z$ and aligned along the $x$-direction, i.e. $\Omega(0,z) \equiv \mathbf{e}_1$. As $z$ varies, the body-orientation rotates uniformly about the $x$-direction with spatial angular frequency $\xi$. As the rotation vector is perpendicular to the direction of variation, \eqref{perptwist} is called a ``perpendicular twist''.  As time evolves, the rotation field is obtained by multiplying on the left the initial perpendicular twist by the rotation ${\mathcal A}(-\omega t, \mathbf{e}_3)$. This means that the whole body frame undergoes a uniform rotation about the $z$-axis with angular velocity $- \omega$. As a consequence, the direction of motion is again independent of $z$. It belongs to the plane orthogonal to $z$ and undergoes a uniform rotation about the $z$-axis. Consequently, the fluid streamlines, which are the integral curves of $c_1 \Omega$, are circles contained in planes orthogonal to $z$ of radius $\frac{c_1}{\omega} = \frac{c_1}{c_4} \frac{1}{\xi}$ traversed in the negative direction if $\xi >0$. These closed circular streamlines motivate the ``milling'' terminology. It can be checked that the MO satisfies: 
$$ \mathbf{r}= \xi \, (\sin(\omega t),\cos(\omega t),0)^\mathrm{T}, \qquad \delta=0.$$
As announced, $\mathbf{r}$ and $\delta$ are uniform but $\mathbf{r}$ depends on time. Actually, $\Omega \times \mathbf{r} = \xi \mathbf{e}_3$ is independent of time. The MO is depicted in Fig. \ref{fig:milling} and its dynamics is visualized in Video \ref{vid:milling_solution} (see Section~\ref{appendix:listvideos}).

Many examples of milling (also known as vortex) solutions have been observed in the collective dynamics literature as well as in biological systems \cite{calovi2014swarming, couzin2003self, vicsek2012collective}. On the modelling side, milling states have not been observed so far in alignment models without the inclusion of an additional process such as an attraction-repulsion force between the agents \cite{carrillo2009double}, a bounded cone of vision \cite{costanzo2018spontaneous} or an anticipation mechanism \cite{gerlee2017impact}. The body-orientation framework is, to the best of our knowledge, a new situation in which milling can be observed just with  alignment assumptions. Milling states can also be found in physical systems. A typical and important example is the motion of a charged particle in a uniform magnetic field, resulting in the formation of so-called cyclotron orbits. Once again, in the body-orientation framework, an external field is not needed and self-induced cyclotron orbits emerge only from the variations of the internal body-orientation. Here, the analog of the magnetic field would be $\Omega \times \mathbf{r}$ and the cyclotron frequency would be $\omega$. Note that $\omega$ is under the control of coefficient $c_4$ which depends on the noise intensity $1/\kappa$.


\subsubsection{Helical traveling wave} 
\label{subsubsec:helical}

We have the following 

\begin{lemma}
The pair $(\rho, {\mathbb A})$ consisting of a constant and uniform density $\rho(t,\mathbf{x}) = \rho_0 =$ constant and the following rotation field:
\begin{eqnarray}
\label{helicalsolution}
{\mathbb A}(t,\mathbf{x}) &=& \tilde {\mathbb A}_{\mbox{\scriptsize htw}}(t,x) \nonumber \\ 
&=& \left(\begin{array}{ccc}
1 &  0 & 0  \\ 
0 & \cos \left( \xi (x-\lambda t) \right) & -\sin \left( \xi (x-\lambda t) \right) \\ 
0 & \sin \left( \xi (x-\lambda t) \right) & \cos \left( \xi (x-\lambda t) \right)
\end{array}\right) \\
&=& {\mathcal A}(\xi (x-\lambda t), \mathbf{e}_1),
\end{eqnarray}
is a solution of the SOHB system~\eqref{equationA'} where $\xi$ and $\lambda$ are defined by \eqref{eq:defxi}. This solution will be referred to as a helical traveling wave (HW). 
\label{lem:HW}
\end{lemma}

\begin{figure}[ht!]
\centering
\subfloat[$t=0$]{\includegraphics[trim={3.5cm 12.cm 2.5cm 4.5cm},clip,width=7cm]{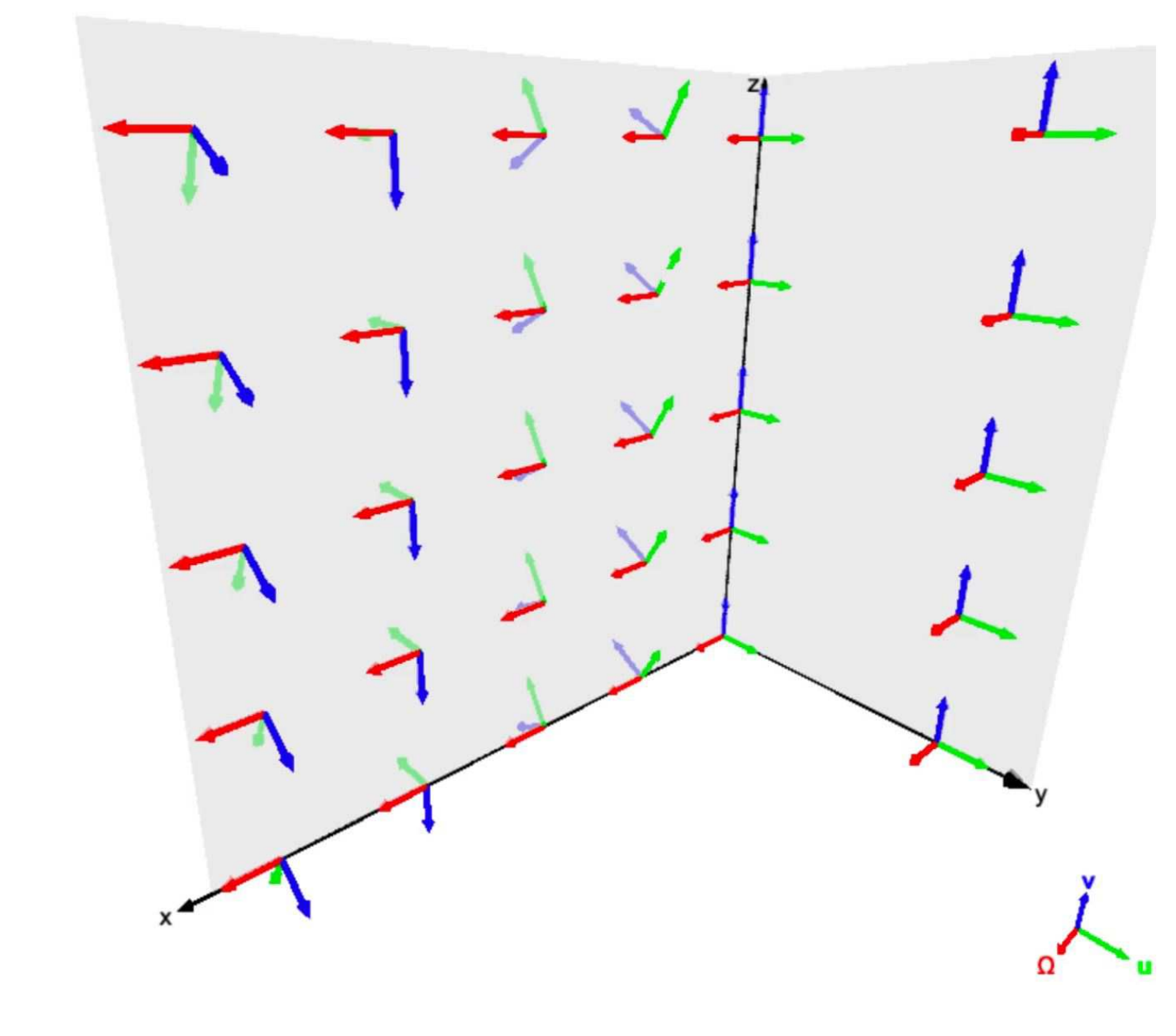}\label{fig:helical0}}
\subfloat[$t>0$]{\includegraphics[trim={3.5cm 12.cm 2.5cm 4.5cm},clip,width=7cm]{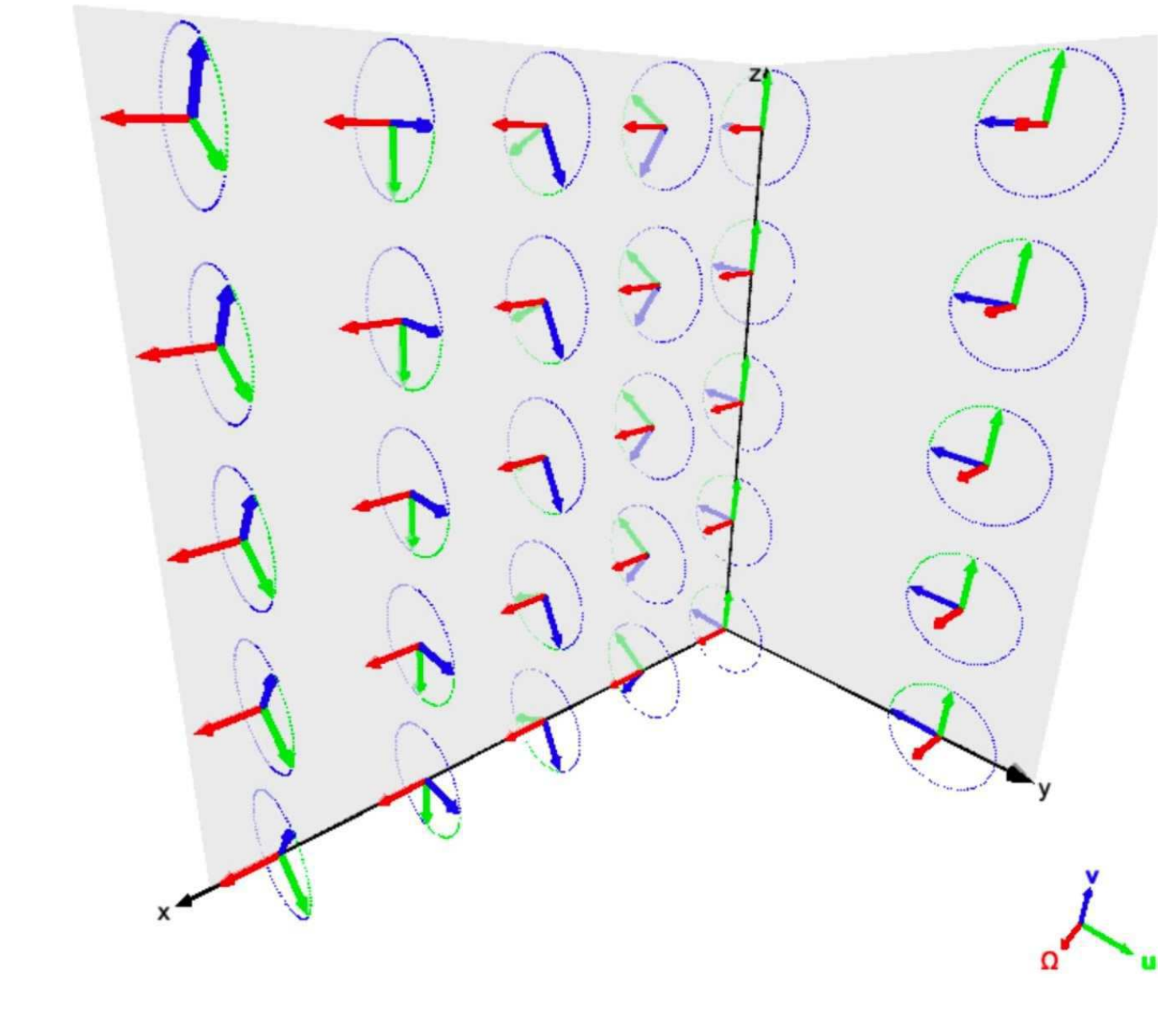}\label{fig:helicalt}}
\caption{Graphical representation of the helical traveling wave (HW) at (a): initial time, and (b): time $t>0$. See Fig. \ref{fig:milling} for captions. See also Section \ref{appendix:listvideos}, Video \ref{vid:helical_solution}.}
\label{fig:helical}
\end{figure}

The proof of this lemma is given in Section \ref{appendix:generalhelical}. The HW is independent of $y$ and $z$. Its initial condition is 
\begin{equation}
\label{paratwist}
{\mathbb A}_{\mbox{\scriptsize htw}} (0,x) = {\mathcal A}(\xi x, \mathbf{e}_1) = \left( \begin{array}{ccc}
1 & 0 & 0\\
0 &\cos(\xi x) & -\sin(\xi x)\\
0 & \sin(\xi x) & \cos(\xi x)
\end{array}\right).
\end{equation}
Here the self-propulsion direction is still independent of $x$ and equal to $\mathbf{e_1}$. Also, the body orientation still rotates uniformly about $\mathbf{e_1}$ with spatial angular frequency $\xi$ but when $x$ is varied instead of $z$. This means that the body orientation is now twisted when varied along the propagation direction. So, this initial condition is called a ``parallel twist''. In the HW, the self propulsion direction $\Omega$ remains constant in time and uniform in space. The initial twist is propagated in time in this direction at speed $\lambda$ and gives rise to a traveling wave 
$$
\tilde {\mathbb A}_{\mbox{\scriptsize htw}}(t,x) = \tilde {\mathbb A}_{\mbox{\scriptsize htw}}(0,x-\lambda t).
$$
Note that the traveling wave speed $\lambda$ depends on the noise intensity $1/\kappa$ and is different from the fluid speed $c_1$. So, the frame carried by a given fluid element followed in its motion is not fixed but rotates in time. Since $\Omega$ does not change, the fluid streamlines are now straight lines parallel to $\mathbf{e}_1$. So, as a fluid element moves, the ends of the frame vectors $\mathbf{u}$ and $\mathbf{v}$ follow a helical trajectory with axis $\mathbf{e}_1$, hence the terminology ``helical traveling waves'' for these solutions. It can be checked that 
$$
\mathbf{r}=0,\qquad \delta=\xi,
$$ 
and again, $\mathbf{r}$ and $\delta$ are spatially uniform as announced. The HW is depicted graphically in Fig. \ref{fig:helical}. Its dynamics is visualized in Video \ref{vid:helical_solution} (see Section \ref{appendix:listvideos}). The HW belongs to a larger class of solutions described in Section \ref{appendix:generalhelical}.

\subsubsection{Generalized topological solutions} 

The three above described classes of solutions can be encompassed by a single family of generalized solutions as stated in the following lemma.  

\begin{lemma}[Generalized solutions]
Let $\xi\in\mathbb{R}$ and $\theta\in[0,\pi]$ be two parameters. Let $\omega\in\mathbb{R}$ and $\tilde{\lambda}\in\mathbb{R}$ be defined by 
\[\omega = c_4\xi,\quad\tilde{\lambda} = c_2\cos\theta.\]
The pair $(\rho, {\mathbb A})$ consisting of a constant and uniform density $\rho(t,\mathbf{x}) = \rho_0 =$ constant and the following rotation field:
\begin{equation}
\mathbb{A}(t,\mathbf{x}) = \mathbb{A}_{\xi,\theta}(t,z) := \mathcal{A}(-\omega t, \mathbf{e}_3)\,\mathcal{A}\left(\theta - \frac{\pi}{2},\mathbf{e}_2\right)\mathcal{A}(\xi(z-\tilde{\lambda} t),\mathbf{e}_1),
\label{eq:generalizedsolutions}
\end{equation}
is a solution of the SOHB system~\eqref{equationA'}. We recall that ${\mathcal A}(\theta, \mathbf{n})$ is the rotation of axis $\mathbf{n} \in {\mathbb S}^2$ and angle $\theta\in\mathbb{R}$. This solution will be referred to as a Generalized topological Solution (GS). 
\label{lemma:generalizedsolutions} 
\end{lemma}

The proof of this lemma is deferred to the Supplementary Material \ref{appendix:generalizedsolutions}. Each of the three previous classes of solutions can be obtained for specific values of the parameters $\xi$ and~$\theta$. 
\begin{itemize}
\item When $\xi=0$, the solution $\mathbb{A}_{0,\theta}$ is constant for any $\theta$, which corresponds to a FS. 
\item When $\theta=\frac{\pi}{2}$ and $\xi\in\mathbb{R}$, then $\tilde{\lambda}=0$ and the rotation with respect to the $y$-axis is equal to the identity: the solution $\mathbb{A}_{\xi,\pi/2}$ is therefore equal to the MO \eqref{millingsolution2}.
\item When $\theta=0$ and $\xi\in\mathbb{R}$ then $\tilde{\lambda}=c_2$ and the solution $\mathbb{A}_{\xi,0}$ is equal to   
\[\mathbb{A}_{\xi,0} = \left(\begin{array}{ccc}
0 & - \sin (\xi(z-\lambda t))& -\cos (\xi(z-\lambda t)) \\
0 & \cos (\xi(z-\lambda t)) & - \sin (\xi(z-\lambda t)) \\
1 &  0 & 0
\end{array}\right),\quad \lambda = c_2+c_4,\]
which is an HW along the $z$-axis. The situation is analogous when $\theta=\pi$.
\end{itemize}
All these solutions have a non-zero gradient in the body-orientation variable which is always along the $z$-axis. This gradient is controlled by the parameter $\xi$. However, in the GS, the direction of motion $\Omega$ (or fluid velocity) is not necessarily parallel nor perpendicular to this gradient. Specifically, $\Omega$ has a constant polar angle equal to the parameter~$\theta$. The behavior of the solution is then a combination of the two previously introduced phenomena: milling around the $z$-axis and a travelling wave of the body-orientation variable along the same axis. The applet accessible at \url{https://www.glowscript.org/#/user/AntoineDiez/folder/MyPrograms/program/BOfield} provides a graphical representation of the GS for arbitrary polar angles using VPython \cite{vpython} and with the same conventions as in Fig. \ref{fig:milling}. 

In the following, we will focus on each of these two elementary behaviors, i.e. the standard milling and helical travelling wave solutions, and in particular on their topological properties. The study of the full continuum of generalized solutions is left for future work. However, we will encounter GS obtained from a perturbed milling solution in Section \ref{sec:robustness}. 


\subsection{Some properties of these special solutions}
\label{sec:properties_solutions}

Clearly, in the definitions of the MO and HW, the choice of reference frame is unimportant. So, in the whole space ${\mathbb R}^3$, such solutions exist in association with any reference frame. In a square domain of side-length $L$ with periodic boundary conditions, periodicity imposes some constraints on the direction of the reference frame. For simplicity, we will only consider the case where the reference frame has parallel axes to the sides of the square and $\xi$ is linked to $L$ by an integrality condition $L \, \xi = 2 \pi \, n$, with $n \in {\mathbb Z} \setminus \{0\}$. 

The study of the stability of the MO and the HW is left for future work. By contrast, the FS is linearly stable as the SOHB system is hyperbolic \cite{degond2021hyperbolicity}. However, there is no guarantee that the FS at the level of the IBM is stable. Indeed, there are strong indications that the FS is not stable for the Vicsek model \cite{chate2008collective} for some parameter ranges and a similar trend is likely to occur here. 

We can now answer the question posed at the end of Section \ref{subsec:ibm_relation} namely whether the inclusion of the full body orientation makes any change in the dynamics of the particle positions and directions compared to the Vicsek model. To this end, we consider the corresponding macroscopic models, i.e. the SOH model \eqref{equation_SOH} for the Vicsek model and the SOHB model \eqref{equationA} for the body-orientation dynamics.  If we initialize the SOH model with uniform initial density $\rho$ and mean direction $\Omega$, inspection of \eqref{equation_SOH} shows that the solution remains constant in time and thus corresponds to a flocking state of the Vicsek model. In the SOHB model, the three classes of solutions described in the previous sections (the FS, MO and HW) also have uniform initial density $\rho$ and mean direction $\Omega$. If the dynamics of the particle positions and directions in the body orientation model was the same as in the Vicsek model, these three classes of solutions should have a constant mean direction $\Omega$. However, it is not the case for the MO, where $\Omega$ changes with time and is subject to a planar rotation. This means that gradients of body attitude do have a non-trivial influence on the direction of motion of the particles and that the body orientation model does not reduce to a Vicsek model for the particle positions and directions. 

There is another, more subtle, difference between the two models concerning the dynamics of $\Omega$. It does not concern the MO and HW but we discuss it here in relation with the previous paragraph. Indeed, Fig. \ref{fig:ci} reveals that the velocities $c_1$ and $c_2$ for the SOHB model crossover at a certain value $\kappa^*$ of the concentration parameter. The coefficients $c_1$ and $c_2$ for the SOH model can be found in \cite{dimarcomotsch16}, Fig. A1(b) and appear to satisfy $c_1 > c_2$ for the whole range of values of $\kappa$, i.e. do not exhibit any crossover. In particular, at large noise, the propagation velocity $c_2$ of $\Omega$ in the SOHB model is larger than the mass transport velocity $c_1$. This means that information (which triggers adjustments in $\Omega$) propagates downstream the fluid by contrast to the Vicsek case where it propagates upstream. While the reason for this difference is unclear at this stage, we expect that it may induce large qualitative differences in the behavior of the system in some cases. This point will be investigated in future work.   

Numerical simulation of the SOHB will be subject to future work. Here, we will restrict ourselves to the MO and HW for which we have analytical formulas. In the next section, using these two special solutions, we verify that the SOHB model and the IBM are close in an appropriate parameter range.


\subsection{Agreement between the models}

In this section we use the MO and HW to demonstrate the quantitative agreement between the SOHB model \eqref{equationA} and the IBM \eqref{deterministicpart}, \eqref{jumppart} in the scaling \eqref{macroscalingnumeric}. In the simulations below, we consider a periodic cube of side-length $L$ and choose 
\begin{equation}
R = 0.025, \quad \nu = 40, \quad c_0 = 1, \quad L = 1, \quad \xi = 2 \, \pi, 
\label{eq:num_val} 
\end{equation}
so that $\frac{R}{L} = \frac{c_0}{\nu \, L} =  0.025 \ll 1$, ensuring that the scaling \eqref{macroscalingnumeric} is satisfied. Furthermore, we see that the choice of $\xi$ is such that the  twists in the MO or HW have exactly one period over the domain size.


\subsubsection{The IBM converges to the macroscopic model as \texorpdfstring{$N \to \infty$}{the number of particles grows to infinity}} 
\label{subsubsec:cvgceNinfty}

In this section, we numerically demonstrate that the solutions of the IBM converge to those of the macroscopic model in the limit $N \to \infty$ and investigate the behavior of the IBM at moderately high values of $N$.   

We sample $N$ particles according to the initial condition \eqref{perptwist} of the MO and simulate the IBM \eqref{deterministicpart}, \eqref{jumppart}. We recall that the average direction $\Omega(t)$ of the exact MO \eqref{millingsolution} is spatially uniform at any time and undergoes a uniform rotation motion about the $z$-axis. So, we will compare $\Omega(t)$ with the average direction $\overline{\Omega}(t)$ of all the particles of the IBM, where $\overline{\Omega}(t)=(\overline{\Omega}^1,\overline{\Omega}^2,\overline{\Omega}^3)^\mathrm{T}$ is defined by:
$$
\overline{\Omega}=\frac{\sum_{k=1}^N \Omega_k(t) }{|\sum_{k=1}^N \Omega_k(t)|}, 
$$
(provided the denominator is not zero, and where we recall that $\Omega_k(t) = A_k(t) \, \mathbf{e}_1$). To ease the comparison, we compute the azimuthal and polar angles of $\overline{\Omega}$ respectively defined by:
\begin{equation}
\label{eq:averageangles}
\bar \varphi := \mathrm{arg}(\overline{\Omega}^1 + i \overline{\Omega}^2) \in[0,2\pi), \quad \bar \theta = \arccos(\overline{\Omega}^3)\in[0,\pi],
\end{equation}
where $\mathrm{arg} (x+iy)$ stands for the argument of the complex number $x+iy$. We note that the corresponding angles $\varphi$ and $\theta$ of $\Omega(t)$ are given by 
\begin{equation}
\label{eq:averageangles_MO}
\varphi(t) = - \omega \, t = - 2 \pi \, c_4(\kappa) \, t, \qquad \theta = \pi/2,
\end{equation}
where we have used \eqref{eq:defxi} and \eqref{eq:num_val} to compute the value of $\omega$.

\begin{figure}[ht!]
\centering
\subfloat[]{\includegraphics[width= 6cm]{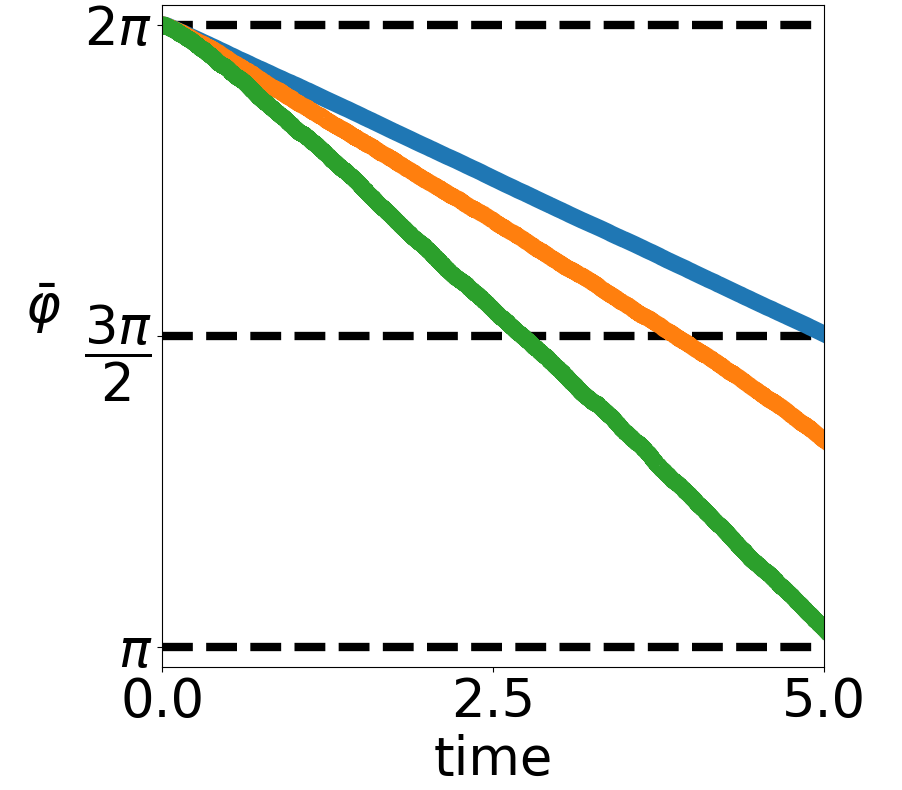}\label{fig:phiN}} 
\hspace{1cm}
\subfloat[]{\includegraphics[width= 6cm]{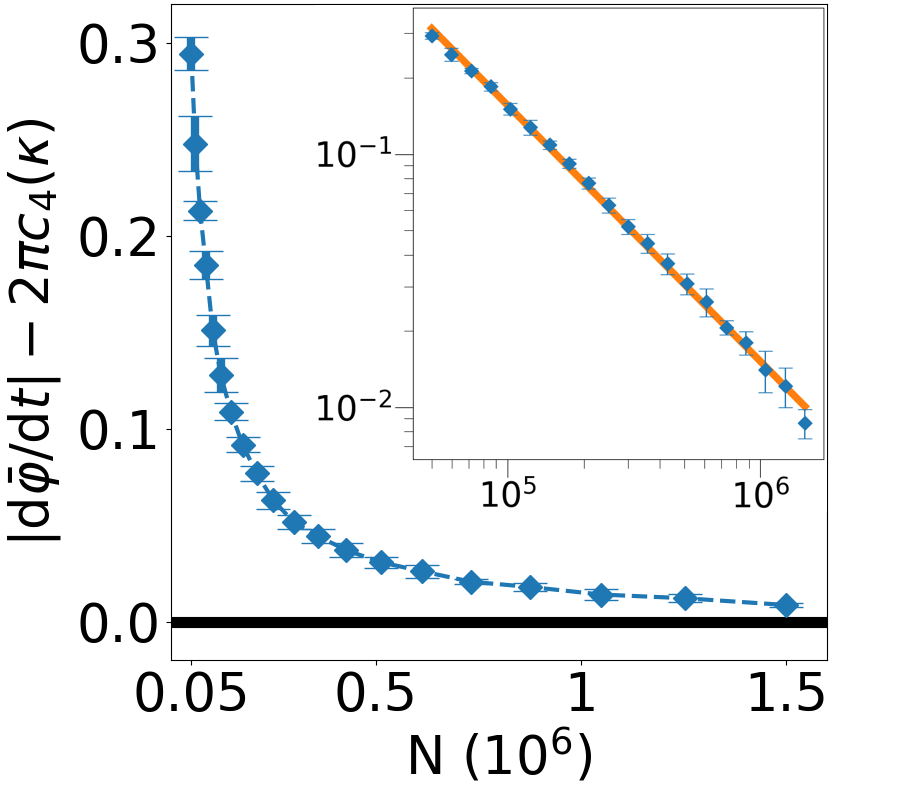}\label{fig:slopeN}}
\caption{(a) Time evolution of the angle $\bar \varphi$ for three values of $N$ : $N=0.05 \, 10^6$ (green curve), $N=0.15 \, 10^6$ (orange curve) and $N=1.5 \, 10^6$ (blue curve). (b) Difference between the measured angular velocity $|\mathrm{d}\bar{\varphi}/\mathrm{d} t|$ and the theoretical value $2\pi c_4(\kappa)$. Each data point (blue dot) is an average of 10 independent simulations with the error bar showing one standard deviation. Solid black horizontal line at 0 for convenience. Inset: same data in log-log scale and regression line (solid orange line). Parameters: $L=1$, $\xi = 2 \pi$, $R=0.025$, $\nu=40$, $c_0=1$, $\kappa =10$.}
\end{figure}

Fig. \ref{fig:phiN} shows the azimuthal angle $\bar \varphi$ as a function of time over 5 units of time, for increasing particle numbers: $N=5 \, 10^4$ (green curve), $N=1.5 \, 10^5$ (orange curve) and $N=1.5 \, 10^6$ (blue curve). Note that for very small values of $N$, the macroscopic model loses its relevance: below a few thousand particles we only observe a noisy behavior, not shown in the figure. For the considered range of particle numbers, we notice that the angle $\bar \varphi$ decreases linearly with time, which shows that the behavior of the IBM is consistent with the exact solution \eqref{eq:averageangles_MO}. However, quantitatively, we see that $|\mathrm{d} \bar \varphi / \mathrm{d} t|$ depends on the particle number and decreases with increasing particle number. We investigate this behavior in more detail in Fig.~\ref{fig:slopeN} where the difference between the measured angular velocity $|\mathrm{d}\bar{\varphi}/\mathrm{d} t|$ and the theoretical prediction $2\pi c_4(\kappa)$ is plotted as a function of $N$. Each data point (blue dot) is an average of 10 independent simulations. This figure confirms that, as $N$ increases, $|\mathrm{d} \bar \varphi / \mathrm{d} t|$ decreases and converges towards $2\pi c_4(\kappa)$. The inset in Fig. \ref{fig:slopeN} shows the same data points in a log-log-scale with the associated regression line (orange solid line). We observe that the error between the measured and theoretical angular velocities behaves like $N^{-\alpha}$ with a measured exponent $\alpha\simeq 1.01$ which is close to the theoretical value $\alpha = 1$ derived in Section \ref{sec:appendix_convergence_rate} of the Supplementary Material.

\subsubsection{Quantitative comparison between the models}
\label{subsubsec:quantitative}

In order to quantitatively confirm the agreement between the IBM and the macroscopic model, we fix a large number $N=1.5 \, 10^6$ of particles and we run the IBM for different values of the concentration parameter $\kappa$ and for the two classes of special solutions, the MO and the HW. To compare the models, we compute the following macroscopic quantities: 
\begin{itemize}
\item For the MO: starting from a sampling of the initial condition \eqref{perptwist}, we measure the angular velocity $|\mathrm{d} \bar \varphi / \mathrm{d} t|$ in a similar way as in the previous section. Given the parameter choice \eqref{eq:num_val}, the theoretical value of $|\mathrm{d}\varphi / \mathrm{d} t|$ predicted by \eqref{millingsolution} is $|\omega|=2\pi c_4(\kappa)$ where the function $c_4$ is given by \eqref{c4}.
\item For the HW, starting from a sampling of the initial condition \eqref{paratwist}, we measure the wave speed. To this aim, using \eqref{projection}, we compute the mean body-orientation ${\mathbb A}$ of the agents in a slice of size $10^{-3}$ along the $x$-axis (which is the global direction of motion) as a function of time. As predicted by \eqref{helicalsolution} the coefficient ${\mathbb A}_{22}$ of the mean orientation is a periodic signal. The inverse of the period of this signal (obtained through a discrete Fourier transform) gives the traveling wave speed of the HW. The theoretical value predicted by \eqref{helicalsolution} is given by $\lambda=c_2(\kappa)+c_4(\kappa)$ where the function $c_2$ is given by \eqref{c2}.
\end{itemize} 

\begin{figure}[ht!]\centering
\subfloat[]{\includegraphics[width= 6cm]{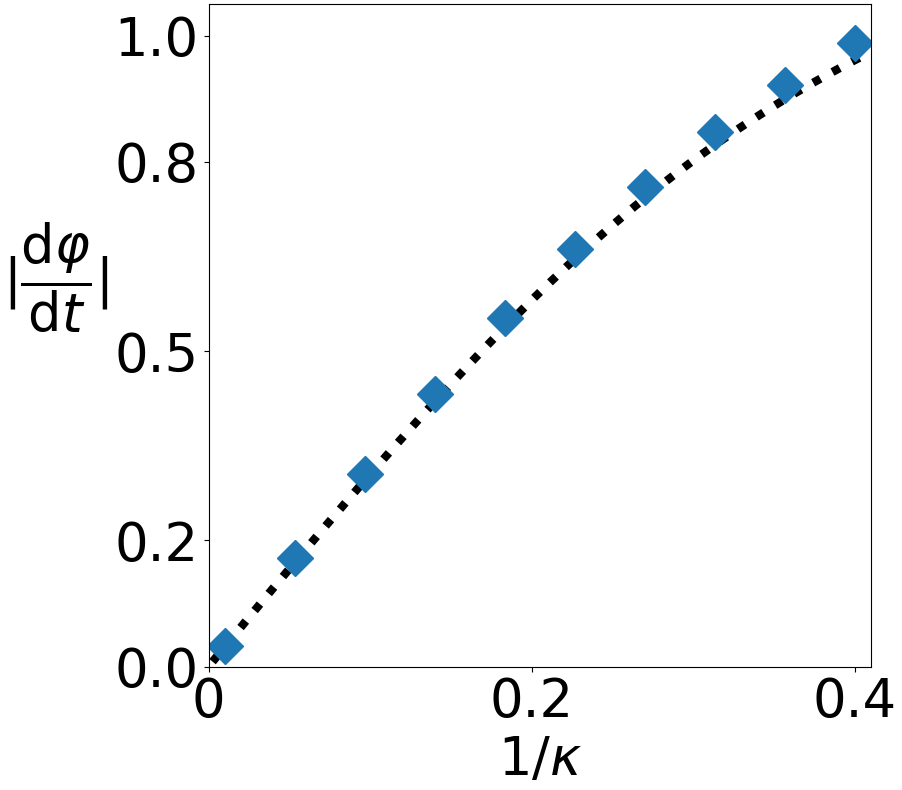}\label{figuregoodagreementa}} \hspace{0.7cm} 
\subfloat[]{\includegraphics[width= 6cm]{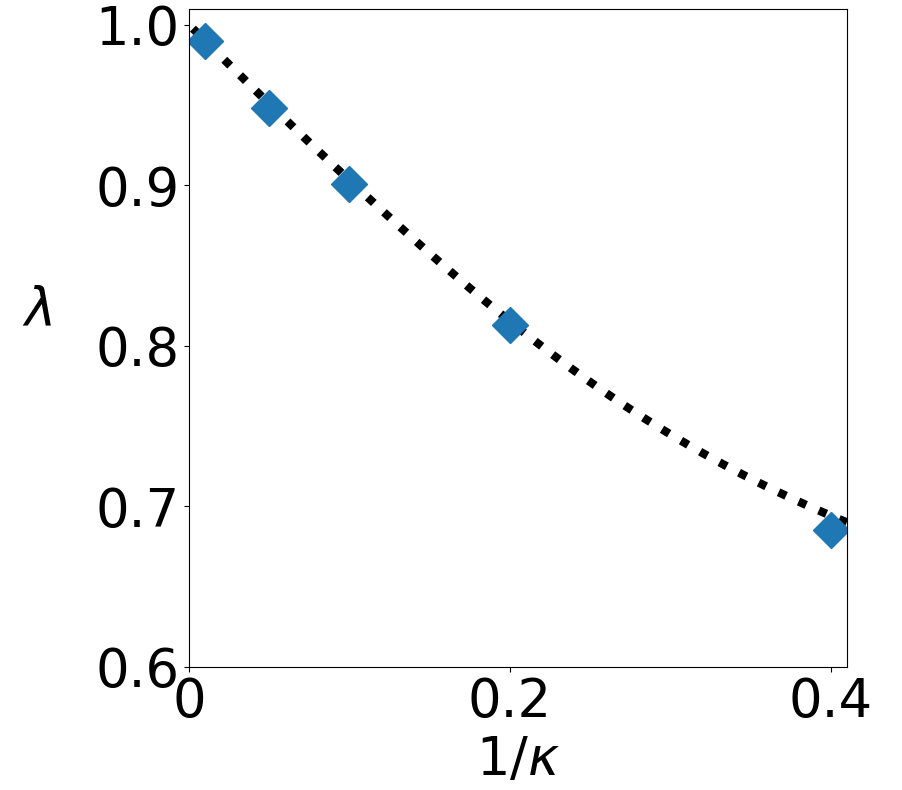}\label{figuregoodagreementb}}
\caption{(a) MO: angular velocity  $|\mathrm{d} \varphi/\mathrm{d} t|$ as a function of $1/\kappa$. (b) HW: traveling wave speed $\lambda$ as a function of $1/\kappa$. Measured values from the IBM at discrete values of $\kappa$ (big blue dots) and theoretical prediction from the SOHB model (dotted black curve). Parameters: $N=1.5 \, 10^6$, $L=1$, $\xi = 2 \pi$, $R=0.025$, $\nu=40$, $c_0=1$.}
\label{figuregoodagreement}
\end{figure}

The output of these simulations is shown in Figs. \ref{figuregoodagreementa} for the MO and \ref{figuregoodagreementb} for the HW. They respectively display the angular velocity and traveling wave speed obtained by running the IBM for a discrete set of values of $\kappa$ (big blue dots). By comparison, the black dotted curves show the theoretical values as functions of $\kappa$.  For the parameters of Fig. \ref{figuregoodagreement}, the order of magnitude of the standard deviation of 10 independent simulations is $10^{-3}$. The relative error between the average measured value and its theoretical prediction varies between 2\% and 5\% on the whole range of concentration parameters considered.

These figures show an excellent agreement between the prediction of the macroscopic SOHB model and the results obtained by running the IBM when the number of particles is large. This confirms that the SOHB model provides an excellent approximation of the IBM, at least during a certain period of time which is a function of the particle number. We will see below that fluctuations induced by the finite number of particles may eventually destabilize the MO and lead to a HW or a FS. As these solutions are associated with different topological structure, these transitions will be analyzed as topological phase transitions in the forthcoming sections.

\subsection{Topology}
\label{sec:topologysolutions}

Both the MO and HW have non-trivial topology: inspecting the perpendicular twist \eqref{perptwist} (see also Fig. \ref{fig:milling0}), we observe that the two-dimensional curve generated by the end of the vector $\mathbf{u}$ in the $(y,z)$-plane as one moves along the $z$-axis is a closed circle. A similar observation can be made on the parallel twist \eqref{paratwist} (see Fig. \ref{fig:helical0}) as one moves along the $x$-axis. Both curves have therefore non-zero winding numbers about the origin. When the domain is ${\mathbb R}^3$, these winding numbers are $\pm \infty$ (where the sign corresponds to that of $\xi$) as these curves make an infinite number of turns. If the domain has finite extension $L$ along the $z$-axis (in the MO case) or the $x$-axis (in the HW case) and, due to the periodic boundary conditions, $L$ is related to $\xi$ by $L = n \, 2 \pi / \xi$ with $n \in {\mathbb Z} \setminus \{0\}$, then the winding numbers are equal to $n$. As observed on Formulas \eqref{millingsolution} and \eqref{helicalsolution} (or on Figs \ref{fig:millingt} and \ref{fig:helicalt}), this initial non-trivial topological structure is propagated in time. 

When we initialize particles by sampling the initial conditions \eqref{perptwist} or \eqref{paratwist}, we expect that the solution of the IBM remains an approximation of the MO \eqref{millingsolution} or HW \eqref{helicalsolution} respectively as evidenced in Section \ref{subsubsec:quantitative}. However, noise induced by both the inherent stochasticity of the IBM and finite particle number effects as explained in Section \ref{subsubsec:cvgceNinfty} may  eventually destabilize the IBM. Then, in most cases, its solution is seen to transition towards an approximation of the FS after some time. This transition implies a change of the topology of the solution which, from initially non-trivial, becomes trivial, since the winding number of the FS is zero. One may wonder whether the evolution towards a FS is slower if the initial state has non-trivial topology and exhibits some kind of ``topological protection'' against noise-induced perturbations. To test this hypothesis quantitatively, we first need to develop appropriate indicators. This is done in the next section.


\section{Order parameters and topological indicators}
\label{sec:tools}

We will use two types of indicators. The first one is the global order parameter which will discriminate between the various types of organization of the system (disorder, MO or HW and FS). The second type of indicators are based on analyzing the roll angle. They will enable a finer characterization of topological phase transitions.


\subsection{Global order parameter}
\label{sec:order}

We first introduce the following scalar binary order parameter which measures the degree of alignment between two agents with body-orientations $A$, $\tilde{A} \in \mathrm{SO}_3({\mathbb R})$ : 
\begin{equation}
\label{eq:binaryop}
\psi(A,\tilde{A}):=\frac{1}{2} \, A\cdot\tilde{A} + \frac{1}{4} .
\end{equation}
In the quaternion framework (see Section \ref{subsec:ibm_num_sim} and \ref{sectionquaternion} for details), we have 
\begin{equation}
\label{eq:binaryop_quat}
\psi(A,\tilde{A}) = (q\cdot\tilde{q})^2, 
\end{equation}
where $q$ and $\tilde{q}$ are two unit quaternions respectively associated to $A$ and $\tilde{A}$, and $q\cdot\tilde{q}$ indicates the inner product of two quaternions. This expression makes it clear that $\psi(A,\tilde{A}) \in [0,1]$. The square exponent in \eqref{eq:binaryop_quat} indicates that $\psi(A,\tilde{A})$ measures the nematic alignment of the two associated unit quaternions, as it should because two opposite quaternions represent the same rotation. We note that $\psi(A,\tilde{A})=1$ if and only if $\tilde{A}=A$. On the other hand, $\psi(A,\tilde{A})=0$ if and only if $A\cdot\tilde{A}=-1/2$, which corresponds to the two rotation axes being orthogonal and one rotation being an inversion about its axis. 

The Global Order Parameter (GOP) of a system of $N$ agents at time $t>0$ is the average of all binary order parameters over all pairs of particles: 
\begin{equation}
\label{eq:orderparameter}
\mbox{GOP}^N(t)  = \frac{1}{N(N-1)} \, \sum_{k \not = \ell } \psi \big( A_k(t),A_{\ell}(t) \big) .
\end{equation}
From \eqref{eq:orderparameter} we have GOP$^N(t) \in [0,1]$. A small GOP$^N$ indicates large disorder and a large one, strong alignment. This is a global measure of alignment, by contrast to a local one where $\psi$ would be averaged over its neighbors only (and the result, averaged over all the particles). This global measure of alignment allows us to separate the MO and HW from the FS as shown below, which would not be possible with a local one. 

The GOP \eqref{eq:orderparameter} can also be defined at the continuum level. As shown in Section~\ref{appendix:derivationmacro}, in the macroscopic limit, the particles become independent and identically distributed over ${\mathbb R}^3\times $SO$_3({\mathbb R})$, with common distribution $\rho \, M_{{\mathbb A}}$ where $(\rho,{\mathbb A})$ satisfies the SOHB system \eqref{equationA} and $M_{{\mathbb A}}$ is the von Mises distribution \eqref{eq:vonmises}. Therefore, the GOP of a solution of the SOHB system $(\rho,{\mathbb A})$ is obtained as \eqref{eq:orderparameter} where the sum is replaced by an integral, $A_k(t)$ is replaced by $A$ distributed according to the measure $(\rho \, M_{{\mathbb A}})(t, \mathbf{x}, A) \, \mathrm{d} \mathbf{x} \, \mathrm{d} A$ and $A_\ell(t)$ is replaced by $\tilde{A}$ distributed according to the same measure, but independently to $A$. Therefore, 
$$
\mbox{GOP}(\rho,{\mathbb A}) := \iint_{({\mathbb R}^3\times \mbox{{\scriptsize SO}}_3({\mathbb R}))^2} \psi(A,\tilde{A}) \, \rho (\mathbf{x}) \, \rho (\tilde{\mathbf{x}}) \, M_{{\mathbb A}(\mathbf{x})} (A) \, M_{{\mathbb A}(\tilde{\mathbf{x}})} (\tilde{A}) \, \mathrm{d} \mathbf{x} \, \mathrm{d} \tilde{\mathbf{x}}  \, \mathrm{d} A \, \mathrm{d} \tilde{A}.
$$
Using \eqref{eq:haar} and \eqref{eq:rodrigues} one can prove that for any ${\mathbb A} \in $SO$_3({\mathbb R})$, we have 
\begin{equation}
\int_{\mbox{{\scriptsize SO}}_3({\mathbb R})} A \, M_{{\mathbb A}}(A) \, \mathrm{d} A = \frac{c_1(\kappa)}{c_0} \, {\mathbb A},
\label{eq:int_AMA}
\end{equation}
with $c_1(\kappa)$ defined by \eqref{c1} and $c_0$ being the particle speed. Using \eqref{eq:binaryop}, we obtain: 
\begin{equation}
\label{eq:globalop}
\mbox{GOP}(\rho,{\mathbb A}) = \frac{1}{2} \left( \frac{c_1(\kappa)}{c_0} \right)^2 \int_{{\mathbb R}^3\times{\mathbb R}^3} {\mathbb A}(\mathbf{x}) \cdot {\mathbb A}(\tilde{\mathbf{x}}) \, \rho(\mathbf{x}) \, \rho(\tilde{\mathbf{x}}) \, \mathrm{d} \mathbf{x} \, \mathrm{d} \tilde{\mathbf{x}} + \frac{1}{4}.
\end{equation}

From now on, we let $\rho$ be the uniform distribution on a square box of side-length~$L$. We can compute the GOP corresponding to each of the three solutions defined in Section \ref{sectionthreesolutions}. For the MO \eqref{millingsolution}, HW \eqref{helicalsolution} and GS \eqref{eq:generalizedsolutions}, for all time $t>0$, in all cases, the GOP remains equal~to:
\begin{equation}
\label{eq:opmilling}
\mbox{GOP}_1 = \frac{1}{4} \, \left( \frac{c_1(\kappa)}{c_0} \right)^2 + \frac{1}{4}.
\end{equation}
For the FS, ${\mathbb A}(\mathbf{x}) \equiv {\mathbb A} = $ constant and the GOP is equal to 
\begin{equation}
\label{eq:opflock}
\mbox{GOP}_2 = \frac{3}{4} \, \left( \frac{c_1(\kappa)}{c_0} \right)^2 + \frac{1}{4}.\end{equation}
Note that the GOP:
\[\mbox{GOP}_0 = \frac{1}{4},\]
corresponds to a disordered state of the IBM where the body-orientations of the particles are chosen independently and randomly uniformly (or equivalently to the SOHB case $\kappa \to 0$ in \eqref{eq:opmilling} and \eqref{eq:opflock}). For the typical value $\kappa = 10$ used in our simulations, one can compute that:
\begin{equation}
\mbox{GOP}_1 \simeq 0.45, \qquad \mbox{GOP}_2 \simeq 0.85.
\label{eq:gap_estimates}
\end{equation}
The GOP values between $\mbox{GOP}_1$ and $\mbox{GOP}_2$ can be reached by generalized HW as shown in Section \ref{appendix:generalisedorderlevels}.


\subsection{Roll angle}

\subsubsection{Definition} 

\begin{figure}[ht!]
\centering
\subfloat[]{\includegraphics[width= 7cm]{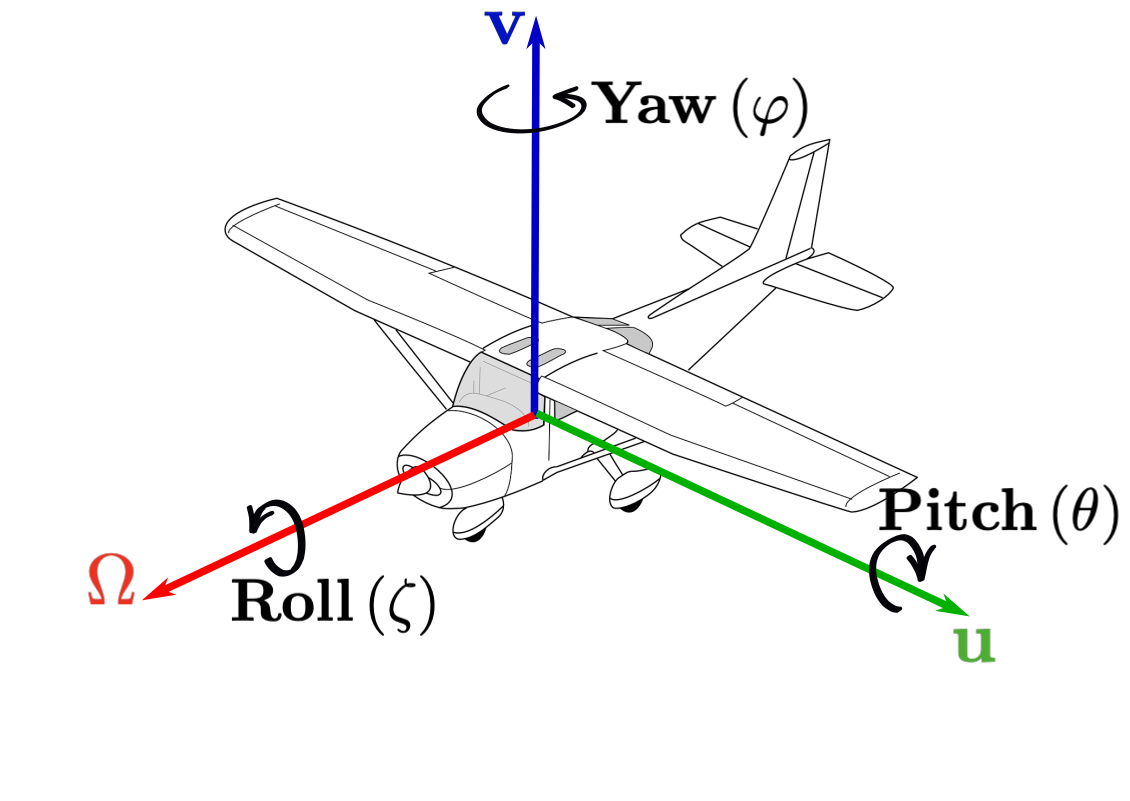}\label{fig:aircraft}}
\subfloat[]{\includegraphics[width= 7cm]{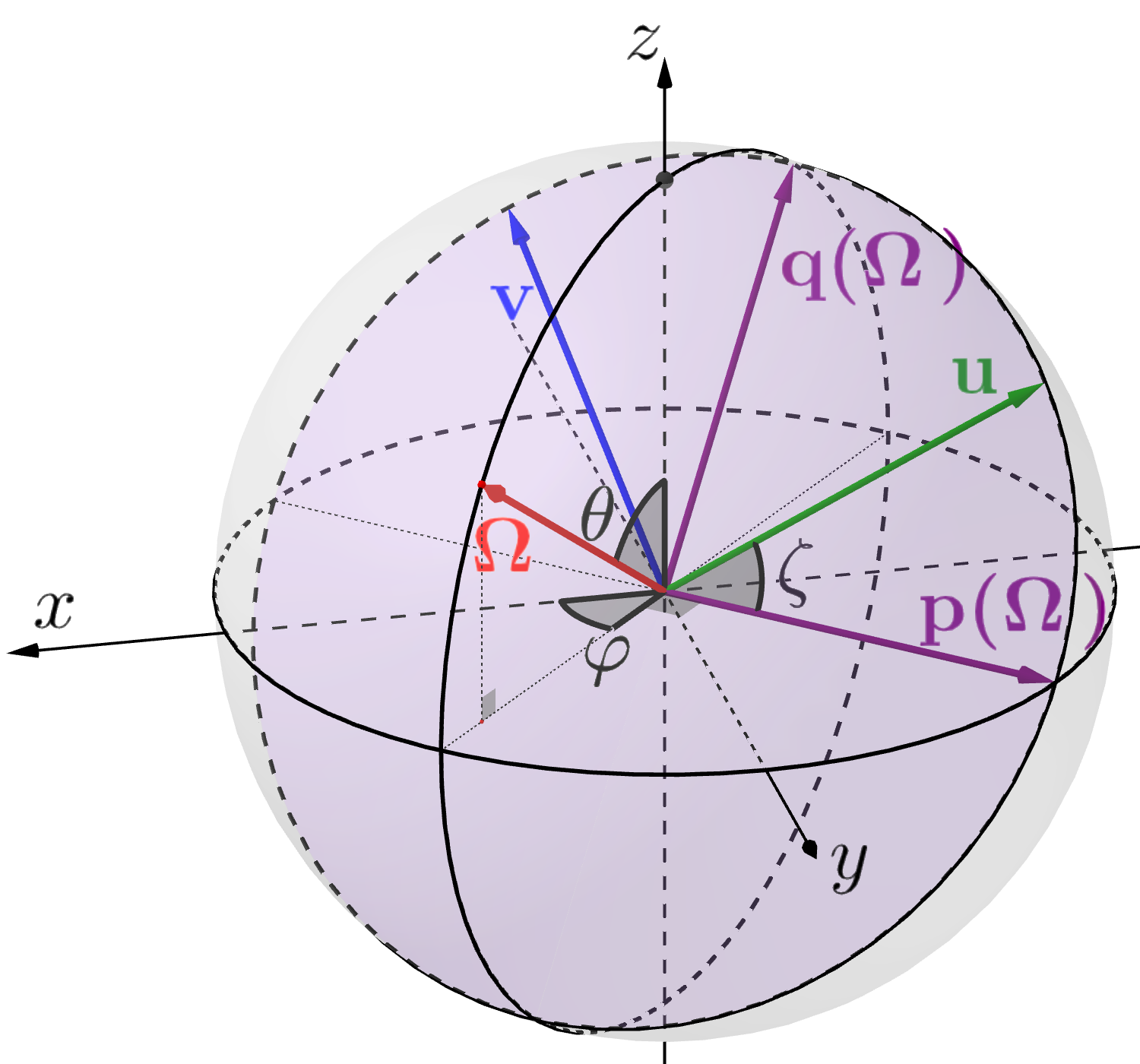} \label{fig:zeta}}
\caption{ (a) Pitch, yaw and roll angles of an aircraft with body orientation $[\Omega,\mathbf{u},\mathbf{v}]$ (original picture released under the Creative Commons CC0 license by \url{https://pixabay.com}).  (b)~Construction of the roll angle of $A= [\Omega, \mathbf{u}, \mathbf{v}]$, where the vectors  $\Omega$, $\mathbf{u}$ and $\mathbf{v}$ are respectively in red, green and blue. The local frame is $(\Omega,\mathbf{p}(\Omega),\mathbf{q}(\Omega))$ where $\mathbf{p}(\Omega)$ and $\mathbf{q}(\Omega))$  and the plane generated by them are in purple. $\mathbf{u}$ and $\mathbf{v}$ belong to this plane. $\zeta$ is the angle between $\mathbf{p}(\Omega)$ and $u$.}
\end{figure}

Let $A = [\Omega, \mathbf{u}, \mathbf{v}] \in $ SO$_3({\mathbb R})$ be a body-orientation. Let $\theta \in [0,\pi]$, $\varphi \in [0,2\pi)$ be the spherical coordinates of $\Omega$ defined by \eqref{eq:averageangles} (omitting the bars). We let $\{\Omega, \mathbf{e}_\theta, \mathbf{e}_\varphi\}$ be the local orthonormal frame associated with the spherical coordinates $(\theta, \varphi)$ and we define $\mathbf{p}(\Omega)=\mathbf{e}_\varphi$ and $\mathbf{q}(\Omega)=-\mathbf{e}_\theta$. Then we define the rotation matrix 
$$
\mathsf{R}(\Omega):= [\Omega, \mathbf{p}(\Omega), \mathbf{q}(\Omega) ] = \left(
\begin{array}{ccc}
\sin \theta \, \cos \varphi & -\sin \varphi &  -\cos \theta \, \cos \varphi \\
\sin \theta \, \sin \varphi &  \cos \varphi  &  -\cos \theta \, \sin \varphi \\ 
\cos \theta                 &              0     &  \sin \theta
\end{array}
\right).
$$ 
Since $\mathbf{u}$ and $\mathbf{v}$ belong to the plane spanned by $\mathbf{p}(\Omega)$ and $\mathbf{q}(\Omega)$, we let $\zeta \in [0, 2\pi)$ be the angle between $\mathbf{p}(\Omega)$ and $\mathbf{u}$. Then, it is an easy matter to show that $A = \mathsf{R}(\Omega) \, {\mathcal A}(\zeta, \mathbf{e}_1)$. In aircraft navigation, $\theta$, $\varphi$ and $\zeta$ are respectively called the pitch, yaw and roll angles: the pitch and yaw control the aircraft direction with respect to the vertical and in the horizontal plane respectively, while the roll controls the plane attitude (see Fig. \ref{fig:aircraft}). These angles are related to the Euler angles. The construction of the roll angle $\zeta$ is summarized in Figure \ref{fig:zeta}. Pursuing the analogy with aircraft navigation, we see from Fig.~\ref{fig:action_F_del} that $\mathbf{F}$ controls variations of pitch and yaw while $\delta$ controls variations of roll.

As an example, we examine the pitch, yaw and roll of the three solutions of the SOHB model \eqref{equationA} described in Section \ref{sectionthreesolutions}. 
\begin{enumerate}
\item FS:  ${\mathbb A}$ is constant and uniform. Then, the pitch, yaw and roll are also constant and uniform. 
\item MO: ${\mathbb A}$ is given by \eqref{millingsolution} (see Figs. \ref{fig:milling}). Using Eq. \eqref{millingsolution2}, we have $\mathsf{R}(\Omega) = {\mathcal A}(- \omega \, t, \mathbf{e}_3)$ and the roll is given by $\zeta = \xi z$. The pitch and yaw are constant and uniform. The roll is constant in time and is also uniform on planes of constant $z$. The non-trivial topology of the MO results from the roll making a complete turn when $z$ increases by the quantity $2 \pi / \xi$. 
\item HW: ${\mathbb A}$ is given by \eqref{helicalsolution} (see Fig. \ref{fig:helical}). Then, we have $\mathsf{R}(\Omega) =$ I$_3$ and $\zeta = \xi \, (x - \lambda \, t)$. The pitch and yaw are constant and uniform while the roll is uniform on planes of constant $x$. It depends on $x$ and time through the traveling phase $x - \lambda \, t$. Here, the non-trivial topology results from the roll making a complete turn when $x$ increases by the quantity $2 \pi / \xi$.
\end{enumerate}
The goal of the next section is to see how we can recover the roll field from the simulation of a large particle system.


\subsubsection{Roll polarization} 

As shown in the last section, the roll of the MO is uniform on planes of constant $z$. When simulating the MO by the IBM, we will use this property to compute an average roll on planes of constant $z$. To cope with the discreteness of the particles, we will rather consider slices comprised between two planes of constant $z$. If the distance $\Delta z$ between the planes is  chosen appropriately, we can access to both the average and the variance of the roll. They will be collected into one single vector, the Roll Polarization in planes of constant $z$ or RPZ. A similar quantity characterizes the HW, the Roll Polarization in planes of constant $x$ or RPX. Below, we detail the construction of the RPZ. Obviously the procedure is the same (changing $z$ into $x$) for the RPX. 

\begin{figure}[ht!]
\centering
\includegraphics[width= 9cm]{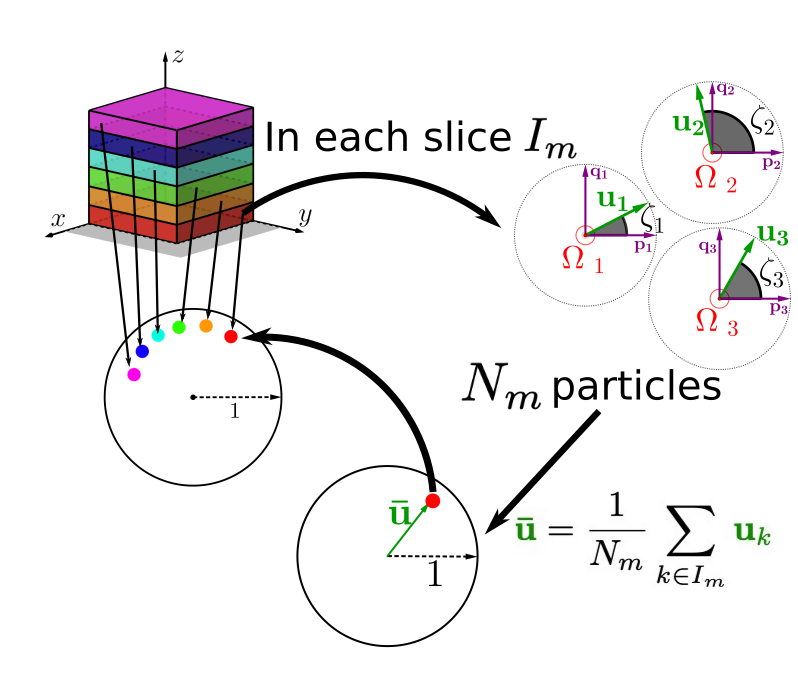}
\caption{Construction of the RPZ and graphical representation. The spatial domain $\mathcal{D}$ is partitioned into $M$ slices represented in different colors (top left). In each slice $S_m$, we have $I_m$ particles with roll $\zeta_k$ each of them plotted in the particle's local plane spanned by $\mathbf{p}(\Omega_k)$, $\mathbf{q}(\Omega_k)$ (top right: we plot $3$ particles in the slice $S_1$). Note that the local planes of different particles of the same slice may not coincide when imbedded in ${\mathbb R}^3$. For this given slice, the RPZ  $\mathbf{\bar{u}}_m$ is computed and plotted in ${\mathbb R}^2$ (bottom right). The RPZ has norm smaller than $1$ and belongs to the unit disk, whose boundary, the unit circle, is plotted for clarity. The RPZ of each slice is then plotted on a single figure in the same color as the slice it corresponds to (bottom left). This collection of points forms a discrete curve (here a fragment of a circle): the RPZ-curve.}
\label{fig:sliceaverage}
\end{figure}

We assume that the domain is a rectangular box of the form $\mathcal{D}: = [0,L_x] \times [0,L_y] \times [0,L_z]$, and $L_z = n \, (2 \pi / \xi)$ with $n \in {\mathbb Z} \setminus \{0\}$.  The domain $\mathcal{D}$ is partitioned into $M$ slices of fixed size across $z$, where $M$ is a fixed integer. For $m \in~\{1,\ldots,M\}$, the slice $S_m$ is defined by: 
$$
S_m :=[0,L_x] \times [0,L_y] \times \left[ \frac{m-1}{M} L_z , \frac{m}{M}L _z \right].
$$ 
Let us consider a system of $N$ agents with positions and body-orientations $(\mathbf{X}_k, A_k)$, indexed by $k \in \{1, \ldots, N\}$. Each body orientation $A_k$ has roll $\zeta_k \in [0,2\pi)$. We define the discrete RPZ for Slice $m$, $\mathbf{\bar{u}}_m$, by
\begin{equation}
\label{eq:uk}
\mathbf{\bar{u}}_m := \frac{1}{N_m} \sum_{k \in I_m} (\cos \zeta_k, \sin \zeta_k)^\mathrm{T}\in{\mathbb R}^2, 
\end{equation}
where $I_m=\{k \in \{ 1, \ldots, N \}, X_k \in S_m \}$ and $N_m$ is the cardinal of $I_m$. Note that the RPZ $\mathbf{\bar{u}}_m$ has norm smaller than one. The unit vector $\mathbf{\bar{u}}_m/ |\mathbf{\bar{u}}_m|$ or equivalently, its angle with the vector $(1,0)^\mathrm{T}$ gives the average roll in $S_m$. The euclidean norm $|\mathbf{\bar{u}}_m|$ is a measure of the variance of the set of roll angles $\{\zeta_k\}_{k \in I_m}$. If this variance is small, then $|\mathbf{\bar{u}}_m| \sim 1$, while if the variance is large, $|\mathbf{\bar{u}}_m| \ll 1$. When plotted in the plane ${\mathbb R}^2$, the set of RPZ $\{ \mathbf{\bar{u}}_m \}_{m=1, \ldots, M}$ forms a discrete curve referred to as the RPZ-curve. It will be used to characterize the topological state of the particle system. A summary of this procedure is shown in Figure \ref{fig:sliceaverage}. 

\subsubsection{Indicators of RPZ-curve morphology}

\begin{figure}[ht!]
\centering
\subfloat[]{\includegraphics[width= 5.2cm]{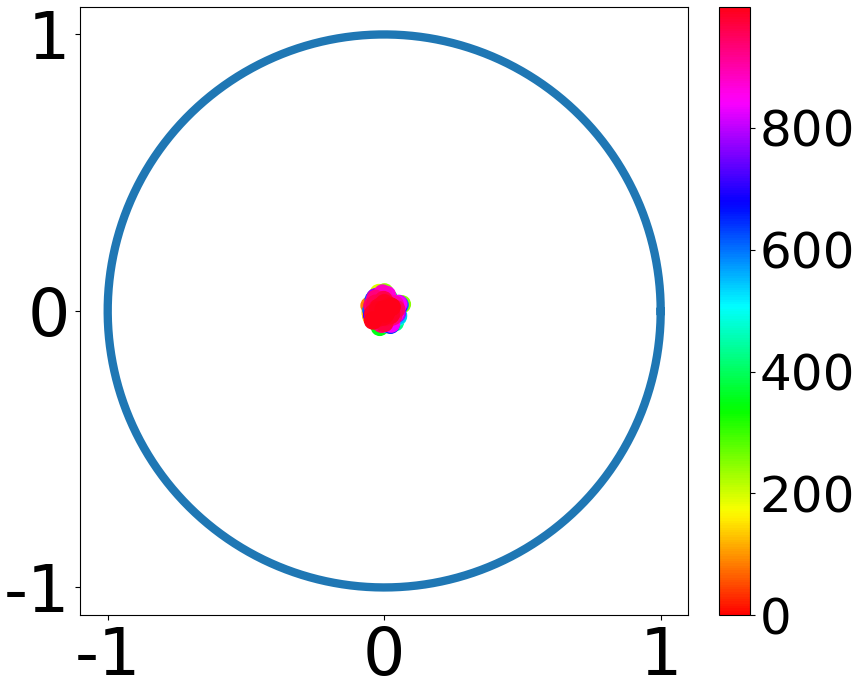}\label{fig:exampleslicedisorder}} \hspace{1cm}
\subfloat[]{\includegraphics[width= 5.2cm]{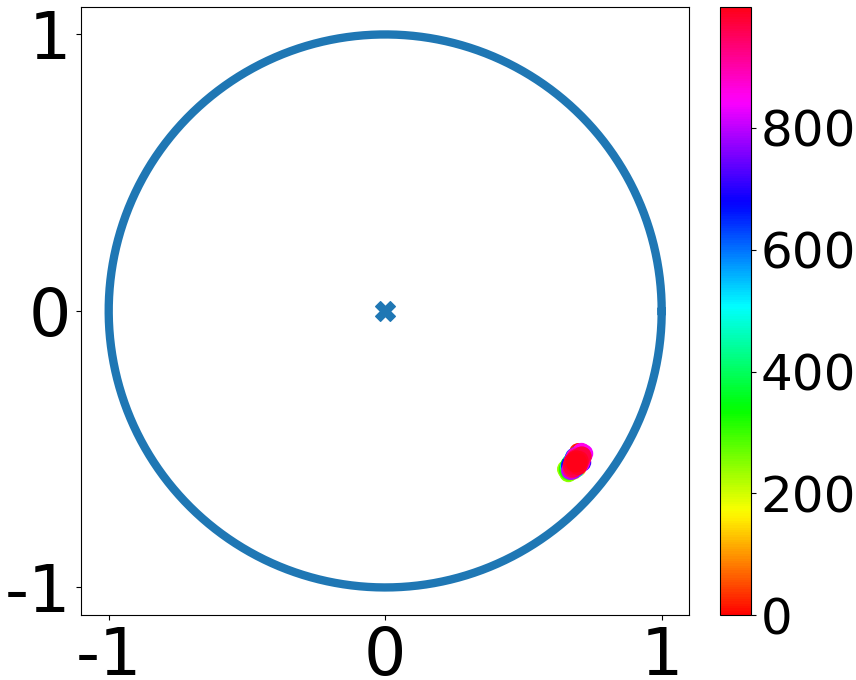}\label{fig:examplesliceflock}} 

\subfloat[]{\includegraphics[width= 5.2cm]{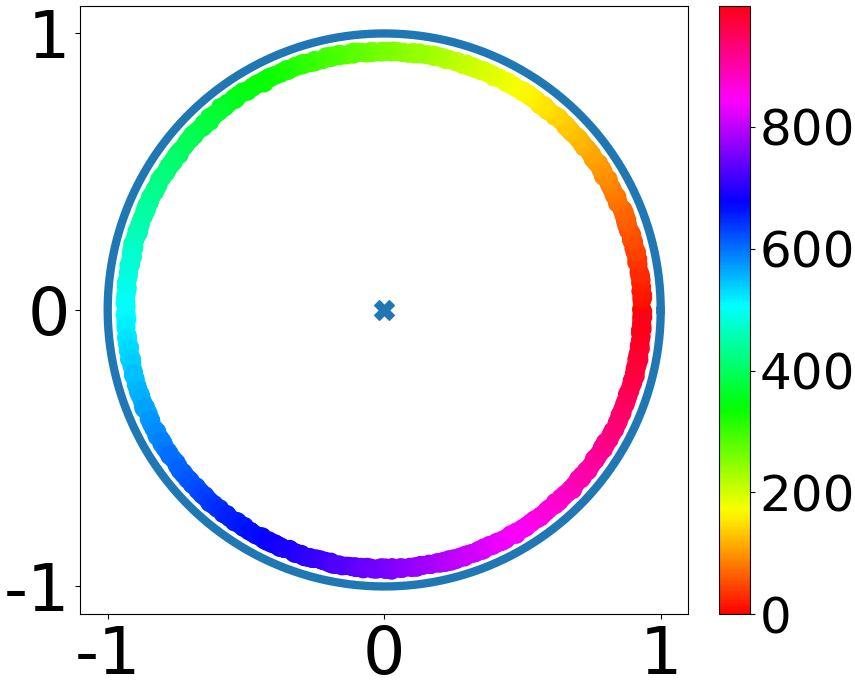}\label{fig:exampleslicetwist}} \hspace{1cm}
\subfloat[]{\includegraphics[width= 5.1cm]{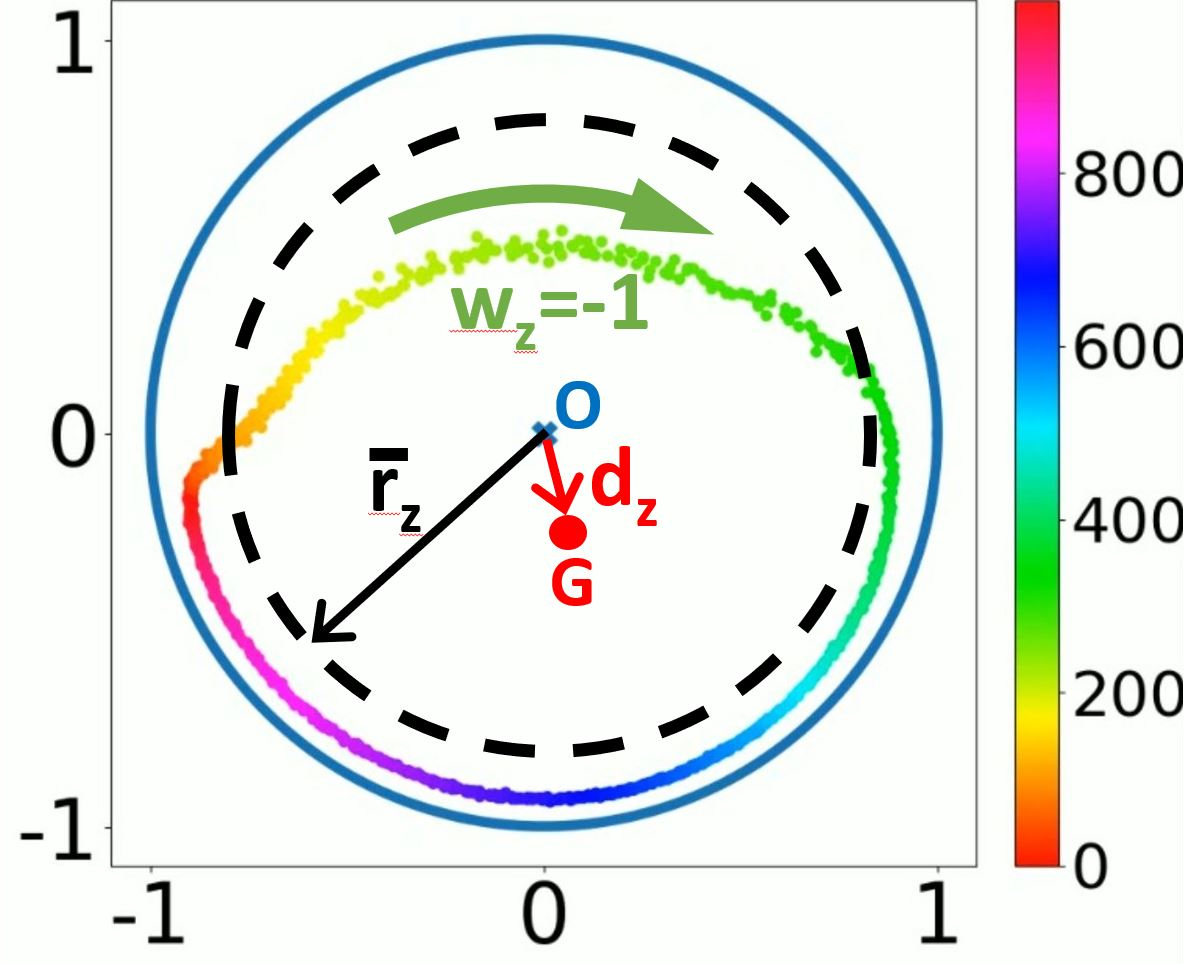}\label{fig:drw}}
\caption{Examples of RPZ-curves: in each figure, the roll Polarization RPZ vectors corresponding to $M=1000$ slices are plotted. The color bar to the right of each figure assigns a unique color to each slice. The same color is used to plot the corresponding RPZ. In each figure the unit circle and its center are represented in blue. (a) Disordered state: all RPZ concentrate near the origin.  (b) FS: all RPZ concentrate on a point close to the unit circle.  (c) MO \eqref{perptwist}: the RPZ-curve is a discrete circle centered at the origin and of radius close to unity. The total number of particles is $N=1.5\cdot10^6$. Note that in Figs. (a) and (b), all RPZ are superimposed and only the last one (in magenta color) is visible. (d) Quantifiers of RPZ curve morphology: point $G$ (in red) is the center-of-mass of the RPZ curve and $d_z$ is its distance to the origin $O$ (shown in blue).  The mean radius $\bar r_z$ of the RPZ curve is illustrated by the circle in black broken line which has same radius. The winding number, which is the number of turns one makes following the spectrum of colors in the same order as in the color bar from bottom to top (the green arrow indicates the direction of progression along the RPZ curve) is $w_z = -1$ in this example.}
\label{fig:examplesliceaverage}
\end{figure}

The RPZ-curve is shown in Figure \ref{fig:examplesliceaverage} (a) to (c), in the three following cases.
\begin{enumerate}
\item \textbf{Disordered state:}  the particles are drawn independently uniformly randomly in the product space $\mathcal{D}\times $ SO$_3({\mathbb R})$. For each $m$, the RPZ \eqref{eq:uk} is an average of uniformly distributed vectors on the circle and its norm is therefore close to 0. The RPZ-curve is thus reduced to the origin, as shown in Figure \ref{fig:exampleslicedisorder}; 
\item \textbf{FS:} the positions of the particles are drawn independently uniformly in $\mathcal{D}$ and their body-orientations independently according to a von Mises distribution $M_{{\mathbb A}_0}$ with a fixed mean body orientation ${\mathbb A}_0 \in $ SO$_3({\mathbb R})$. In this case, for all slices, the corresponding RPZ \eqref{eq:uk} is an average of identically distributed vectors on the circle whose distribution is peaked around the same point of the unit circle, and the peak is narrower as $\kappa$ is larger. Therefore, the RPZ vectors \eqref{eq:uk} concentrate on a point near the unit circle (Figure~\ref{fig:examplesliceflock}). The RPZ-curve reduces to a single point different from the origin; 
\item \textbf{MO:} the positions of the particles are drawn independently uniformly in $\mathcal{D}$. Then for a particle at position $\mathbf{x}$, its body-orientation is drawn independently according to a von Mises distribution $M_{{\mathbb A}_{\mbox{\scriptsize mill}} (0,z)}$ with  ${\mathbb A}_{\mbox{\scriptsize mill}} (0,z)$ defined by \eqref{perptwist} (with $\xi = 2 \pi / L_z$). This time, the von Mises distribution is peaked around a point which depends on $z$. For each slice, the position of the RPZ \eqref{eq:uk} depends on $m$. Since ${\mathbb A}_{\mbox{\scriptsize mill}} (0,z)$ is $L_z$-periodic, the RPZ-curve is a discrete closed circle (Figure~\ref{fig:exampleslicetwist}). Note that the RPX-curve of a HW is similar. 
\end{enumerate}

From Figure \ref{fig:examplesliceaverage}, we realize that three quantities of interest can be extracted from the RPZ-curve: 
\begin{enumerate}
\item the distance of its center of mass to the origin $d_z$: 
\begin{equation}
d_z = \Big| \frac{1}{M} \sum_{m=1}^M \mathbf{\bar{u}}_m \Big|, 
\label{eq:dz}
\end{equation}
\item its mean distance to the origin $\bar r_z$:
\begin{equation}
\bar r_z = \frac{1}{M} \sum_{m=1}^M |\mathbf{\bar{u}}_m|, 
\label{eq:rz}
\end{equation}
\item its winding number about the origin $w_z$: for $m \in \{1, \ldots, M \}$, let $\beta_m = \mathrm{arg} \big( (\mathbf{\bar{u}}_m)^1 + i (\mathbf{\bar{u}}_m)^2 \big) \in [0, 2 \pi)$ (with $\mathbf{\bar{u}}_m = ((\mathbf{\bar{u}}_m)^1, (\mathbf{\bar{u}}_m)^2)^\mathrm{T}$) and $\delta \beta_{m+1/2} \in [- \pi, \pi)$ be such that $\delta \beta_{m+1/2} \equiv \beta_{m+1} - \beta_m$ modulo $2 \pi$, where we let $\beta_{M+1} = \beta_1$. Then:
$$ 
w_z = \frac{1}{2 \pi} \sum_{m=1}^M \delta \beta_{m+1/2}, 
$$
(see e.g. \cite[p. 176]{hughes2012computer}). 
\end{enumerate}
The subscript $z$ indicates that the slicing has been made across $z$. Similar quantities with an index '$x$' will correspond to the slicing made across $x$. Fig. \ref{fig:drw} provides a graphical illustration of the triple $(d_z, \bar r_z, w_z)$. For the examples given above, this triple has the following values: 
\begin{eqnarray}
&& \mbox{Disordered state:} \, (d_z, \bar r_z, w_z) = (0, 0, \mbox{ND}), \, \mbox{where ND stands for ``undefined''},  
\\
&&  \mbox{FS:} \, (d_z, \bar r_z, w_z) \approx (1,1,0), \label{eq:flock_char} \\
&&  \mbox{MO:} \, (d_z, \bar r_z, w_z) \approx (0,1,w), \, \mbox{with} \, \,  w \not = 0. \label{eq:twist_char}
\end{eqnarray}
We have a similar conclusion with $(d_x, \bar r_x, w_x)$ for a disordered state or an FS. For an HW, we have $(d_x, \bar r_x, w_x) \approx (0,1,w)$ with $w \not = 0$. Thus, monitoring either or both triples (according to the situation) will give us an indication of the state of the system in the course of time. In particular, non-trivial topological states are associated with non-zero winding numbers $w_x$ or $w_z$. In practice, we will use the nonzero-rule algorithm to compute the winding numbers numerically \cite[p. 176]{hughes2012computer}.


\section{Topological phase transitions: are the MO and HW topologically protected?}
\label{sectionnumericalexperiments}

As pointed out in Section \ref{sec:topologysolutions}, for the IBM, the MO and HW are only metastable: they typically persist for a finite time before degenerating into a FS. This is in stark contrast with the macroscopic model for which they persist for ever. The transition of a MO or HW to a FS implies a topological change. To analyze whether the MO or HW are more robust due to their non-trivial topological structure (i.e. are topologically protected), we will compare them with similar but topologically trivial initial conditions (Sections \ref{subsec:setting}, \ref{sec:topoprotec} and \ref{subsec:repro}). We also test their robustness against perturbed initial conditions and show that, in this case, MO may transition to GS (Section \ref{sec:robustness}). In the Supplementary Material \ref{sec:rareevents}, we investigate rarer events, where an MO does not transition directly to an FS but through a HW.

%
%

\subsection{Initial conditions}
\label{subsec:setting}

In Section \ref{sec:topoprotec}, we will compare the solutions of the IBM with different initial conditions using the perpendicular or parallel twists as building blocks. Some will have a non-trivial topology and the others, a trivial one. Specifically we define the following initial conditions. 

\subsubsection{Milling orbit}

Let $\mathcal{D}=[0,L]\times[0,L]\times[0,2L]$ be a rectangular domain with periodic boundary conditions and let $\xi = 2 \pi / L$. We consider the following two initial conditions:

\begin{itemize} 
\item \textbf{Double mill} initial condition MO1: 
\begin{equation}
{\mathbb A}_{m,1}(0,z) = {\mathcal A}(\xi \, z, \mathbf{e}_1), \quad z \in [0, 2L],
\label{eq:inimill_nt}
\end{equation}
where we recall again that ${\mathcal A}(\theta, \mathbf{n})$ is the rotation of axis $\mathbf{n} \in {\mathbb S}^2$ and angle $\theta \in {\mathbb R}$ defined by \eqref{eq:rodrigues}. This initial condition has non-trivial topology: the curve generated by the end of the vector $\mathbf{u}$ in the $(y,z)$-plane as $z$ ranges in $[0, 2L]$ makes two complete turns around the origin in the same direction. Thus, this initial condition has winding number equal to $2$. 

\item \textbf{Opposite mills} initial condition MO2: 
\begin{equation}
{\mathbb A}_{m,2}(0,z) = \left\{ \begin{array}{ll}
{\mathcal A}(\xi \, z, \mathbf{e}_1),& \quad z \in [0, L], \\
{\mathcal A}(- \xi \, z, \mathbf{e}_1),& \quad z \in [L, 2L]. 
\end{array} \right.
\label{eq:inimill_t}
\end{equation}
This initial condition has trivial topology: starting from $z=0$, the curve generated by the end of the vector $\mathbf{u}$ makes one complete turn around the origin in the counterclockwise direction until it reaches $z=L$ but then reverses its direction and makes a complete turn in the clockwise direction until it reaches $z = 2L$. Thus, this initial condition has winding number equal to $0$ and has trivial topology. 

\item \textbf{Perturbed double mill} initial condition MO3:
\begin{equation}
{\mathbb A}_{m,3}(0,z) = {\mathcal A}(\xi \, z + \sqrt{\sigma}B_z, \mathbf{e}_1), \quad z \in [0, 2L],
\label{eq:inimill_perturb}
\end{equation}
where $(B_z)_z$ is a given one-dimensional standard Brownian motion in the $z$ variable and $\sigma>0$ is a variance parameter which sets the size of the perturbation. The Brownian motion is subject to $B_0 = B_{2L} = 0$ (i.e. it is a Brownian bridge). Similarly to the initial condition MO1 \eqref{eq:inimill_nt}, this initial condition has a nontrivial topology, in this case a winding number equal to 2. 

\end{itemize}

\subsubsection{Helical traveling wave}

Let now $\mathcal{D}=[0,2L]\times[0,L]\times[0,L]$. Compared to the previous case, the domain has size $2L$ in the $x$-direction instead of the $z$-direction. Let again $\xi = 2 \pi / L$. We consider now the following two initial conditions:

\begin{itemize} 
\item \textbf{Double helix} initial condition HW1: 
\begin{equation}
{\mathbb A}_{h,1}(0,x) = {\mathcal A}(\xi \, x, \mathbf{e}_1), \quad x \in [0, 2L],
\label{eq:inihelic_nt}
\end{equation}
This initial condition has non-trivial topology and has winding number equal to $2$ by the same consideration as for initial condition MO1. 

\item \textbf{Opposite helices} initial condition HW2: 
\begin{equation}
{\mathbb A}_{h,2}(0,x) = \left\{ \begin{array}{ll}
{\mathcal A}(\xi \, x, \mathbf{e}_1),& \quad x \in [0, L], \\
{\mathcal A}(- \xi \, x, \mathbf{e}_1),& \quad x \in [L, 2L]. 
\end{array} \right.
\label{eq:inihelic_t}
\end{equation}
Again, by the same considerations as for MO2, this initial condition has trivial topology, i.e.  winding number equal to $0$. 
\end{itemize}


\subsection{Observation of topological phase transitions}
\label{sec:topoprotec}

\begin{figure}[ht!]
\centering
\subfloat[]{\includegraphics[height=4cm]{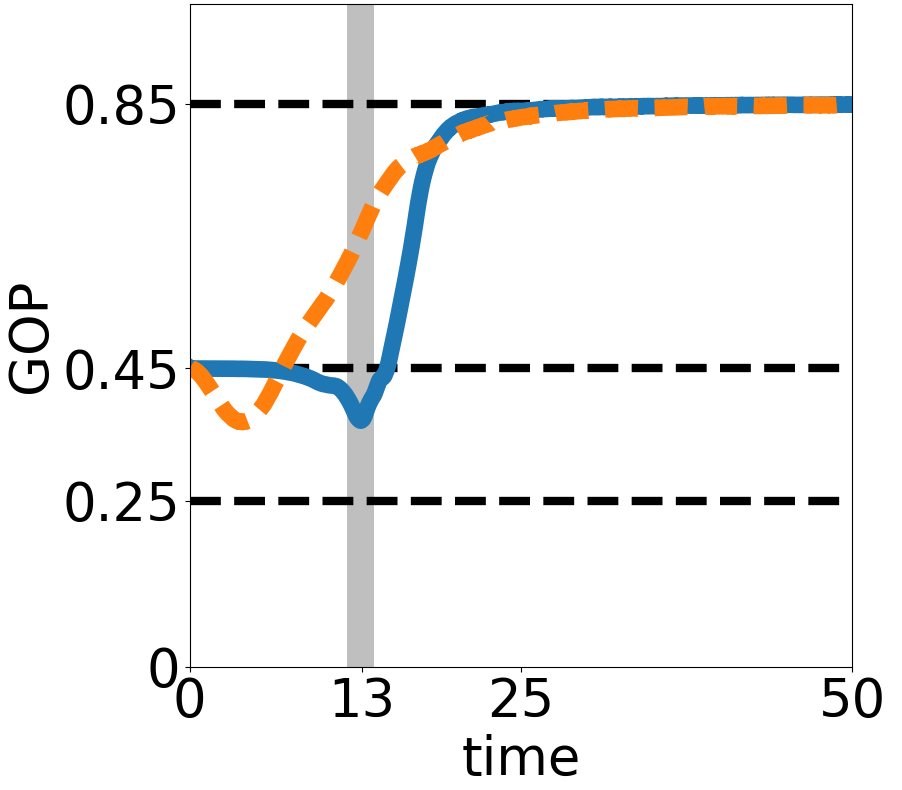}\label{subfig:topo_protec_perptwist}}
\subfloat[]{\includegraphics[height= 4cm]{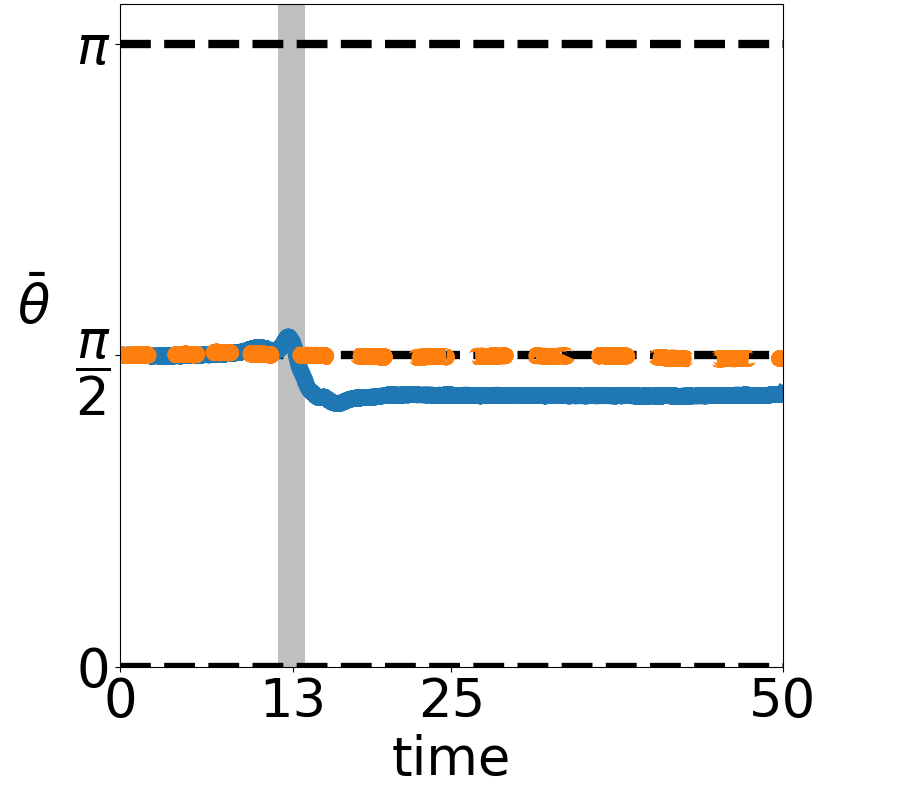}\label{subfig:milling_to_flocking_theta}} 
\subfloat[]{\includegraphics[height= 4cm]{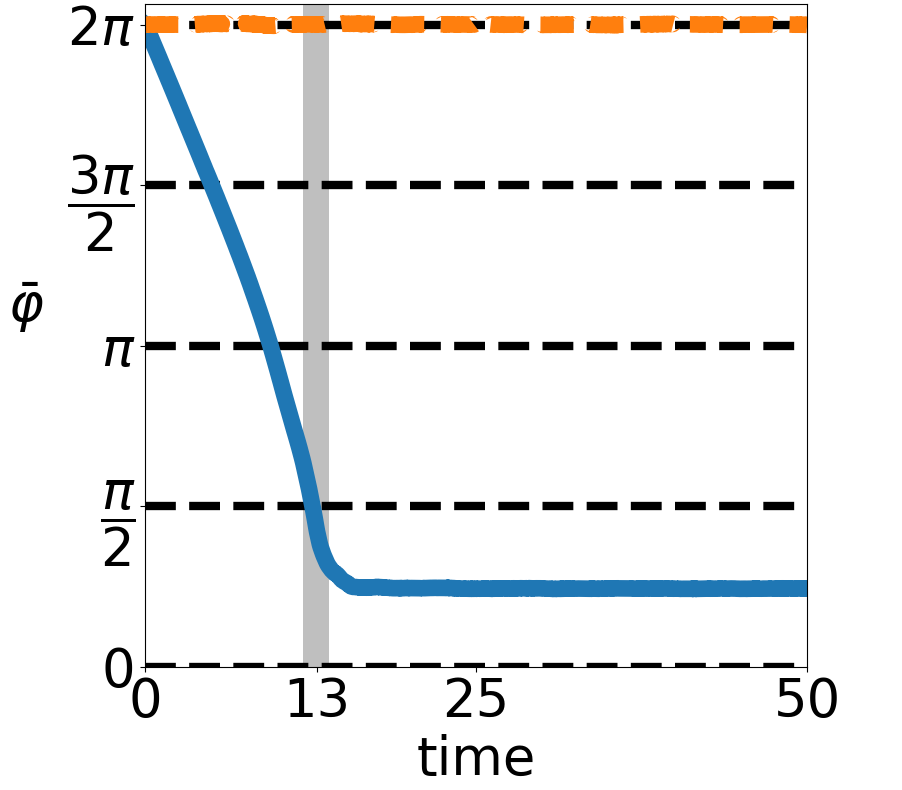}\label{subfig:milling_to_flocking_phi}}

\subfloat[]{\includegraphics[height= 4cm]{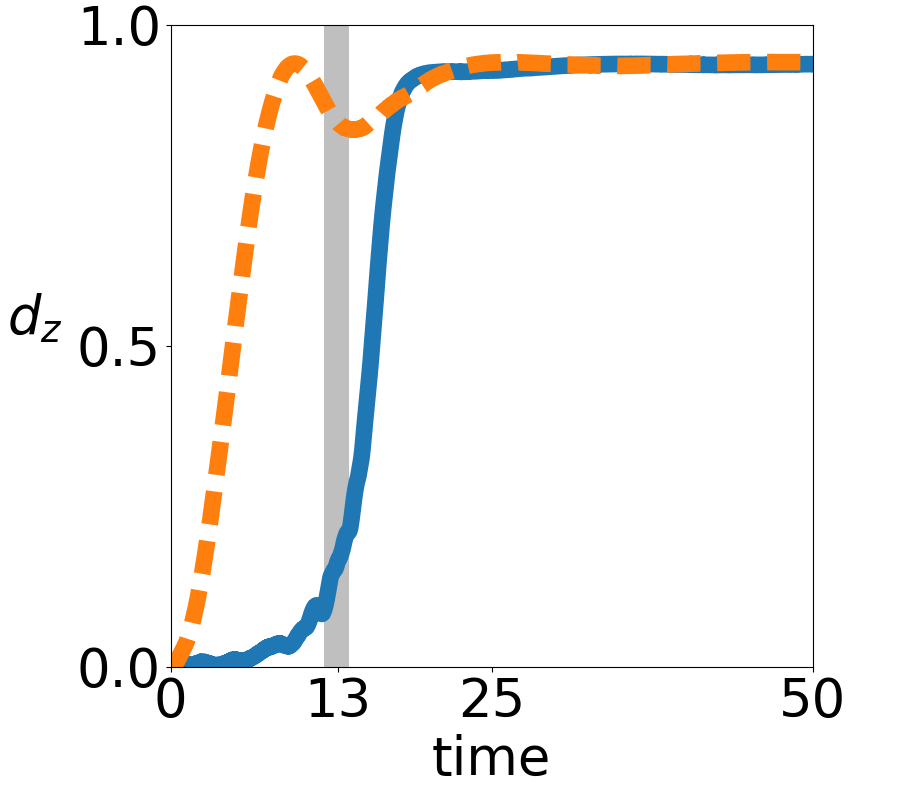}\label{subfig:double_milling_d_z}}
\subfloat[]{\includegraphics[height=4cm]{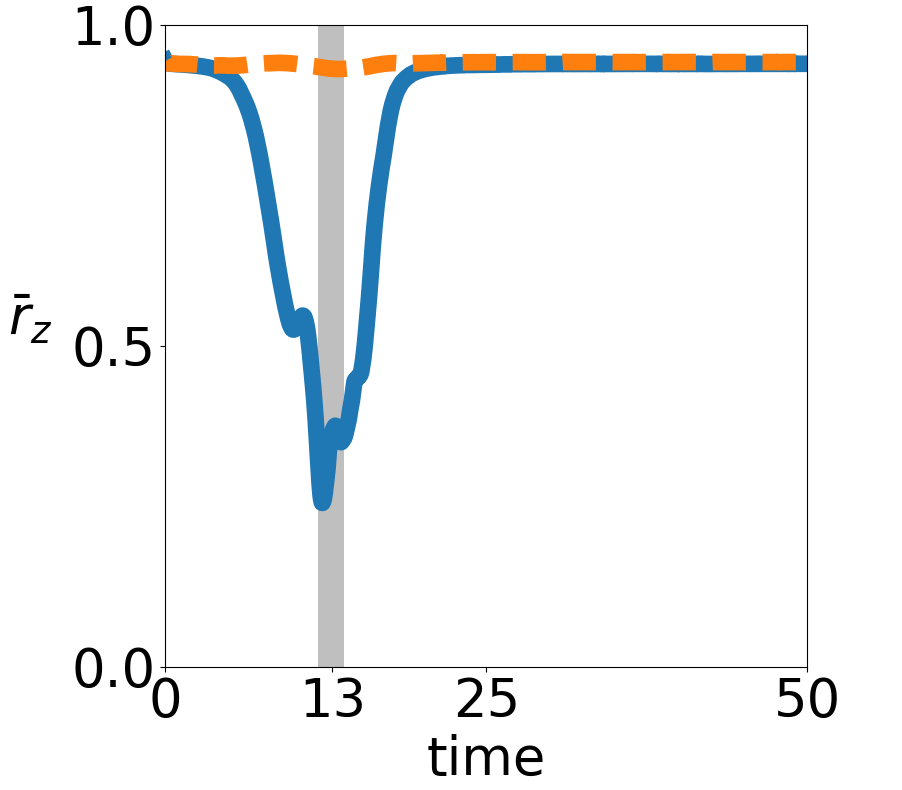}\label{subfig:double_milling_r_z}}
\subfloat[]{\includegraphics[height=4cm]{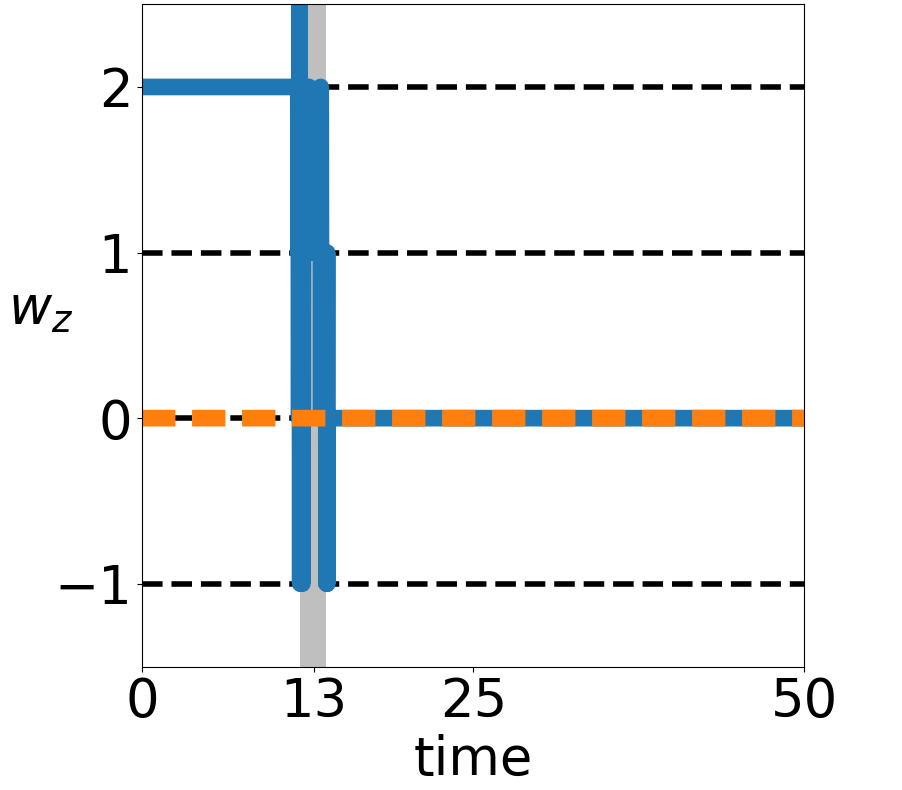}\label{subfig:double_milling_wn_z}}
\caption{Examples of solutions of the IBM for initial conditions sampled from the double mill MO1 (plain blue curves) and the opposite mills MO2 (boken orange curves). The following indicators are plotted as functions of time:  
(a) Global Order Parameter (GOP) (see Eq. \eqref{eq:orderparameter}). Horizontal lines at GOP values $0.25$, $0.45$ and $0.85$ materialize the special values GOP$_0$, GOP$_1$ and GOP$_2$ respectively corresponding to totally disordered states, MO or HW, and FS (see Eqs. \eqref{eq:opmilling}-\eqref{eq:gap_estimates}). 
(b) Pitch angle $\bar \theta$ of the global particle average direction $\bar \Omega$ (see \eqref{eq:averageangles}). 
(c)~Yaw $\bar \varphi$ of $\bar \Omega$. 
(d) Distance of center of mass of RPZ curve to the origin $d_z$ (see \eqref{eq:dz}). 
(e)~Mean distance of RPZ curve to the origin $\bar r_z$ (see \eqref{eq:rz}). 
(f) Winding number of RPZ curve $w_z$ (see \eqref{eq:dz}). 
Gray shaded zones highlight a small region around the time of minimal GOP for the MO1 solution. 
Parameters: $N=3 \, 10^6$, $R=0.025$, $\kappa = 10$, $\nu=40$, $c_0=1$, $L=1$, $\xi = 2 \pi$. See also Videos \ref{vid:milling_to_flocking_particles} to \ref{vid:trivial_milling_RPZ} in Section \ref{appendix:listvideos}.}
\label{fig:topo_protec_graphs_mills}
\end{figure}

\begin{figure}[ht!]
\centering
\subfloat[]{\includegraphics[height= 4cm]{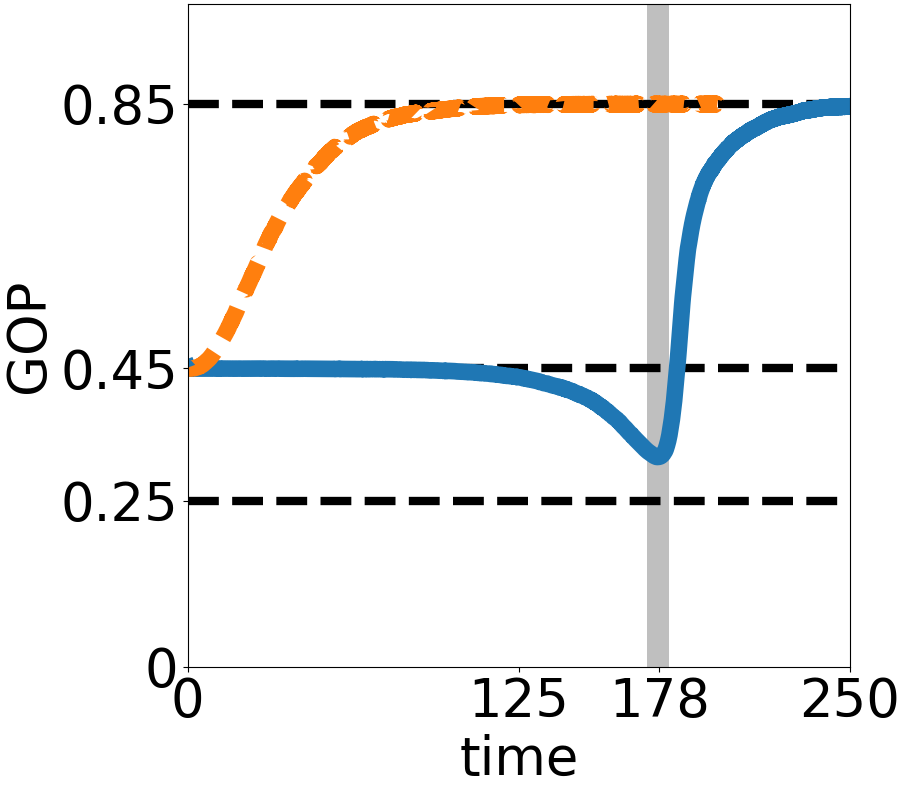}\label{subfig:topo_protec_paratwist}}
\subfloat[]{\includegraphics[height= 4cm]{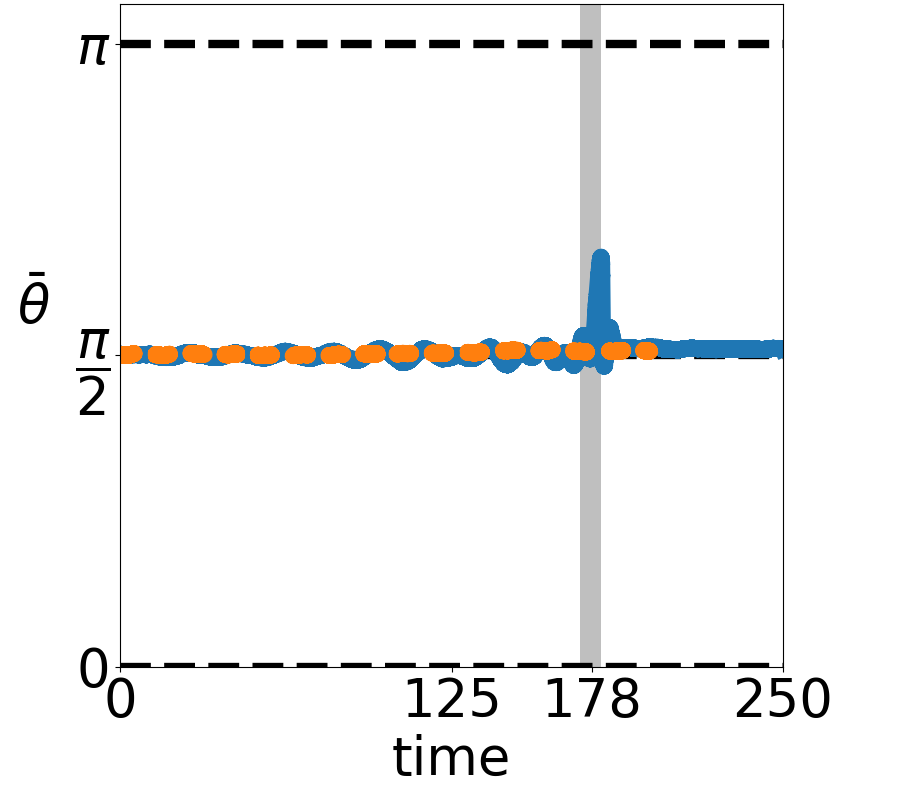}\label{subfig:helical_to_flocking_theta}} 
\subfloat[]{\includegraphics[height= 4cm]{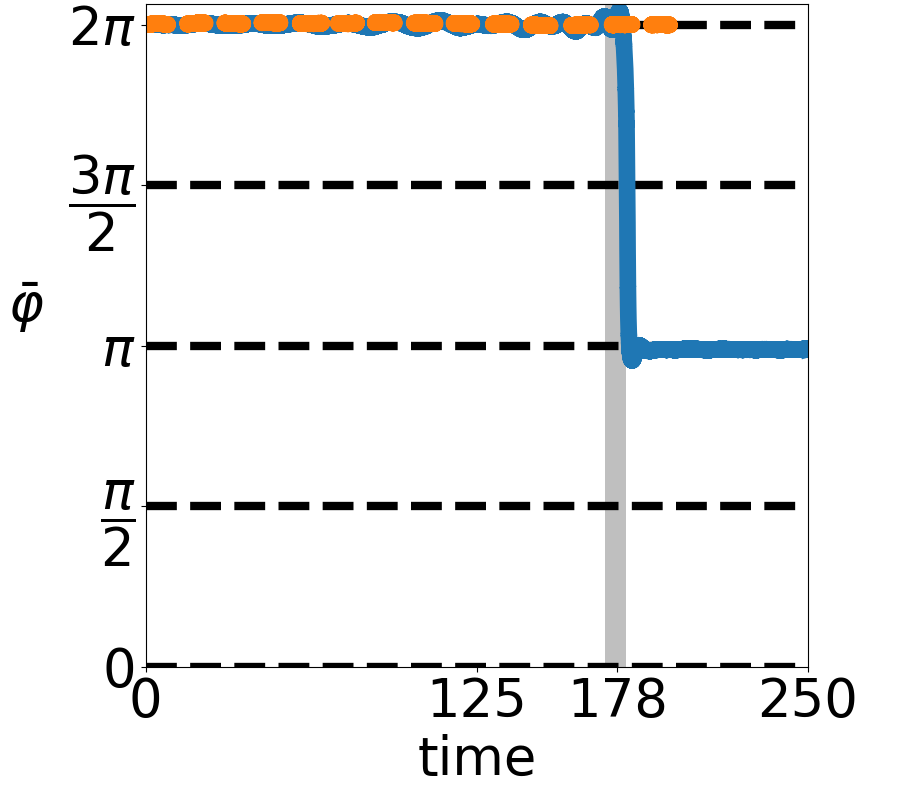}\label{subfig:helical_to_flocking_phi}}

\subfloat[]{\includegraphics[height= 4cm]{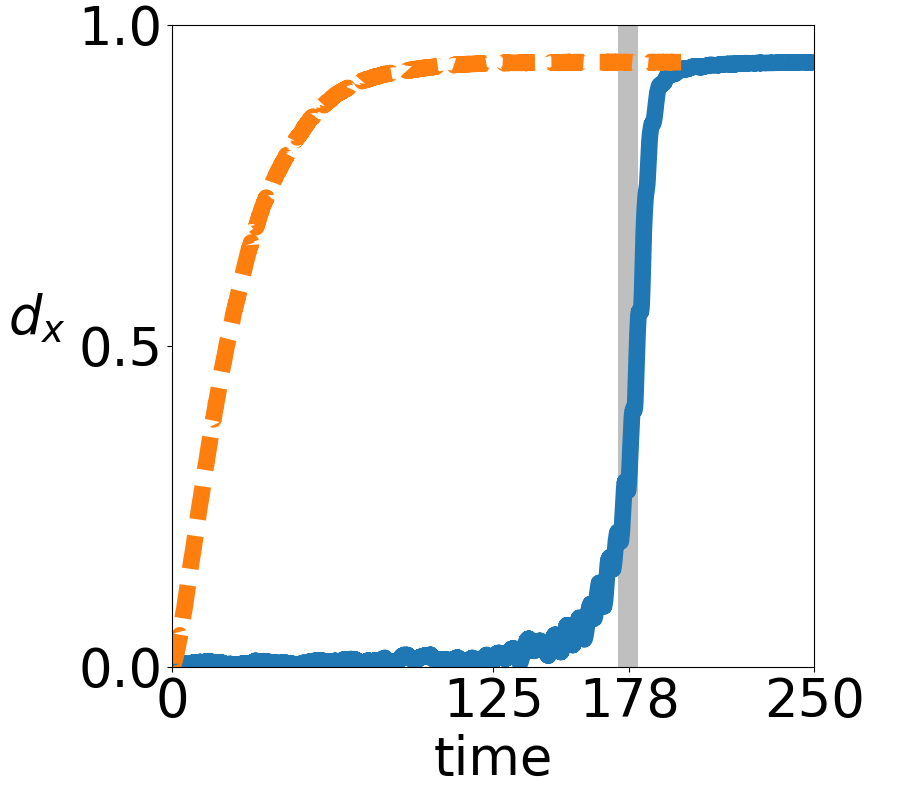}\label{subfig:double_helical_d_x}}
\subfloat[]{\includegraphics[height=4cm]{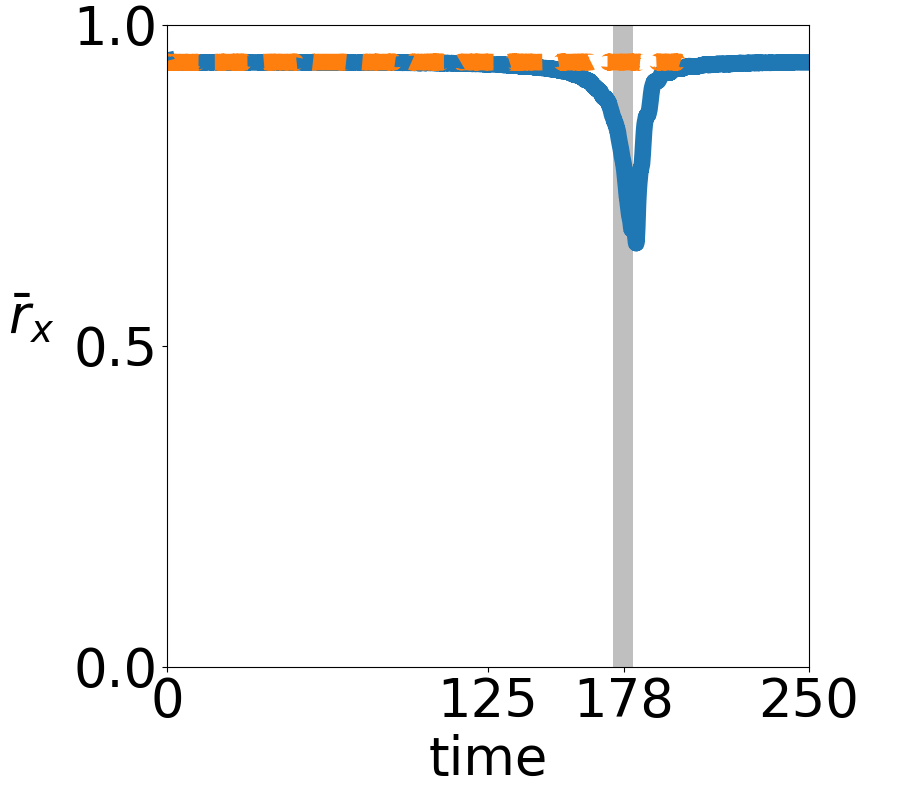}\label{subfig:double_helical_r_x}}
\subfloat[]{\includegraphics[height=4cm]{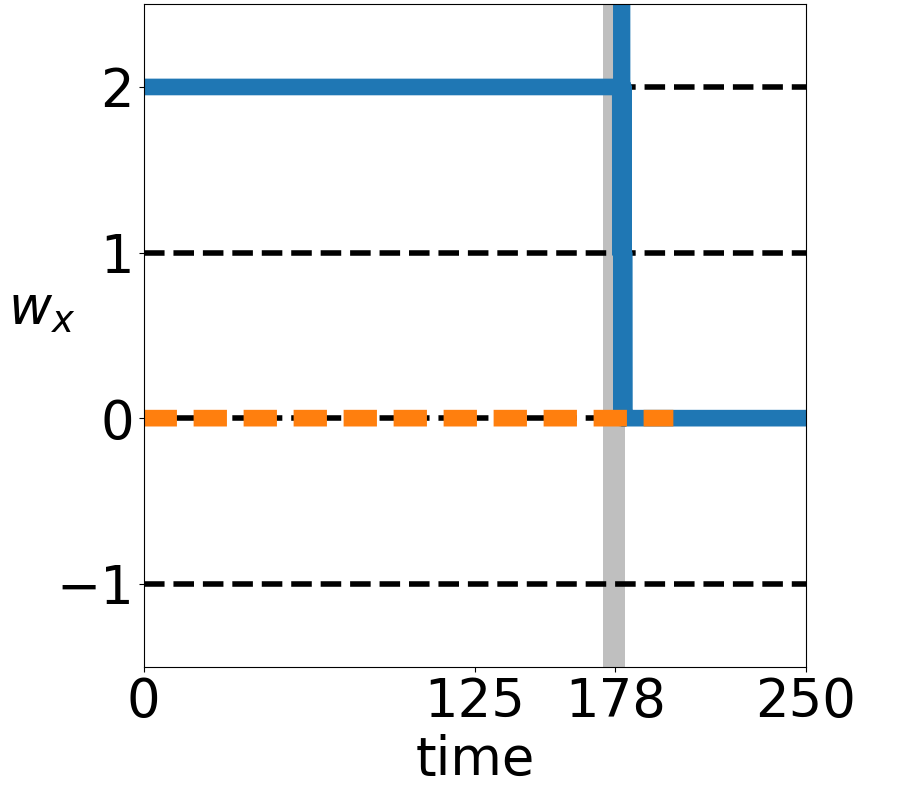}\label{subfig:double_helical_wn_x}}
\caption{Examples of solutions of the IBM for initial conditions sampled from the double helix HW1 (plain blue curves) and the opposite helices HW2 (broken orange curves). The following indicators are plotted as functions of time:  
(a) Global Order Parameter (GOP). 
(b)~Pitch angle $\bar \theta$ of $\bar \Omega$. 
(c) Yaw $\bar \varphi$ of  $\bar \Omega$. 
(d) Distance of center of mass of RPX curve to the origin $d_x$. 
(e) Mean distance of RPX curve to the origin $\bar r_x$. 
(f) Winding number of RPX curve $w_x$. 
Gray shaded zones highlight a small region around the time of minimal GOP for the HW1 solution. 
The HW2 and HW1 solutions are computed during 200 and 250 units of time respectively. The two simulations have reached equilibrium by their final time. 
Parameters: $N=3 \, 10^6$, $R=0.025$, $\kappa = 10$, $\nu=40$, $c_0=1$, $L=1$, $\xi = 2 \pi$. See caption of Fig. \ref{fig:topo_protec_graphs_mills} for further indications. See also Videos \ref{vid:helical_to_flocking_particles} to \ref{vid:trivial_helical_RPX} in Section \ref{appendix:listvideos}.
}
\label{fig:topo_protec_graphs_helices}
\end{figure}

We initialize the IBM by drawing $N$ positions independently uniformly randomly in the spatial domain and  $N$ body-orientations independently from the von Mises distribution $M_{{\mathbb A}(0,\mathbf{x})}$ where $ {\mathbb A}(0,\mathbf{x})$ is one of the initial conditions MO1 or MO2. Then, we run the IBM and record the various indicators introduced in Section \ref{sec:tools} as functions of time. The results are plotted in Fig. \ref{fig:topo_protec_graphs_mills}, as plain blue lines for the solution issued from MO1 (the topologically non-trivial initial condition), and as broken orange lines for that issued from MO2 (the topologically trivial one). We proceed similarly for the two initial conditions HW1 and HW2 and display the results in Fig. \ref{fig:topo_protec_graphs_helices}. See also Videos~\ref{vid:milling_to_flocking_particles} to \ref{vid:trivial_milling_RPZ} in Section \ref{appendix:listvideos} supplementing Fig. \ref{fig:topo_protec_graphs_mills} and Videos \ref{vid:helical_to_flocking_particles} to \ref{vid:trivial_helical_RPX} supplementing Fig. \ref{fig:topo_protec_graphs_helices}.

Figs. \ref{subfig:topo_protec_perptwist} and \ref{subfig:topo_protec_paratwist} display the GOP. We observe that, for all initial conditions, the GOP has initial value GOP$_1$, which is consistent with the fact that the initial conditions are either MO or HW. Then, again, for all initial conditions, at large times, the GOP has final value GOP$_2$ which indicates that the final state is a FS. This is confirmed by the inspection of the second line of figures in Figs. \ref{fig:topo_protec_graphs_mills} and \ref{fig:topo_protec_graphs_helices} which provide the triplet of topological indicators $(d_z, \bar r_z, w_z)$ for MO solutions and $(d_x, \bar r_x, w_x)$ for HW solutions. Specifically, $d_z$ and $d_x$ are given in Figs.~\ref{subfig:double_milling_d_z} and~\ref{subfig:double_helical_d_x} respectively, $\bar r_z$  and $\bar r_x$ in Figs. \ref{subfig:double_milling_r_z} and \ref{subfig:double_helical_r_x},  and $w_z$ and $w_x$ in Figs.~\ref{subfig:double_milling_wn_z} and~\ref{subfig:double_helical_wn_x}. Initially both triplets corresponding to MO1 or HW1 solutions have value $(0,1,2)$ as they should (see~\eqref{eq:twist_char}). Their final value is $(1,1,0)$ which indicates a FS (see \eqref{eq:flock_char}). The fact that the final state is a FS implies, for MO1 and HW1, first that the IBM has departed from the MO and HW exact solutions of the macroscopic model described in Sections~\ref{subsubsec:milling} and~\ref{subsubsec:helical}, and second, that a topological phase transition has taken place, bringing the topologically non-trivial MO1 and HW1 to a topologically trivial FS. For the topologically trivial MO2 and HW2 initial conditions, no topological phase transition is needed to reach the FS. The differences in the initial topology of the solutions induce strong differences in the trajectories followed by the system. 

For the topologically non-trivial initial conditions MO1 or HW1, the system remains in the MO or HW state for some time; hence it follows the macroscopic solution during this phase. Indeed, the GOP displays an initial plateau at the value GOP$_1$, while the triplet of topological indicators stays at the value $(0,1,2)$, which characterize the MO or HW state. For MO1, this is also confirmed by the yaw $\bar \varphi$ (Fig. \ref{subfig:milling_to_flocking_phi}, blue curve), which varies linearly in time and by the pitch $\bar \theta$ (Fig. \ref{subfig:milling_to_flocking_theta} blue curve) which is constant in time, consistently with the MO solution of the macroscopic model (Section \ref{subsubsec:milling}) (see also Fig. \ref{fig:phiN} for the linear variation of the yaw). The duration of this initial phase, also referred to as the persistence time, is significantly longer for HW1 than for MO1. In our experiments, the former can reach several hundred units of time and sometimes be infinite (up to our computational capacity). By contrast, the latter is random and of the order of ten units of time. After this initial plateau, the GOP decreases until it reaches a minimum at a time highlighted in Figs. \ref{fig:topo_protec_graphs_mills}, \ref{fig:topo_protec_graphs_helices} and subsequent figures by a gray shaded zone, showing that the system passes through a state of maximal disorder. Around that time, $\bar r$ has a sharp drop which is another confirmation of an increased disorder. The topological transition precisely occurs at this time with a transition of the winding number from $2$ to $0$ through a short sequence of oscillations. However, $\bar r$ has not reached $0$ and $d$ has already started to increase, which suggests that disorder is not complete. At this time also, the linear variation of $\bar \varphi$ suddenly stops and $\bar \varphi$ remains constant afterward, while $\bar \theta$ shows a small oscillation and jump. For HW1, $\bar\theta$ and $\bar\varphi$ are initially plateauing with small oscillations. At the time when the system leaves the HW state (around $t\simeq178$), we observe a sudden drop of $\bar\varphi$ from $2\pi$ to $\pi$ which indicates that the system suddenly reverses its average direction of motion. The GOP starts to decrease significantly before this time so we can infer that during the time period between $t\simeq125$ and $t\simeq178$, even though the mean direction of motion $\bar\Omega$ remains constant, groups of particles of almost similar proportions are moving in opposite directions, which preserves the average direction of motion (and may explain the oscillations during the initial persistence phase). This is confirmed by Video \ref{vid:helical_to_flocking_particles} (see description in Section \ref{appendix:listvideos}). Then, once this minimum is reached, the GOP increases quickly to finally reach the value GOP$_2$ of the FS. Likewise, $\bar r$ and $d$ quickly reach the value~$1$ while the winding number stays at the value $0$. 

By contrast to the previous case, the system immediately leaves the topologically trivial initial conditions MO2 or HW2 as shown by the GOP immediately leaving the value GOP$_1$. For HW2  the GOP increases right after initialization and smoothly reaches the value GOP$_2$, at a much earlier time than HW1. The trend is different for MO2. In this case, the GOP first decreases. Then, after a minimum value, it increases again and smoothly reaches the value GOP$_2$ at a time similar to MO1. The initial decay of the GOP for the MO2 solution can be explained by the fact that the macroscopic direction $\Omega$ turns in opposite directions for the two opposite mills, thus decreasing the global order. For HW2, the macroscopic direction stays constant and uniform. So, it is the same for the two opposite helices, giving rise to a larger GOP. The mean radii $\bar r_z$ and $\bar r_x$ stay constant it time, showing that the evolutions of MO2 and HW2 do not involve phases of larger disorder. The quantity $d_x$ increases monotonically towards the value $1$ while $d_z$ is subject to some oscillations close to convergence. This is due to the fact that the RPZ or RPX curves stay arcs of circles with decreasing arc length for the RPX and  with some arc length oscillations for the RPZ as displayed in Videos \ref{vid:trivial_milling_RPZ} and \ref{vid:trivial_helical_RPX}. Of course, the winding number stays constant equal to $0$ as it should for topologically trivial solutions. In both the MO2 and HW2 cases, $\bar \theta$ and $\bar \varphi$ remain constant throughout the entire simulation. In the MO2 case, this is the consequence of the two counter-rotating mills which preserve the direction of motion on average. In the HW2 case, this is due to the fact that there is no variation of the direction of motion for HW solutions in general (see also Video~\ref{vid:trivial_milling_particles} and Video \ref{vid:trivial_helical_particles}). Again, we observe that the convergence towards the FS takes more time for HW2 than for MO2.  This points towards a greater stability of the HW-type solutions compared to the MO ones.

\subsection{Reproducibility}
\label{subsec:repro}

\begin{figure}[ht]
\centering
\includegraphics[height= 5cm]{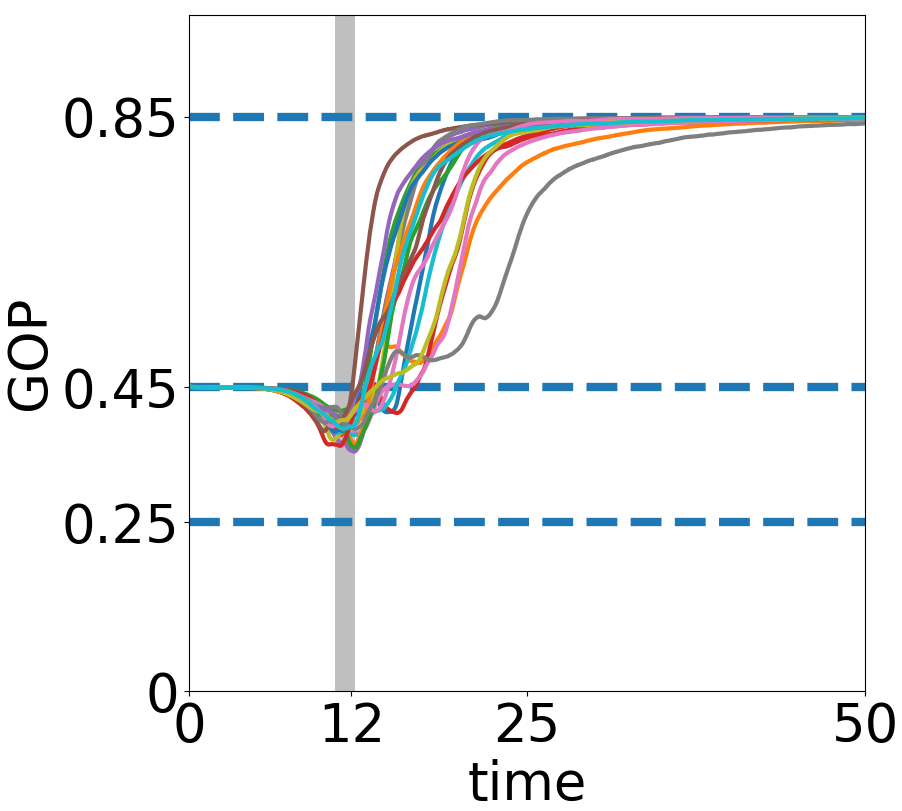}
\caption{GOP as a function of time for 20 independent simulations of the transition from a MO to a FS starting from MO1. The parameters are the same as the ones on Figure~\ref{fig:topo_protec_graphs_mills}.
}
\label{fig:double_mill_many}
\end{figure}

Since the IBM is a stochastic model, one may wonder whether Figs.~\ref{fig:topo_protec_graphs_mills} and~\ref{fig:topo_protec_graphs_helices} are representative of a typical solution. In Fig. \ref{fig:double_mill_many}, the GOP is plotted as a function of time for 20 independent simulations with MO1 initial conditions and the same parameters as in Fig. \ref{fig:topo_protec_graphs_mills} (blue curves). The same features as in Fig. \ref{fig:topo_protec_graphs_mills} are observed, namely: (i) an initial stable milling phase which lasts about 10 units of time; (ii) a decrease of the GOP between approximately 10 to 15 units of time; (iii) a subsequent increase of the GOP which reaches the value $\mathrm{GOP}_2$ of the FS. A similar reproducibility of the results has been observed for the other initial conditions (MO2, HW1, HW2) (not shown). 

\subsection{Robustness against perturbations of the initial conditions}\label{sec:robustness}

\begin{figure}[ht!]
\centering
\subfloat[]{\includegraphics[height= 3.5cm]{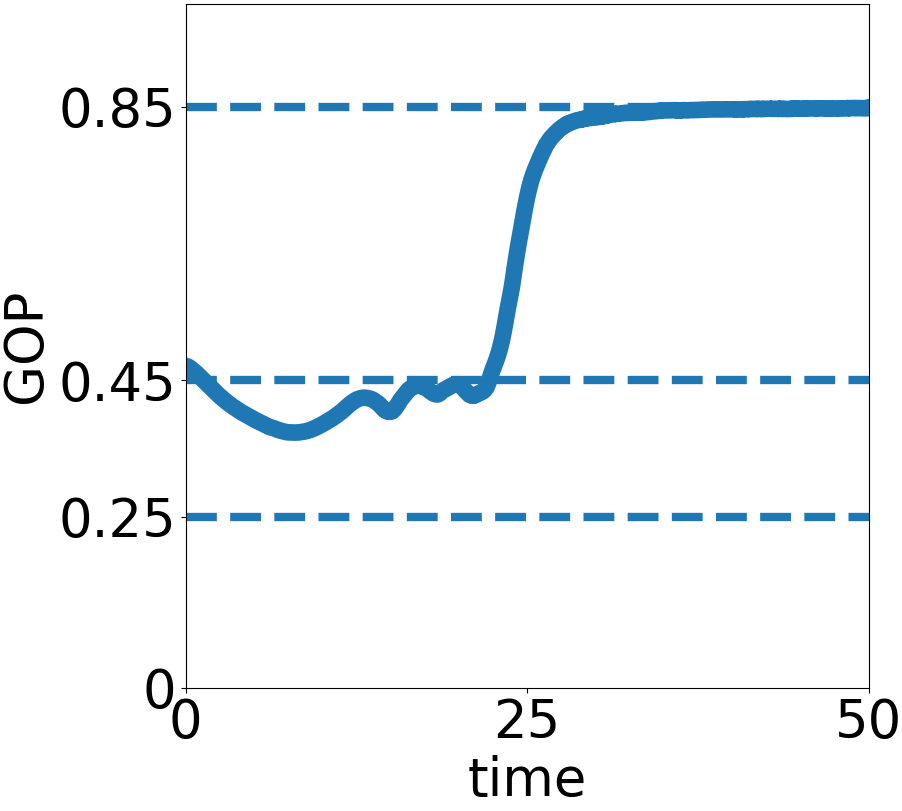}\label{fig:perturb_134_op}}
\subfloat[]{\includegraphics[height= 3.5cm]{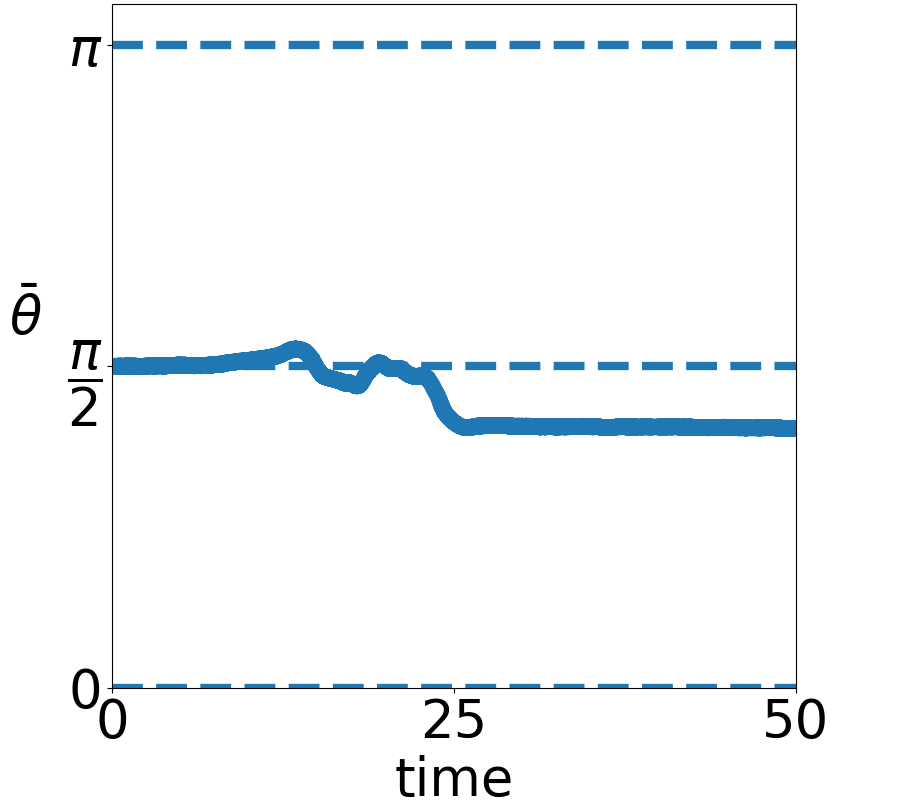}\label{fig:perturb_134_theta}}\subfloat[]{\includegraphics[height= 3.5cm]{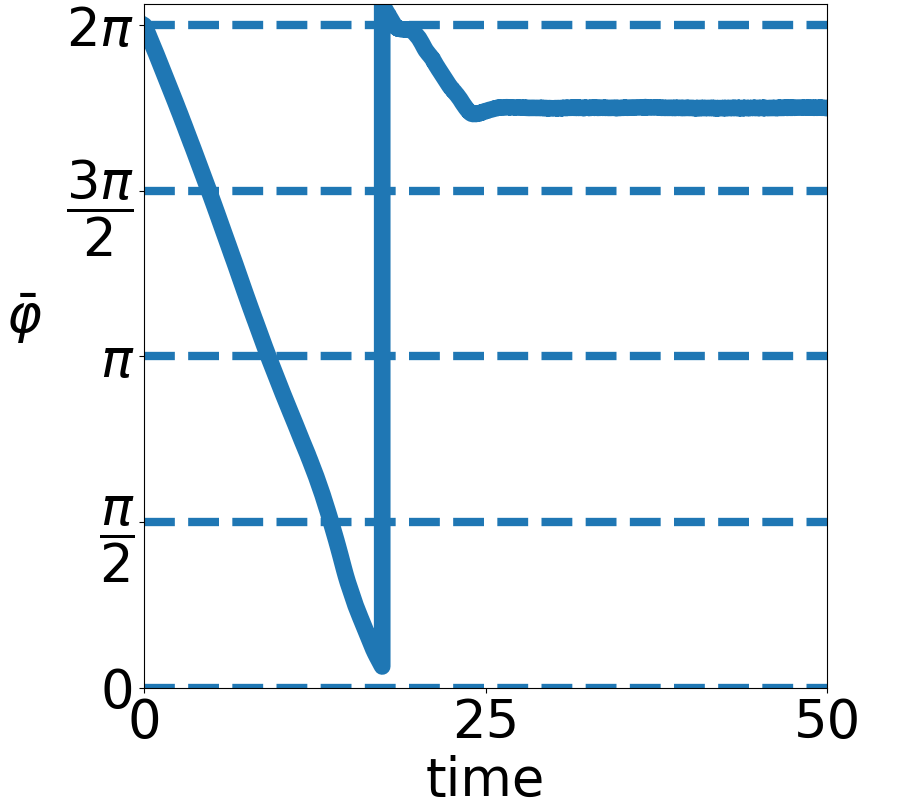}\label{fig:perturb_134_phi}}
\subfloat[]{\includegraphics[height= 3.5cm]{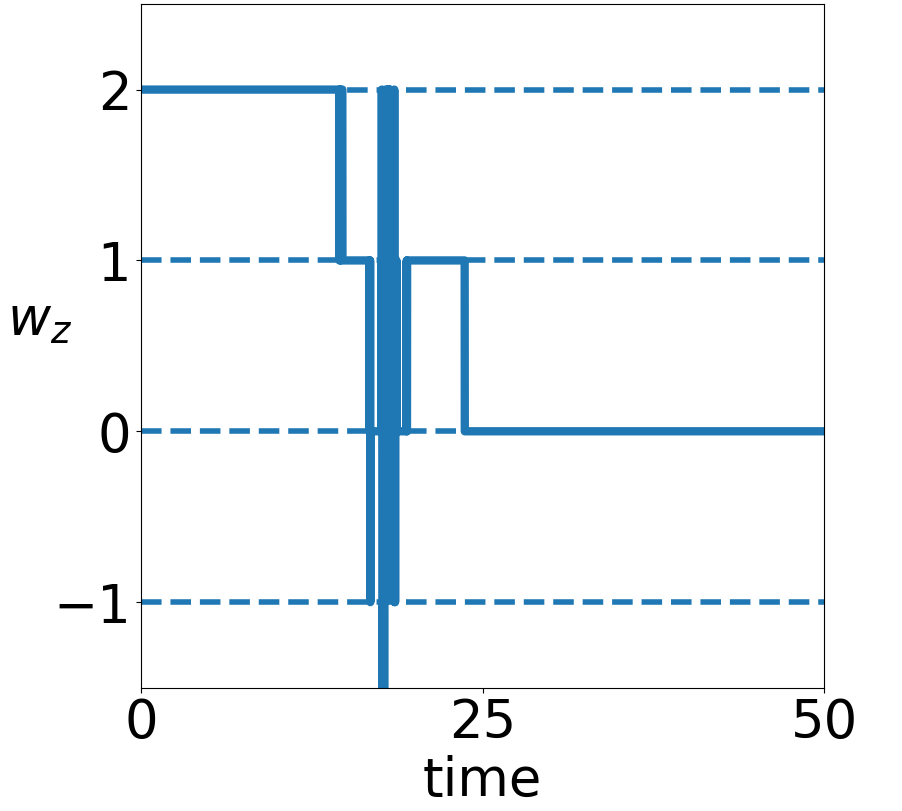}\label{fig:perturb_134_wn}} 

\subfloat[]{\includegraphics[height= 3.5cm]{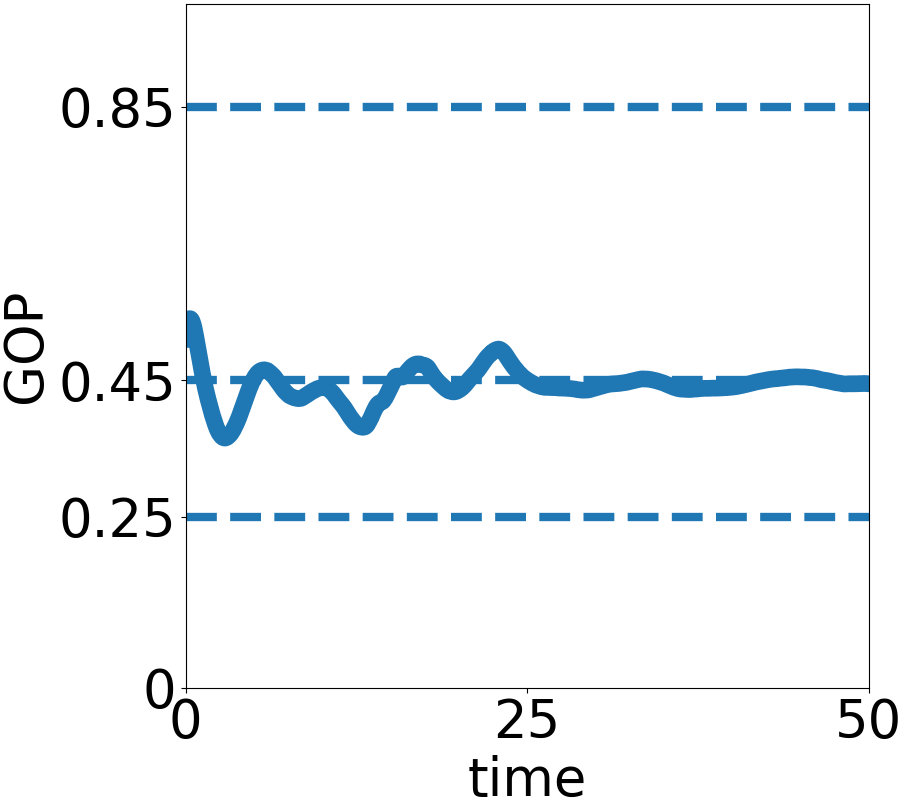}\label{fig:perturb_753_op}}
\subfloat[]{\includegraphics[height= 3.5cm]{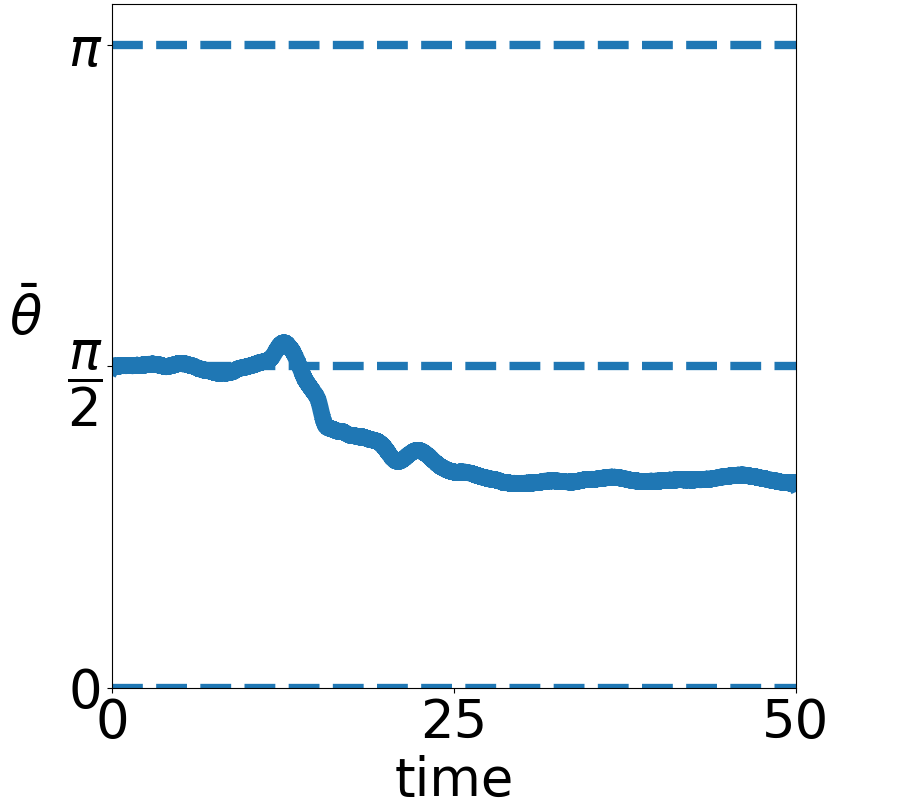}\label{fig:perturb_753_theta}}
\subfloat[]{\includegraphics[height= 3.5cm]{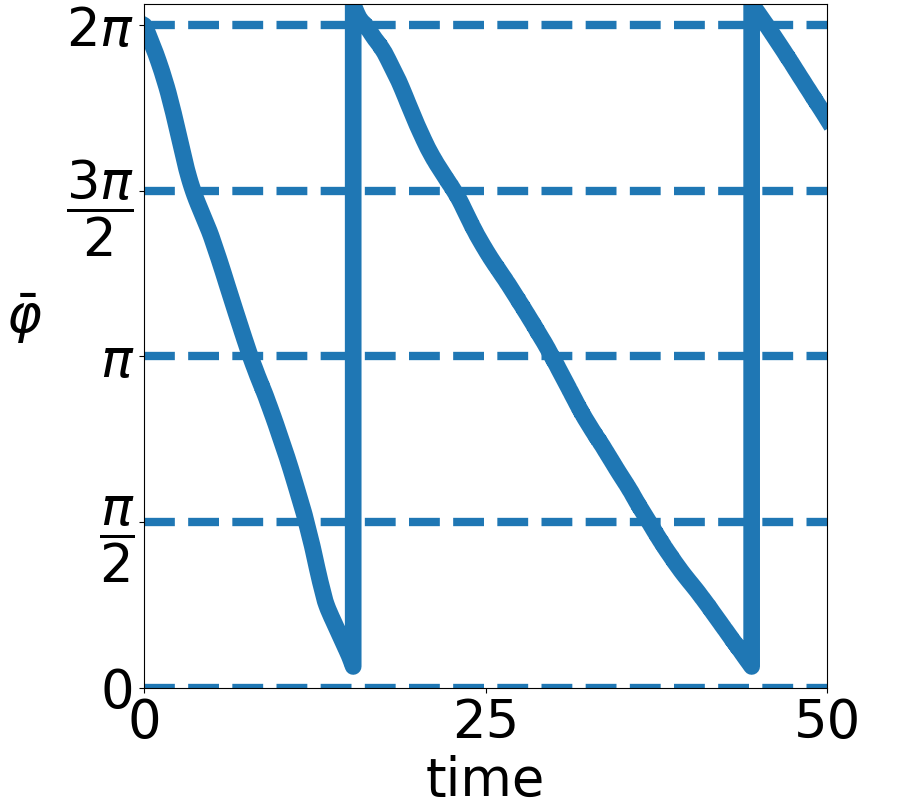}\label{fig:perturb_753_phi}}
\subfloat[]{\includegraphics[height= 3.5cm]{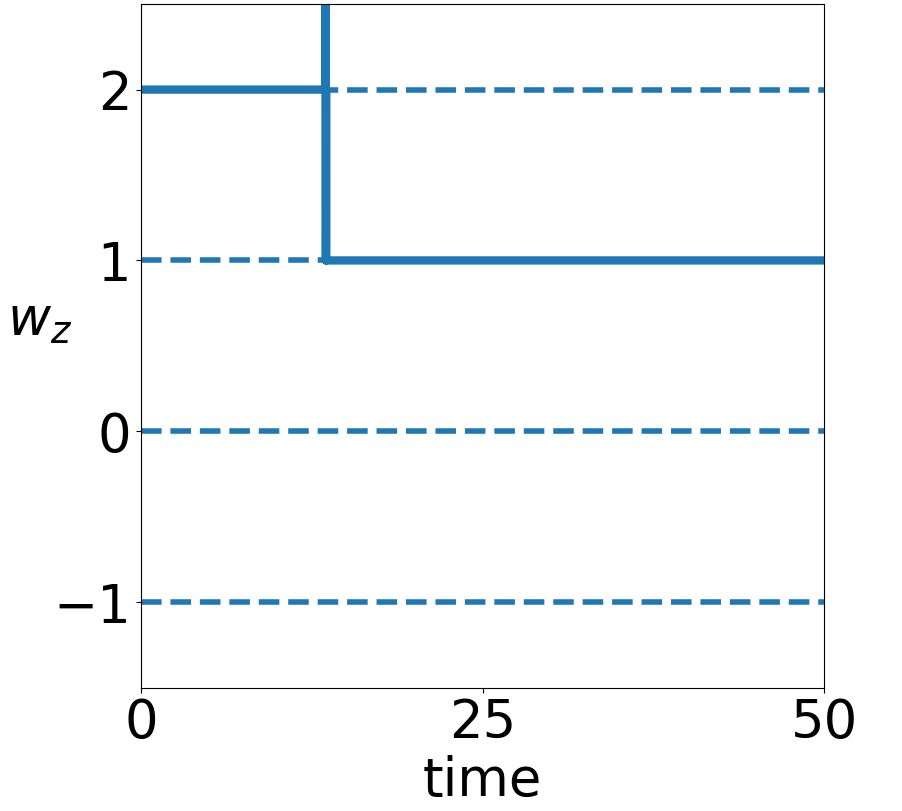}\label{fig:perturb_753_wn}} 

\subfloat[]{\includegraphics[height= 3.5cm]{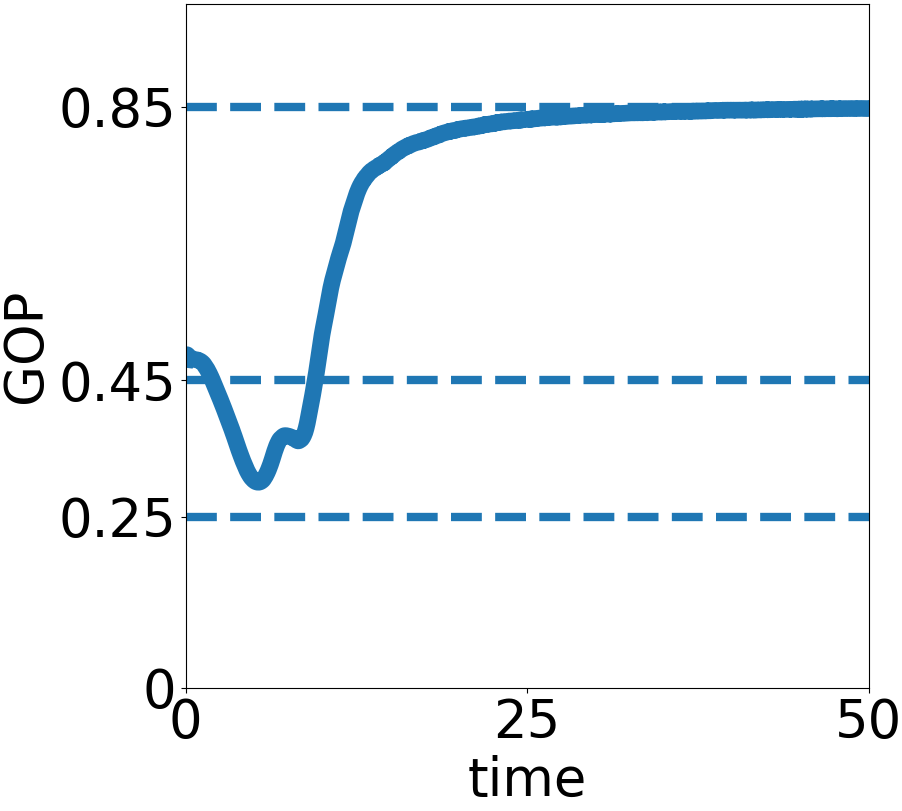}\label{fig:perturb_1000_op}}
\subfloat[]{\includegraphics[height= 3.5cm]{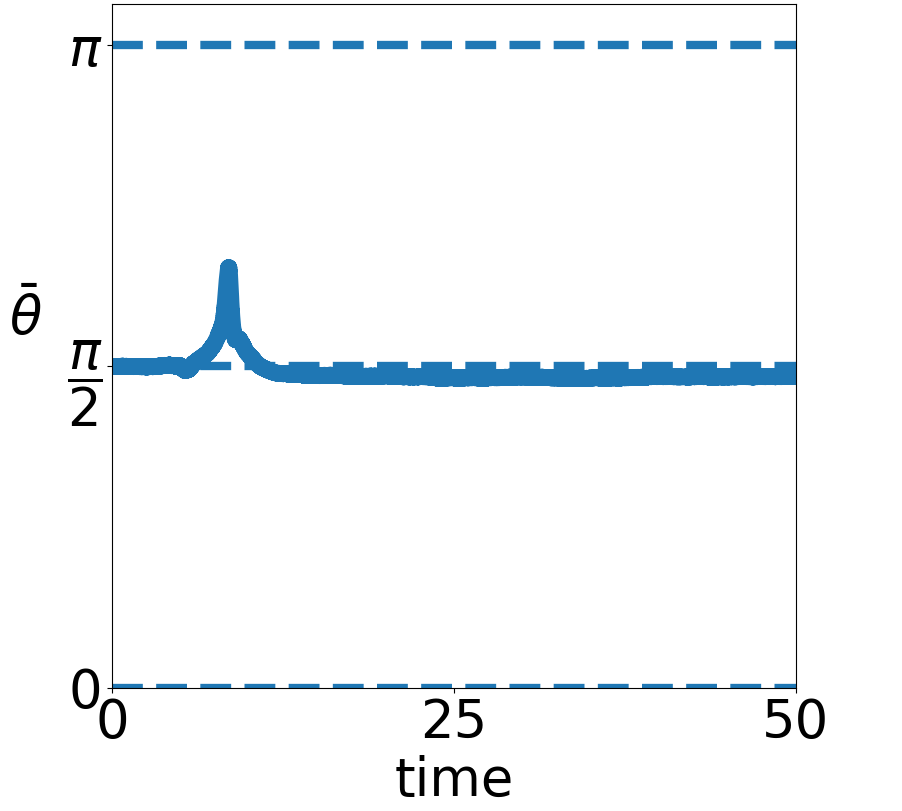}\label{fig:perturb_1000_theta}}
\subfloat[]{\includegraphics[height= 3.5cm]{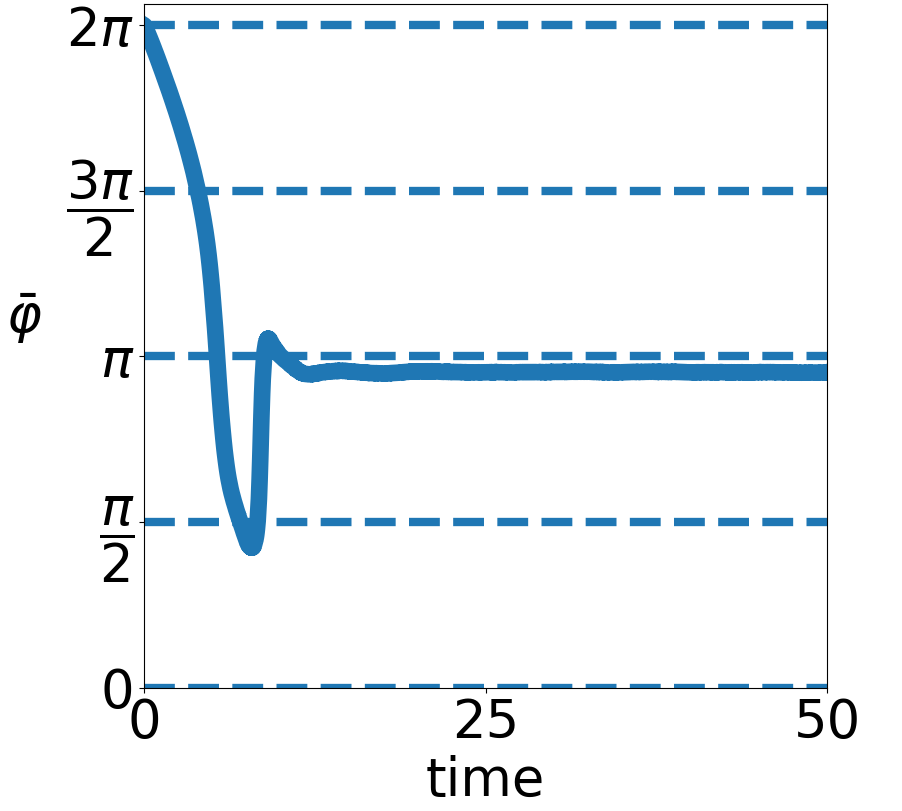}\label{fig:perturb_1000_phi}}
\subfloat[]{\includegraphics[height= 3.5cm]{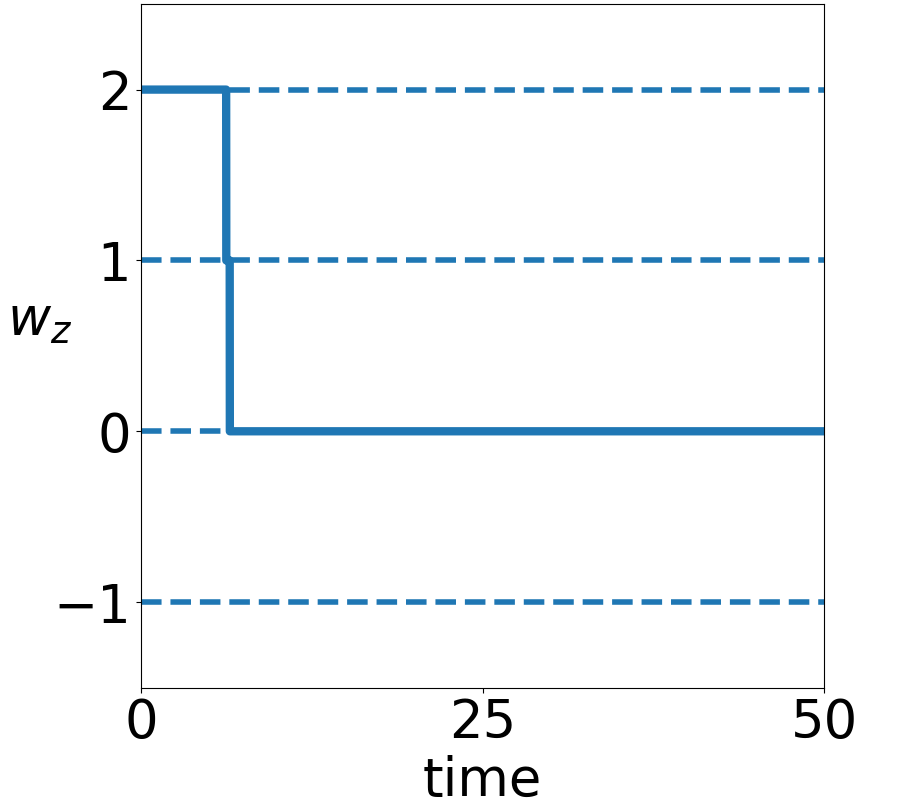}\label{fig:perturb_1000_wn}} 
\caption{Different outcomes of the simulation of the IBM starting from perturbed initial MO. Only the four main indicators are shown: from left to right, the GOP, the mean polar angle (or pitch) $\bar \theta$, the mean azimuthal angle (or yaw) $\bar \varphi$ and the winding number $w_z$. (a)-(d) For $\sigma=134$, the system stays a MO for a long time ($t\simeq20$) but eventually converges to a FS; (e)-(h) for $\sigma=753$, the system converges towards a generalized solution with a polar angle not equal to $\pi/2$ and a winding number equal to 1 along the $z$-axis;  (i)-(l) for $\sigma=1000$, the MO is quickly disrupted (at $t\simeq5$) and converges almost immediately towards a FS. Parameters: $N=3 \, 10^6$, $R=0.025$, $\kappa = 10$, $\nu=40$, $c_0=1$, $L=1$, $\xi = 2 \pi$.}
\label{fig:topo_protec_double_mill_perturb}
\end{figure}

In this section, we study the robustness of the MO when the initial condition is randomly perturbed as described by the initial condition MO3 \eqref{eq:inimill_perturb}. Three typical outcomes for three different values of the perturbation size~$\sigma$ are shown in Fig.~\ref{fig:topo_protec_double_mill_perturb}. For each value of $\sigma$, the temporal evolution of the four main indicators are shown: the GOP (Figs. \ref{fig:perturb_134_op}, \ref{fig:perturb_753_op}, \ref{fig:perturb_1000_op}), the mean polar angle or pitch (Figs. \ref{fig:perturb_134_theta}, \ref{fig:perturb_753_theta}, \ref{fig:perturb_1000_theta}), the mean azimuthal angle or yaw (Figs. \ref{fig:perturb_134_phi}, \ref{fig:perturb_753_phi}, \ref{fig:perturb_1000_phi}) and the winding number along the $z$-axis (Figs.~\ref{fig:perturb_134_wn}, \ref{fig:perturb_753_wn}, \ref{fig:perturb_1000_wn}).  For small to moderate values (approximately $\sigma<100$), the outcomes of the simulation are the same as in Fig.~\ref{fig:topo_protec_graphs_mills} and are not shown. However, they demonstrates the robustness of the topological solutions. 
When $\sigma$ increases and crosses this threshold, the behavior becomes different. Around this threshold (for $\sigma=134$), in Fig.~\ref{fig:perturb_134_op}, we observe that the GOP does not remain initially constant (contrary to the un-perturbed case shown in Fig.~\ref{subfig:topo_protec_perptwist}) but immediately decreases, then increases and oscillates around the value $\mathrm{GOP}_1$ before transitioning towards the value $\mathrm{GOP}_2$ corresponding to a FS. In Figs.~\ref{fig:perturb_134_phi} and \ref{fig:perturb_134_wn}, we observe that the MO is preserved during a comparable, slightly longer, time than in Figs.~\ref{subfig:milling_to_flocking_phi} and \ref{subfig:double_milling_wn_z}  (around~20 units of time) before degenerating into a~FS. 

Passed this threshold, when $\sigma$ increases again and up to another threshold value around $\sigma\simeq1000$, a new topological phase transition is observed from a MO with winding number 2 to a GS \eqref{eq:generalizedsolutions} with winding number 1. For $\sigma=753$, the GOP shown in Fig. \ref{fig:perturb_753_op} initially strongly oscillates around the value $\mathrm{GOP}_1$ before stabilizing, still around this value, which is in stark contrast with the previous experiments. The winding number shown in Fig. \ref{fig:perturb_753_wn} reveals that this final steady behavior is linked to a winding number equal to 1 after a transition around $t\simeq12$. Consequently, a milling behavior is observed in Fig. \ref{fig:perturb_753_phi} for the mean azimuthal angle. This angle evolves linearly but with a slower speed, approximately divided by 2, after the transition, as expected since the winding number has dropped from 2 to 1. However, the final mean polar angle $\bar{\theta}$ shown in Fig. \ref{fig:perturb_753_theta} is not equal to $\pi/2$. Since the gradient in body-orientation is along the $z$-axis, this indicates that the final state corresponds to a GS rather than a standard MO. This demonstrates that the family of generalized topological solutions enjoys some greater stability. The transition between~MO and GS has not been observed when starting from a non-perturbed initial state. However, starting with perturbed initial conditions, the MO and GS with winding number 1 seem stable during several tens of units of time. 

The transition between MO and GS with different winding numbers happens when the perturbation size is large enough and seems to be the typical behavior: out of 6 independent simulations for values of $\sigma$ evenly spread between 258 and 876, 5 simulations led to a MO or a GS with winding number 1 stable during more than 50 units of time. The other one led to a FS. We can think that the perturbation brings the system to a state closer to the MO with winding number 1, in particular due to the stochastic spatial inhomogeneities of the perturbation. On the particle simulations, we observe that the density of agents does not remain uniform, which creates different milling zones with possibly different milling speeds depending on the local gradient of body-orientations. The denser region then seems to attract the other particles before expanding into the full domain. The global direction of motion is not necessarily preserved during this process. In comparison, starting from an unperturbed MO with winding number 2, the density remains uniform and the system is globally subject to numerical errors which homogeneously degrade the topology up to the point that the system becomes closer to a FS. The situation is analogous when the size of the perturbation is too large as shown in Figs. \ref{fig:perturb_1000_op}, \ref{fig:perturb_1000_phi}, \ref{fig:perturb_1000_wn} for $\sigma=1000$ : the MO is preserved during less than 5 units of time and after an immediate drop of the GOP, the system quickly reaches a FS.

\subsection{Critique}

The existence of a persistence time for the MO1 and HW1 solutions suggests that they enjoy some kind of topological protection against the noisy perturbations induced by the IBM and that MO2 and HW2 do not have such protection. However, since explicit solutions of the SOHB model for the initial conditions MO2 and HW2 are not available, it is not possible to assess the role of noise in the observed evolutions of the MO2 and HW2 solutions. So, further investigations are needed to confirm that non-trivial topology actually provides increased robustness against perturbations. Moreover, the MO1 is robust against perturbed initial conditions. The MO and GS with winding number 1 seem to be much more more stable than with winding number 2. 



\section{Discussion and conclusion}\label{sectiondiscussion}

An Individual Based Model describing the alignment of body-orientations in 3D and its macroscopic limit have been presented. The model involves new kinds of internal degrees of freedom involving geometrical constraints, here due to the manifold structure of SO$_3({\mathbb R})$, leading to new types of self-organized phenomena. In particular, the macroscopic model has been shown to host special solutions with non-trivial topological structures. Corresponding solutions of the Individual Based Model have been computed and their non-trivial topological structure, shown to persist for a certain time before being destroyed by noise-induced fluctuations. Quantitative estimates of the agreement between the Individual Based Model and the Macroscopic model have been given. This study provides one more evidence of the role of geometry and topology in the emergence of self-organized behavior in active particle systems. The model presented in this article opens many new research directions. Some of them are listed below. 

\begin{enumerate}
\item The stability of the MO \eqref{millingsolution}, HW \eqref{helicalsolution} and GS \eqref{eq:generalizedsolutions} solutions as well as those of the generalized HW solutions described in Section \ref{appendixgeneralsolutions} is an open problem. It would enable us to investigate the potential link between topological structure and stability. 

\item Numerical simulations have been carried out in a periodic setting. Real systems though are confined by solid walls. To model the influence of confinement, it is necessary to explore wider classes of boundary conditions.

\item Most topological states in physical systems consist of linear perturbations of bulk states that propagate on the edges of the system (edge states). It would be interesting to determine whether linear perturbations of the MO or HW solutions could host such edge states.  

\item Beyond the mean-field limit $N \to \infty$, it would be interesting to quantify the fluctuation about the mean-field, for instance through a large deviation approach (see e.g. \cite{barre2019gamma, berlyand2019continuum, bortolussi2016mean, dawson1987large, fernandez1997hilbertian, lancellotti2009fluctuations}). 

\item Direct numerical simulations of the macroscopic model need to be developed to answer some of the questions raised by the study of topological protection (see Section \ref{sectionnumericalexperiments}). 

\item It is desirable to develop more sophisticated topological indicators to gain better insight into the topological structure of the solutions.

\item The multiscale approach developed here could be extended to other geometrically structured systems involving e.g. a wider class of manifolds which would enlarge the applicability of the models. 

\end{enumerate}

\newpage

\appendix

\begin{center}
\LARGE 
\textbf{Supplementary Material}
\bigskip
\end{center}

\section{List of supplementary videos}
\label{appendix:listvideos}

This article is supplemented by several videos which can be accessed by following this link: \url{https://figshare.com/projects/Bulk_topological_states_in_a_new_collective_dynamics_model/96491}. They are listed and described below. 

\begin{video}
\label{vid:example}
It supplements Fig. \ref{figureexample} of Section \ref{subsec:ibm_num_sim} and provides a visualization of the time evolution of the system considered in this figure.
\end{video}

\begin{video}
\label{vid:milling_solution}
It supplements Fig. \ref{fig:milling} of Section \ref{subsubsec:milling}: it provides a visualization of the time evolution of a MO. Several frames ${\mathbb A}=(\Omega,\mathbf{u},\mathbf{v}) \in $ SO$_3({\mathbb R})$ are placed at various locations of space and evolve according to \eqref{millingsolution} (with arbitrary chosen parameters). The vectors $\Omega$, $\mathbf{u}$ and~$\mathbf{v}$ are displayed respectively in red, green and blue. 
\end{video}

\begin{video}
\label{vid:helical_solution}
It supplements Fig. \ref{fig:helical} of Section \ref{subsubsec:helical}: it provides a visualization of the time evolution of a HW. See caption of Video \ref{vid:milling_solution} for details on the graphical representation.
\end{video}

\begin{video}
\label{vid:milling_to_flocking_particles}
It supplements Fig. \ref{fig:topo_protec_graphs_mills} in Section \ref{sec:topoprotec}. It shows the time-evolution of the particles for the initial condition MO1 \eqref{eq:inimill_nt}. For clarity, only a sample of 5000 particles are shown. We refer to Fig. \ref{fig:tetra} for details on the representation of the body orientation using four-colored tetrahedra. We notice the ensemble rotation of the particle directions about the $z$ axis until an instability disrupts the body orientation twist along the $z$ axis (around time $t \approx 13$) and eventually drives the system to a FS.
\end{video}

\begin{video}
\label{vid:milling_to_flocking_RPZ}
It supplements Fig. \ref{fig:topo_protec_graphs_mills} in Section \ref{sec:topoprotec}. It provides the time-evolution of the RPZ curve for the initial condition MO1 \eqref{eq:inimill_nt}. The RPZ curve remains a circle until time $t \approx 8$ where its radius shrinks down. Then, the RPZ-curve shows a fairly chaotic dynamics during which the topology is lost. This happens around time $t \approx 13$ which is the first time when the RPZ-curve passes through the origin; at this time, the winding number is not defined. Then, the RPZ-curve slowly migrates towards the unit circle while shrinking to a single point which signals a FS. From time $t \approx 15$ on, it remains a single immobile point. 
\end{video}

\begin{video}
\label{vid:trivial_milling_particles}
It supplements Fig. \ref{fig:topo_protec_graphs_mills} in Section \ref{sec:topoprotec}. It shows the time-evolution of the particles for the initial condition MO2 \eqref{eq:inimill_t}. For clarity, only a sample of 5000 particles are shown (see Fig. \ref{fig:tetra} for details on the representation of the body orientation). We notice the counter-rotation of the particle directions about the $z$ axis in the bottom and top halves of the domain, corresponding to the opposite mills. These two counter-rotations gradually dissolve while the solution approaches the FS.
\end{video}

\begin{video}
\label{vid:trivial_milling_RPZ}
It supplements Fig. \ref{fig:topo_protec_graphs_mills} in Section \ref{sec:topoprotec}. It provides the time-evolution of the RPZ curve for the initial condition MO2 \eqref{eq:inimill_t}. The circle formed by the initial RPZ curve immediately opens. The opening width constantly increases, until the arc is reduced to a single point opposite to the opening point at time $t \approx 10$. Then there is a bounce and the arc forms again and increases in size until it reaches a maximum and decreases again. Several bounces are observed with decreasing amplitudes. These bounces result in the non-monotonous behavior of the quantity $d_z$ displayed on Fig. \ref{subfig:double_milling_d_z}. 
\end{video}

\begin{video}
\label{vid:helical_to_flocking_particles}
It supplements Fig. \ref{fig:topo_protec_graphs_helices} in Section \ref{sec:topoprotec}. It shows the time-evolution of the particles for the initial condition HW1 \eqref{eq:inihelic_nt} (see Fig. \ref{fig:tetra} for details on the representation of the body orientation). For clarity, only a sample of 5000 particles are shown. Before time $t \simeq 125$, we observe a steady HW state. Then, after time $t \approx 125$, the particles show an undulating wave-like behavior, with slowly increasing frequency and amplitude, which causes the decrease of the GOP. Around time $t \approx 178$, the particles are divided into two groups with pitch angles $\theta \simeq 0$ and $\theta \simeq \pi$, which suddenly reverses the global direction of motion. After time $t \approx 178$, the particles quickly adopt the same body-orientation. Shortly after time $t=178$, the particles still have an undulating behavior but it quickly fades away until a FS is reached. 
\end{video}

\begin{video}
\label{vid:helical_to_flocking_RPX}
It supplements Fig. \ref{fig:topo_protec_graphs_helices} in Section \ref{sec:topoprotec}. It shows the time-evolution of the RPX-curve for the initial condition HW1. Unlike in the MO case, the RPX curve does not shrinks to the center of the circle before migrating to its limiting point. In this case, the limiting point near the unit circle towards which the RPX curve is converging attracts the RPX. During this transition, the circular shape of the RPX curve is preserved until it becomes a point. \end{video}

\begin{video}
\label{vid:trivial_helical_particles}
It supplements Fig. \ref{fig:topo_protec_graphs_helices} in Section \ref{sec:topoprotec}. It shows the time-evolution of the particles for the initial condition HW2 \eqref{eq:inihelic_t}. For clarity, only a sample of 5000 particles are shown (see Fig. \ref{fig:tetra} for details on the representation of the body orientation). At the beginning, we see two opposite alternations of the three side colors of the tetrahedra (green-blue-magenta followed by green-magenta-blue), which signals a double parallel twist. Then, gradually, the green color is eaten up by the blue and magenta ones and only one alternation of the blue and magenta colors remains. Then the color alternation shades away and gives room to a homogeneous color showing that the body orientations have stopped rolling and a FS is attained. 
\end{video}

\begin{video}
\label{vid:trivial_helical_RPX}
It supplements Fig. \ref{fig:topo_protec_graphs_helices} in Section \ref{sec:topoprotec}. It provides the time-evolution of the RPX curve for the initial condition HW2 \eqref{eq:inihelic_t}. The circle formed by the initial RPX curve immediately opens. The opening width constantly increases, although at a slower pace than for MO2 (see Video \ref{vid:trivial_milling_RPZ}). Here, also contrasting with the MO2 case, the monotonous opening of the arc results in a monotonously increasing quantity $d_x$ as shown in  Fig. \ref{subfig:double_helical_d_x}. 
\end{video}

\begin{video}
\label{vid:milling_to_helical_particles}
It supplements Fig. \ref{fig:milling_to_helical_graphs} in Section \ref{sec:millingtohelical}. It shows the time-evolution of the particles for a MO initial condition \eqref{eq:inimill_nt} in a rare case where it evolves into a HW. For clarity, only a sample of 5000 particles are shown (see Fig. \ref{fig:tetra} for details on the representation of the body orientation). It starts like Video \ref{vid:milling_to_flocking_particles} with the ensemble rotation of the particle directions about the $z$ axis until an instability initiated at time $t \approx 10$ gradually disrupts this organization. However, the disruption does not drive the system to an FS, but rather to a HW as shown by the alternations of blue, green and magenta colors propagating along the particle orientations. 
\end{video}  

\begin{video}
\label{vid:milling_to_helical_RPZ}
It supplements Fig. \ref{fig:milling_to_helical_graphs} in Section \ref{sec:millingtohelical}. It provides the time-evolution of the RPZ curve for a MO initial condition \eqref{eq:inimill_nt} in a rare case where it evolves into a HW. The behavior is essentially the same as in Video \ref{vid:milling_to_flocking_RPZ} except that the RPZ-curve shrinks to a single point far away from the unit circle. This shows that the end state of the RPZ-curve is closer to disorder than for a milling to flocking transition. Before that, the non-trivial topology across~$z$ is lost following a similar scenario as for the milling-to-flocking transition. 
\end{video}

\begin{video}
\label{vid:milling_to_helical_RPX}
It supplements Fig. \ref{fig:milling_to_helical_graphs} in Section \ref{sec:millingtohelical}. It provides the time-evolution of the RPX curve for a MO initial condition \eqref{eq:inimill_nt} in a rare case where it evolves into a HW. Initially, the RPX-curve is reduced to the origin, showing total disorder across the $x$ direction. Then, after some chaotic transient, a closed curve enclosing the origin is formed. This curve initially stays close to the origin, still showing strong disorder. But gradually, the radius of the curve increases and  approaches the unit circle. Thus, across $x$, the topology is initially undefined, but when it builds up, it shows its non-trivial character, the emerging RPX-curve having non-zero winding number about the origin. 
\end{video}

\begin{video}
\label{vid:milling_to_helical_then_flock_particles}
It supplements Fig. \ref{fig:milling_to_helical_then_flock_graphs} in Section \ref{sec:millingtoflockviahelical}. It shows the time-evolution of the particles for a MO initial condition \eqref{eq:inimill_nt} in a rare case where it evolves into a FS through a transient HW. For clarity, only a sample of 5000 particles are shown (see Fig. \ref{fig:tetra} for details on the representation of the body orientation). The point of view is changed from Video~\ref{vid:milling_to_helical_particles} to better visualize the transient HW moving along the diagonal, appearing around time $t \approx 16$.  At the beginning we witness the ensemble rotation of the particles and its disruption by an instability. After some chaotic behavior, the transient HW establishes as shown by the alternations of blue, green and magenta colors propagating along the diagonal. But after some time, the HW structure is disrupted again and the system eventually establishes a FS.  
\end{video}  

\begin{video}
\label{vid:milling_to_helical_then_flock_RPZ}
It supplements Fig. \ref{fig:milling_to_helical_then_flock_graphs} in Section \ref{sec:millingtoflockviahelical}. It provides the time-evolution of the RPZ curve for a MO initial condition \eqref{eq:inimill_nt} in a rare case where it evolves into a FS through a transient HW. The behavior is essentially the same as in Video \ref{vid:milling_to_flocking_RPZ} except that the RPZ-curve undergoes a longer-lasting chaotic dynamics before shrinking to a point which migrates towards the unit circle. 
\end{video}


\section{Quaternion framework}
\label{sectionquaternion}

Despite its formal simplicity, the SO$_3({\mathbb R})$-framework used in the definition of the Individual Based Model is not well suited to numerical simulations due to the high computational cost required to store and manipulate rotation matrices. A more efficient representation of rotations in ${\mathbb R}^3$ is the quaternion representation based on the group isomorphism 
$$
\begin{array}{rcl}
\Phi : \mathbb{H}/\pm 1 & \longrightarrow & \mbox{SO}_3({\mathbb R})\\
q & \longmapsto &\Phi(q) : \mathbf{w} \in{\mathbb R}^3 \mapsto \{q[\mathbf{w}]q^*\} \in {\mathbb R}^3, 
\end{array}
$$
where the 3-dimensional vector $\mathbf{w}=(w_1,w_2,w_3)^\mathrm{T}\in{\mathbb R}^3$ is identified with the pure imaginary quaternion denoted by $[\mathbf{w}]=iw_1+jw_2+kw_3$ and $q^*$ denotes the conjugate quaternion to~$q$. Conversely, the pure imaginary quaternion $q=iq_1+jq_2+kq_3$ is identified with the 3-dimensional vector denoted by $\{q\}:=(q_1,q_2,q_3)^\mathrm{T}$. Note that for any quaternion $q$ and any vector $\mathbf{w}\in{\mathbb R}^3$, the quaternion $q[\mathbf{w}]q^*$ is a pure imaginary quaternion. The group of unit quaternions is denoted by $\mathbb{H}$ and is homeomorphic to the sphere $\mathbb{S}^3\subset{\mathbb R}^4$. 

We refer the reader to \cite[Section 2]{degond2018alignment} and  \cite[Appendix A]{degondfrouvellemerinotrescases18} where details about the equivalence between the two representations can be found. Note that \cite{degondfrouvellemerinotrescases18} studies a model in a full quaternion framework. Table \ref{tab:quater_corresp} below summarizes how the different objects can be computed in either of the two representations.  

\renewcommand{\arraystretch}{2}
\begin{table}[ht!]
\small
\begin{center}\begin{tabular}{|p{3cm}|l|p{6.5cm}|} \hline  & Matrix & Quaternion \\ \hline \hline 
Orientation & $A \in \mbox{SO}_3 ({\mathbb R}) $ & $ q \in \mathbb{H} / \pm 1$ such that $\Phi(q) = A$ \\ \hline 
Flux & $J_k = \sum_j K (\mathbf{X}_k-\mathbf{X}_j) A_j$ & $Q_k = \sum_j K (\mathbf{X}_k-\mathbf{X}_j) \,  (q_j \otimes q_j-1/4 \mbox{I}_4)$  \\ \hline 
Mean orientation & ${\mathbb A} = \mbox{arg\,max} \{ A \mapsto A \cdot J \}$ & $\bar q \in \mathbb{H}$ eigenvector associated to the largest eigenvalue of $Q$ \\ \hline
 Von Mises distribution & $\displaystyle{ M_{\mathbb A} (A) = \frac{\exp( \kappa {\mathbb A} \cdot A)}{\mathcal{Z} }} $ & $\displaystyle{ M_{\overline{q}} (q) = \frac{\exp (2 \kappa  (\overline{q}\cdot q)^2)}{\mathcal{Z}}}$ \\ \hline  \end{tabular} \caption{Matrix \textit{vs} quaternion formulation}
\label{tab:quater_corresp}
\end{center}
\end{table}


\section{Numerical methods}
\label{sectionsimulation}

The IBM \eqref{deterministicpart}, \eqref{jumppart} has been discretized within the quaternion framework using the time-discrete algorithm described in Table \ref{tab:algo} below. This table shows one iteration of the algorithm during which the positions $\mathbf{X}_k^n \in {\mathbb R}^3$ and orientations $q_k^n \in \mathbb{H}$ for $k \in \{1,\ldots,N\}$ are updated into $\mathbf{X}_k^{n+1}$ and $q_k^{n+1}$ respectively.
 
\begin{table}[ht!]
\small
\begin{center}
\begin{tabular}{|p{14.5cm}|}
\hline
\textbf{Algorithm:} Iteration $n\to n+1$ of the time-discrete algorithm \\
\hline
\textbf{1. Update the positions:} for $k \in \{1\ldots,N\}$, set $\mathbf{X}_k^{n+1} = \mathbf{X}_k^n + c_0 \, \{q_k^n[\mathbf{e}_1] (q_k^n)^*\} \, \Delta t$ \\ 
\textbf{2. Draw a subset $I \subset \{1,\ldots,N\}$ of jumping agents:} for each agent $k \in \{1\ldots,N\}$, draw a random number $r_k$ uniformly in $[0,1]$. If $r_k >  \exp(- \nu \, \Delta t)$, then $k \in I$.  \\
\textbf{3. Compute the local flux:} for $k \in I$, compute\\
\hspace{3cm }$\overline{Q}_k^n = \frac{1}{N}\sum_{j=1}^N K(\mathbf{X}_k^n-\mathbf{X}_j^n) \, (q_j^n \otimes q_j^n -\frac{1}{4} \mbox{I}_4 ).$ \\
\textbf{4. Update the orientations:} for $k \in I$ compute one unit eigenvector $\overline{q}_k^n$ of $Q_k^n$ of maximal eigenvalue and draw $q_k^{n+1} \sim M_{\overline{q}_k^n}$.\\ \hline
\end{tabular}
\caption{One iteration of the time-discrete algorithm}
\label{tab:algo}
\end{center}
\end{table}
\renewcommand{\arraystretch}{1}

At step 2, the Poisson process is discretized with a time step $\Delta t$ during which the indices of the jumping agents are recorded. In the simulations $\Delta t$ has to be chosen small enough so that the event that an agent jumps twice or more during a time interval of size $\Delta t$ is negligible. In all the simulations, we take $\Delta t$ such that $\nu \, \Delta t=10^{-2}$.

At step 3, a random quaternion $q$ sampled from a von Mises distribution with prescribed mean orientation $\bar{q}$ can be obtained as $q=\bar{q}r$ where $r\in\mathbb{H}$ is sampled from a von Mises distribution with mean orientation 1 (see \cite[Proposition 9]{degond2018alignment}). An efficient rejection algorithm to sample von Mises distributions can be found in \cite{kent2018new}. 

All the simulations in this paper take place in a periodic box of size $L=(L_x,L_y,L_z)$. The observation kernel $K$ is the indicator of the ball centered at $0$ and of radius $R>0$. The six parameters of the simulations are summarized in Table \ref{tab_parameters_IBM}. 

Finally, we would like to stress that the quaternion formulation is not only a convenient numerical trick. The equivalence it provides between body-orientation models and models of nematic alignment of polymers in dimension four has been exploited in \cite{degond2019phase} to study phase transitions in the body alignment model.


\section{Derivation of the macroscopic model}
\label{appendix:derivationmacro}

The derivation of the continuum theory presented in Section \ref{sectionmacro} has been achieved in \cite{degond2018alignment} (see also \cite{degond2019phase}) following earlier works \cite{degondfrouvellemerino17, degondfrouvellemerinotrescases18}. It consists of two steps. The first step is the derivation of a mean-field kinetic model in the limit $N \to \infty$ showing that the system satisfies the propagation of chaos property: the agents, seen as random variables in ${\mathbb R}^3\times$ SO$_3({\mathbb R})$ become independent and identically distributed. Their law is given by the kinetic particle distribution $f$ which satisfies the following PDE: 
\[\partial_t f + c_0 \, A \mathbf{e}_1 \cdot \nabla_\mathbf{x} f = \nu \, (\rho_f \, M_{{\mathbb A}_{K*f}}-f),\]
where $\rho_f \equiv \rho_f(t,\mathbf{x})$ is the local spatial density:
$$
\rho_f(t,\mathbf{x}) =\int_{\mbox{{\scriptsize SO}}_3({\mathbb R})} f(t,\mathbf{x},A) \, \mathrm{d} A, 
$$
and ${\mathbb A}_{K*f} \equiv {\mathbb A}_{K*f}(t,\mathbf{x})$ is the local average body-attitude defined by
$$
{\mathbb A}_{K*f} (t,\mathbf{x}) := \mbox{arg\,max}_{A \in \mbox{\scriptsize{SO}}_3({\mathbb R})} A \cdot J_{K*f}(t,\mathbf{x}), 
$$
computed from the local flux:
$$
J_{K*f} \equiv J_{K*f} (t,\mathbf{x}): = \iint_{{\mathbb R}^3\times \mbox{{\scriptsize SO}}_3({\mathbb R})} K(\mathbf{x}-\mathbf{y}) \, A \, f(t,\mathbf{y},A) \, \mathrm{d}\mathbf{y} \, \mathrm{d} A.
$$
From a mathematical point of view, the probability distribution $f\equiv f(t,\mathbf{x},A)$ is obtained as the limit in law of the empirical measure of the $N$-particle system. We refer to \cite{diez2019propagation} where a rigorous proof of this result is presented for a similar model, and to \cite{bolley2012mean} for a related work on the Vicsek model. 

In the macroscopic regime the agent interactions become strong, which is expressed by the following hydrodynamic scaling:
\[\varepsilon \sim \frac{c_0}{\nu \, L} \sim \frac{R}{L}  \ll 1,\]
where $L$ is a typical macroscopic length-scale of the system (such as the typical size of the flock). 
We define $\tilde c_0 = \varepsilon \nu L = {\mathcal O}(1)$ and $c'_0 = c_0/\tilde c_0$. Then, defining dimensionless time and space variables $t'$ and $\mathbf{x}'$ such that $\mathbf{x} = L \mathbf{x}'$ and $t = (L/\tilde c_0) t'$, we obtain (dropping the primes for simplicity): 
\begin{equation}
\label{bgkepsilon}
\partial_t f^\varepsilon + c_0 \, A \mathbf{e}_1 \cdot \nabla_\mathbf{x} f^\varepsilon = \frac{1}{\varepsilon} \, (\rho_{f^\varepsilon} \, M_{{\mathbb A}_{f^\varepsilon}}-f^\varepsilon) + \mathcal{O}(\varepsilon), 
\end{equation}
where 
$$
{\mathbb A}_{f^\varepsilon} \equiv {\mathbb A}_{f^\varepsilon} (t,\mathbf{x}) :=\mbox{arg\,max}_{A \in \mbox{\scriptsize{SO}}_3({\mathbb R})} A \cdot J_{f^\varepsilon}(t,\mathbf{x}), 
$$
and 
$$
J_{f^\varepsilon} \equiv J_{f^\varepsilon}(t,\mathbf{x}): = \int_{\mbox{{\scriptsize SO}}_3({\mathbb R})} \, A \, f^\varepsilon(t,\mathbf{x},A) \,  \mathrm{d} A.
$$
This last expression is obtained by Taylor expanding $J_{K*f^\varepsilon} = J_{f^\varepsilon} + \mathcal{O}(\varepsilon^2)$ and means that the interactions between the agents become spatially localized in the macroscopic regime. 

The macroscopic model is obtained by formally taking the limit $\varepsilon \to 0$ in \eqref{bgkepsilon}. If such a limit exists, it is necessarily of the form 
\begin{equation}
\label{limitepsilon}
f^\varepsilon \, \underset{\varepsilon\to0}{\longrightarrow} \, \rho \,  M_{{\mathbb A}}
\end{equation}
where $\rho \equiv \rho(t,\mathbf{x})$ and ${\mathbb A} \equiv {\mathbb A}(t,\mathbf{x})$ depend on $t$ and $\mathbf{x}$. Thus, the limiting distribution is fully described by the spatial density of agents and their average orientation. To obtain a system of equations for $(\rho,{\mathbb A})$, we first use the local conservation of mass: integrating \eqref{bgkepsilon} over SO$_3({\mathbb R})$ and noting the right-hand side vanishes, it holds that, 
$$
\partial_t\int_{\mbox{{\scriptsize SO}}_3({\mathbb R})} f^\varepsilon \, \mathrm{d} A + c_0 \, \int_{\mbox{{\scriptsize SO}}_3({\mathbb R})} A \, \mathbf{e}_1 \cdot \nabla_\mathbf{x} f^\varepsilon \, \mathrm{d} A = \mathcal{O}(\varepsilon). 
$$
When $\varepsilon\to0$, assuming \eqref{limitepsilon} and using \eqref{eq:int_AMA}, we obtain \eqref{equationrhoA}. 

To obtain an equation for ${\mathbb A}$, it could be tempting to pursue this approach and multiply \eqref{bgkepsilon} by $A$ before integrating it over SO$_3({\mathbb R})$. However, the term resulting from the right-hand side of \eqref{bgkepsilon} does not vanish but equals (using \eqref{eq:int_AMA} again):
$$
\frac{1}{\varepsilon} \int_{\mbox{{\scriptsize SO}}_3({\mathbb R})} A \, (\rho_{f^\varepsilon} \, M_{{\mathbb A}_{f^\varepsilon}} - f^\varepsilon) \, \mathrm{d} A = \frac{1}{\varepsilon} \Big( \frac{c_1}{c_0} \, \rho_{f^\varepsilon} \, {\mathbb A}_{f^\varepsilon} - J_{f^\varepsilon} \Big) \ne 0.
$$
Due to the factor $\varepsilon^{-1}$, its limit as $\varepsilon \to 0$ is unknown. An easy fix can be found if, instead of multiplying Eq. \eqref{bgkepsilon} by $A$  before integrating it over SO$_3({\mathbb R})$, we multiply it by the quantity $\psi_{{\mathbb A}_{f^\varepsilon}}(A) := {\mathbb A}_{f^\varepsilon}^\mathrm{T}A-A^\mathrm{T}{\mathbb A}_{f^\varepsilon}$. The rationale for using this quantity is because we aim to find an equation for the time-derivative of ${\mathbb A}$. Such a derivative must lie in the tangent space to SO$_3({\mathbb R})$ at ${\mathbb A}$, denoted by $T_{\mathbb A}$. This suggests to multiply \eqref{bgkepsilon} by an element of $T_{\mathbb A}$. Given an arbitrary matrix $A$, a natural way to obtain an element of $T_{\mathbb A}$ is to take its orthogonal projection on $T_{\mathbb A}$, which is given by $\frac{1}{2} (A - {\mathbb A} A^\mathrm{T} {\mathbb A})$. We could therefore choose to multiply \eqref{bgkepsilon} by this quantity. But a further simplification is possible by noting that this quantity is equal to $\frac{1}{2} {\mathbb A} \, \psi_{{\mathbb A}}(A)$ and that $\frac{1}{2} {\mathbb A}$ does not depend on $A$ and so can be factored out of the integral with respect to $A$. These considerations naturally lead to the choice of the antisymmetric matrix $\psi_{{\mathbb A}_{f^\varepsilon}}(A)$ as a multiplier. Because ${\mathbb A}_{f^\varepsilon}$ is obtained as the polar decomposition of $J_{f^\varepsilon}$, there exists a symmetric matrix $S$ such that $J_{f^\varepsilon} ={\mathbb A}_{f^\varepsilon} S$. Using this remark and \eqref{eq:int_AMA}, we easily find that 
$$
\frac{1}{\varepsilon} \int_{\mbox{{\scriptsize SO}}_3({\mathbb R})} \psi_{{\mathbb A}_{f^\varepsilon}}(A) \, (\rho_{f^\varepsilon} \, M_{{\mathbb A}_{f^\varepsilon}} - f^\varepsilon) \, \mathrm{d} A = 0.
$$
Then, multiplying \eqref{bgkepsilon} by $\psi_{f^\varepsilon}$, taking the limit $\varepsilon \to 0$ and assuming \eqref{limitepsilon} leads to: 
$$
\int_{\mbox{{\scriptsize SO}}_3({\mathbb R})} (\partial_t (\rho M_{\mathbb A}) + c_0 \, A \mathbf{e}_1 \cdot \nabla_\mathbf{x} (\rho M_{\mathbb A})) \,  \psi_{{\mathbb A}}(A) \, \mathrm{d} A = 0.
$$
Eq. \eqref{equationbigA} of the SOHB model follows from this equation through tedious but straightforward computations detailed in \cite{degondfrouvellemerino17, degondfrouvellemerinotrescases18}. 

Note that the simple form of the multiplier $\psi_{{\mathbb A}_f}$ is due to a particular simple expression of the collision operator. In more general cases, the obtention of the multiplier (referred to as the generalized collision invariant in \cite{degond2008continuum}) is more involved (see e.g.  \cite{degondfrouvellemerino17, degondfrouvellemerinotrescases18, degond2018alignment}). A rigorous convergence result for the limit $\varepsilon \to 0$ is not available to date. In the case of the Vicsek model, such a rigorous result has been proved in \cite{jiang2016hydrodynamic}.


\section{Alternate expressions of \texorpdfstring{$\delta$}{the coefficient delta}}
\label{section_delta_alternate}

The following lemma provides alternate expressions for $\delta$:
\begin{lemma}
We have 
\begin{eqnarray}
\delta &=& -  \big\{ [(\mathbf{u} \cdot \nabla_\mathbf{x})\, \Omega] \cdot \mathbf{v} + [(\mathbf{v}  \cdot \nabla_\mathbf{x}) \mathbf{u} ] \cdot\Omega + [(\Omega  \cdot \nabla_\mathbf{x}) \mathbf{v}] \cdot \mathbf{u} \big\} \label{eq:delta_equiv1}\\
&=& - \frac{1}{2} \big\{ ( \nabla_\mathbf{x} \times \Omega) \cdot \Omega + ( \nabla_\mathbf{x} \times \mathbf{u}) \cdot \mathbf{u} + ( \nabla_\mathbf{x} \times \mathbf{v}) \cdot \mathbf{v} \} . \label{eq:delta_equiv2}
\end{eqnarray}
\end{lemma} 

\noindent
\begin{proof} Eq. \eqref{eq:delta_equiv1} follows from inserting the formula 
$$   0 = \nabla_\mathbf{x} (\Omega \cdot \mathbf{u}) = (\Omega \cdot \nabla_\mathbf{x}) \mathbf{u} + (\mathbf{u} \cdot \nabla_\mathbf{x}) \Omega + \Omega \times (\nabla_\mathbf{x} \times \mathbf{u}) + \mathbf{u} \times (\nabla_\mathbf{x} \times \Omega), $$
and similar formulas after circular permutation of $\{\Omega, \mathbf{u} , \mathbf{v}\}$into \eqref{eq:def_delta}. Eq. \eqref{eq:delta_equiv2} follows from taking the half sum of \eqref{eq:def_delta} and \eqref{eq:delta_equiv1} and applying the formula 
$$ \nabla_\mathbf{x} \times \mathbf{v} = \nabla_\mathbf{x} \times (\Omega \times \mathbf{u}) = (\nabla_\mathbf{x} \cdot \mathbf{u}) \, \Omega - (\nabla_\mathbf{x} \cdot \Omega) \, \mathbf{u} + (\mathbf{u} \cdot \nabla_\mathbf{x}) \Omega - (\Omega \cdot \nabla_\mathbf{x}) \mathbf{u}, $$
and similar formulas after circular permutation of $\{\Omega, \mathbf{u} , \mathbf{v}\}$. 
\end{proof}


\section{MO, HW, GS and generalized HW solutions}
\label{appendixgeneralsolutions}

In this section, we provide proofs of Lemmas~\ref{lem:MO}, \ref{lem:HW} and \ref{lemma:generalizedsolutions}. The prototypical helical traveling wave (HW) presented in Lemma~\ref{lem:HW} belongs to a more general class of solutions called generalized HW solutions described in Section \ref{appendix:generalhelical} below.


\subsection{Proof of Lemma \ref{lem:MO}}

Starting from the initial condition \eqref{perptwist}, we are looking for solutions of \eqref{equationbigA} of the form 
$$
{\mathbb A}(t, \mathbf{x}) = \left(
\begin{array}{ccc}
\cos(\omega t) & u_1(t,z) & v_1(t,z) \\
 -\sin(\omega t) & u_2(t,z) & v_2(t,z) \\ 
 0 & u_3(t,z) & v_3(t,z)
\end{array}
\right), 
$$
where $\omega\in{\mathbb R}$ is an angular velocity which will be related to the parameters of the problem later and where the basis vectors $\mathbf{u}=(u_1,u_2,u_3)^\mathrm{T}$ and $\mathbf{v}=(v_1,v_2,v_3)^\mathrm{T}$ depend only on the $z$ variable and time. In this situation, Equation \eqref{equationrhoA} is trivially satisfied which means that the system stays homogeneous in space. Solutions of this form have to satisfy three geometrical constraints which ensure that ${\mathbb A} \in$ SO$_3({\mathbb R})$. The first two ones are $\Omega\times\mathbf{u}=\mathbf{v}$ and $\mathbf{v}\times\Omega = \mathbf{u}$, which lead to
\begin{equation}
\label{eq:lambdagmillling}
{\mathbb A}(t, \mathbf{x}) = \left(
\begin{array}{ccc}
\cos(\omega t) & \sin(\omega t)v_3(t,z) & - \sin(\omega t)u_3(t,z) \\
 -\sin(\omega t) & \cos(\omega t)v_3(t,z)& -\cos(\omega t)u_3(t,z) \\ 
 0 & u_3(t,z) & v_3(t,z)
\end{array} \right).
\end{equation}
The third one is a normalization constraint: 
\begin{equation}
\label{eq:normalisationuv}
\forall t>0, \quad \forall z \in {\mathbb R}, \qquad u_3(t,z)^2+v_3(t,z)^2=1.
\end{equation}
Using \eqref{eq:normalisationuv}, we define a function $\alpha\equiv\alpha(t,z)$ such that 
$$
u_3(t,z) = \sin(\alpha(t,z)),\qquad v_3(t,z)=\cos(\alpha(t,z)).
$$
A direct computation shows that for ${\mathbb A}$ of the form \eqref{eq:lambdagmillling}, we have
$$
\mathbf{r} = (\partial_z u_3)\, \mathbf{u} + (\partial_z v_3) \, \mathbf{v}, \qquad\delta = 0.
$$
Therefore, Eq. \eqref{equationbigA} can be rewritten more concisely into: 
\begin{equation}
\label{eq:sohbgmilling}
\partial_t {\mathbb A} + c_4 \, [\Omega \times \mathbf{r}]_\times {\mathbb A} = 0,
\end{equation}
where we recall Eq. \eqref{eq:def[]x} for the definition of $[ \, ]_\times$. A direct computation shows that 
\begin{equation}
\label{eq:omegatimesr}
\Omega \times \mathbf{r} = (v_3 \, \partial_z u_3 - u_3 \, \partial_z v_3) \, \mathbf{e}_3 = (\partial_z\alpha) \, \mathbf{e}_3.
\end{equation}
Inserting this in \eqref{eq:sohbgmilling} implies that $u_3(t,z)\equiv u_3(z)$ and $v_3(t,z)\equiv v_3(z)$ are independent of time. We then observe that: 
\begin{equation}
{\mathbb A}(t,\mathbf{x}) = {\mathcal A}(- \omega t, \mathbf{e}_3) \, {\mathcal A}(\alpha(z), \mathbf{e}_1), 
\label{eq:A=AA}
\end{equation}
where we recall Eq. \eqref{eq:rodrigues} for the meaning of ${\mathcal A}$. Therefore, using \eqref{eq:sohbgmilling} and \eqref{eq:omegatimesr}, we obtain: 
$$
-\omega\, [\mathbf{e}_3]_\times {\mathbb A} + c_4 \, (\partial_z\alpha) \, [\mathbf{e}_3]_\times {\mathbb A}=0,
$$
from which we deduce that ${\mathbb A}$ satisfies \eqref{equationbigA} if and only if $\alpha$ and $\omega$ satisfy: 
$$
c_4 \, \partial_z \alpha = \omega,
$$
which implies
\begin{equation}
\label{eq:alphamilling}
\alpha(z) = \frac{\omega}{c_4} \, z + \bar \alpha,
\end{equation}
where $\bar \alpha$ is a constant, which can be interpreted as the phase at the origin $z=0$. To recover Eq.~\eqref{millingsolution}, we just need to take $\bar \alpha = 0$ and define $\xi = \omega / c_4$. Eq. \eqref{millingsolution2} follows from \eqref{eq:A=AA}.


\subsection{Generalized HW and proof of Lemma \ref{lem:HW}}
\label{appendix:generalhelical}

Starting from the initial condition~\eqref{paratwist}, we are looking for solutions of \eqref{equationbigA} of the form
$$
{\mathbb A}(t,\mathbf{x}) = \left(
\begin{array}{ccc}
1 & 0 & 0 \\
0 & \cos(\alpha(t,x)) & -\sin(\alpha(t,x)) \\ 
0 &  \sin(\alpha(t,x)) & \cos(\alpha(t,x))
\end{array}
\right), 
$$
for a real-valued function $\alpha$ of the $t$ and $x$ variables only. In this case, $\Omega$ is a constant vector and Equation \eqref{equationrho} is trivially satisfied. Moreover a direct computation shows that: 
$$
\mathbf{r} = 0, \qquad \delta = (\partial_x \alpha)(t,x).
$$
As a consequence, Eq. \eqref{equationOmega} is trivially satisfied and straightforward computations show that Eq. \eqref{equationbigA} reduces to 
$$
\partial_t \alpha + (c_2+c_4) \, \partial_x \alpha = 0.
$$
This last equation is a linear transport equation with velocity $c_2+c_4$, the solutions of which are given by
\begin{equation}
\label{eq:alphahelical}
\alpha(t,x) = \alpha_0(x-(c_2+c_4)t) 
\end{equation}
for any initial condition $\alpha_0 \in L^1_\text{loc}({\mathbb R})$. In the case of \eqref{paratwist}, $\alpha_0(x)=\xi \, x$. However, we see that there are as many different solutions as functions in $L^1_\text{loc}({\mathbb R})$. Such general solutions are called ``generalized HW''.


\subsection{Proof of Lemma \ref{lemma:generalizedsolutions}}
\label{appendix:generalizedsolutions} 

The three rotation matrices are given by 
\[\mathcal{A}(-\omega t, \mathbf{e}_3) = \left(
\begin{array}{ccc}
\cos(\omega t) & \sin(\omega t) & 0 \\
-\sin(\omega t) & \cos(\omega t) & 0 \\ 
0 & 0 & 1
\end{array}\right), \]
\[\mathcal{A}(\theta - \pi/2, \mathbf{e}_2) = \left(
\begin{array}{ccc}
\sin\theta & 0 & -\cos\theta \\
0 & 1 & 0 \\ 
\cos\theta & 0 & \sin\theta
\end{array}\right), \]
\[\mathcal{A}(\xi(z-\tilde{\lambda t}), \mathbf{e}_1) = \left(
\begin{array}{ccc}
1 & 0 & 0 \\
0 & \cos(\xi(z-\tilde{\lambda t})) & -\sin(\xi(z-\tilde{\lambda t})) \\ 
0 & \sin(\xi(z-\tilde{\lambda t})) & \cos(\xi(z-\tilde{\lambda t}))
\end{array}\right), \]
and a direct computation shows that the three column vectors $\Omega$, $\mathbf{u}$ and $\mathbf{v}$ of the matrix $\mathbb{A}_{\xi,\theta}$ are given by 
\[\Omega = \left(\begin{array}{c} \sin\theta\cos(\omega t) \\ -\sin\theta\sin(\omega t) \\ \cos\theta\end{array}\right), \]
\[\mathbf{u} = \left(\begin{array}{c}
-\cos\theta\sin(\xi(z-\tilde{\lambda t}))\cos(\omega t ) + \cos(\xi(z-\tilde{\lambda t}))\sin(\omega t) \\
\cos\theta\sin(\xi(z-\tilde{\lambda t}))\sin(\omega t) + \cos(\xi(z-\tilde{\lambda t}))\cos(\omega t) \\ 
\sin\theta\sin(\xi(z-\tilde{\lambda t}))
\end{array}\right),\]
\[\mathbf{v} = \left(\begin{array}{c}
-\cos\theta\cos(\xi(z-\tilde{\lambda t}))\cos(\omega t ) - \sin(\xi(z-\tilde{\lambda t}))\sin(\omega t) \\
\cos\theta\cos(\xi(z-\tilde{\lambda t}))\sin(\omega t) - \sin(\xi(z-\tilde{\lambda t}))\cos(\omega t) \\ 
\sin\theta\cos(\xi(z-\tilde{\lambda t}))
\end{array}\right).\]
Then we compute 
\begin{align*}
\mathbf{r} &= \xi \sin\theta \cos(\xi(z-\tilde{\lambda t})) \mathbf{u} - \xi\sin\theta\sin(\xi(z-\tilde{\lambda t}))\mathbf{u} = \xi\sin\theta (\sin(\omega t), \cos(\omega t), 0 )^\mathrm{T}, \\
\delta &= \cos \theta \partial_z \mathbf{u} \cdot \mathbf{v} + u_3\delta_z \mathbf{v} \cdot \Omega = \xi\cos\theta,\end{align*}
where we have used that $\partial_z \mathbf{u} = \xi \mathbf{v}$ and $\partial_z\mathbf{v}  = -\xi\mathbf{u}$. It remains to check that Eq. \eqref{equationbigA} holds true. We split this equation into three equations, one for each vector $\Omega$, $\mathbf{u}$ and $\mathbf{v}$. The first equation on $\Omega$ reads 
\[(\partial_t  + c_2(\Omega\cdot\nabla_\mathbf{x})) \Omega + c_4 P_{\Omega^\perp} \mathbf{r} = 0.  \]
This equation holds true because 
\[\partial_t\Omega = -\omega\left(\begin{array}{c}\sin\theta\sin(\omega t) \\ \sin\theta\cos(\omega t) \\ 0 \end{array}\right),\quad (\Omega\cdot\nabla_\mathbf{x})\Omega = 0, \quad P_{\Omega^\perp}\mathbf{r} = \mathbf{r} - (\mathbf{r}\cdot\Omega)\Omega = \xi\sin\theta\left(\begin{array}{c}\sin(\omega t) \\ \cos(\omega t) \\ 0 \end{array}\right),\]
and $\omega = c_4\xi$. The second equation on $\mathbf{u}$ reads 
\[(\partial_t  + c_2(\Omega\cdot\nabla_\mathbf{x})) \mathbf{u} - c_4(\mathbf{u}\cdot\mathbf{r})\Omega +c_4\delta\mathbf{v} = 0. \]
Because $\tilde{\lambda} = c_2\cos\theta$, we have 
$$ \partial_t  + c_2 \Omega\cdot\nabla_\mathbf{x}  = \partial_t  + c_2 \cos \theta \partial_z = \partial_t + \tilde \lambda \partial_z \quad \textrm{and} \quad \partial_t + \tilde \lambda \partial_z (z - \tilde \lambda t) = 0. $$
Thus 
$$
(\partial_t  + c_2(\Omega\cdot\nabla_\mathbf{x})) \mathbf{u} = \omega \left(\begin{array}{c}
\cos\theta\sin(\xi(z-\tilde{\lambda t}))\sin(\omega t ) + \cos(\xi(z-\tilde{\lambda t}))\cos(\omega t) \\
\cos\theta\sin(\xi(z-\tilde{\lambda t}))\cos(\omega t) - \cos(\xi(z-\tilde{\lambda t}))\sin(\omega t) \\ 
0
\end{array}\right),
$$
and using $\omega=c_4\xi$, it can be checked that 
\[(\partial_t  + c_2(\Omega\cdot\nabla_\mathbf{x})) \mathbf{u}  - c_4(\mathbf{u}\cdot\mathbf{r})\Omega = -c_4\xi \cos\theta \mathbf{v} = -c_4\delta\mathbf{v},\]
which yields the result. The equation on $\mathbf{v}$ is analogous. 

\subsection{GOP of the MO and generalized HW}
\label{appendix:generalisedorderlevels}

The GOP (given by Eq. \eqref{eq:globalop}) of the MO and HW do not depend on time and only depend on the function $\alpha$ defined respectively by \eqref{eq:alphamilling} and \eqref{eq:alphahelical}. Using Eq. \eqref{eq:globalop}, we can compute that the GOP is equal to: 
$$
\mbox{GOP} = \frac{1}{2} \left( \frac{c_1(\kappa)}{c_0} \right)^2  \big( 1 + 2 \, |\langle \mathbf{u} \rangle|^2 \big) + \frac{1}{4},
$$
where $\langle\mathbf{u}\rangle$ denotes the spatial average of the vector $\mathbf{u}$ with respect to $\rho$ (here the with respect to the uniform measure on the domain since $\rho$ is constant and uniform). With the previous notations, we obtain 
$$
|\langle \mathbf{u} \rangle|^2 = \langle \cos \alpha \rangle^2 + \langle \sin \alpha \rangle^2,
$$
For the generalized HW, depending on the choice of $\alpha$, the GOP can take any value between GOP$_1$ and GOP$_2$, these two extreme values being attained respectively when $|\langle \mathbf{u} \rangle|=0$ and $|\langle \mathbf{u} \rangle|=1$.


\section{Convergence rate of \texorpdfstring{$|\mathrm{d} \bar \varphi / \mathrm{d} t|$ as $N \to \infty$}{the milling speed as the number of particles grows to infinity}}
\label{sec:appendix_convergence_rate}

\begin{figure}[ht!]
\centering
\includegraphics[width= 6cm]{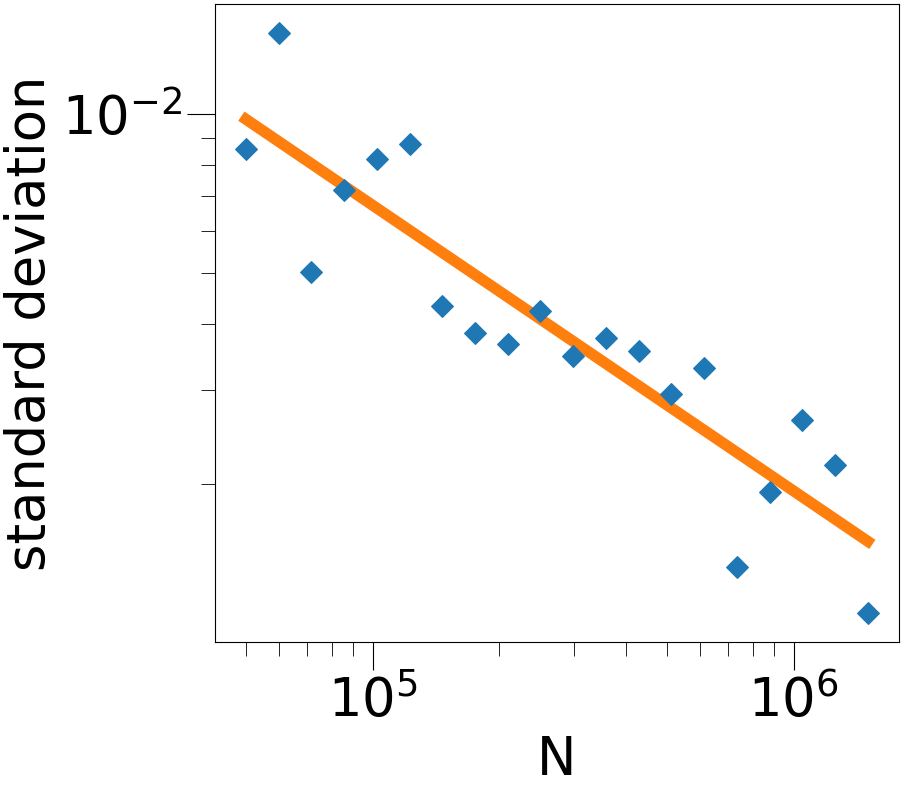}
\caption{Standard deviation of the 10 independent simulations as a function of $N$ (blue dots) and regression line (solid orange line) in log-log scale. Parameters: $L=1$, $\xi = 2 \pi$, $R=0.025$, $\nu=40$, $c_0=1$, $\kappa =10$.}
\label{fig:stddevN}
\end{figure}

The fact that the convergence rate of $|\mathrm{d} \bar \varphi / \mathrm{d} t|$ is close to $N^{-1}$ agrees with previously documented observations in spherical statistics. Indeed, it has been shown in \cite[Theorem 3(e)]{schou1978estimation} that the estimation of the concentration parameter of a (spherical) von Mises distribution obtained from a crude averaging procedure from $N$ independent samples produces a biased estimator with a (nonnegative) bias of order $N^{-1}$ (see also \cite[Section 10.3]{mardia2009directional}). In the present case, a similar reasoning can be applied, which we now briefly develop. The key observation is that all the measured quantities are functions of empirical averages of the form \eqref{eq:flux}. Under the chaos assumption (see Section \ref{appendix:derivationmacro}), when $N$ is large, the body-orientations of the particles behave as $N$ independent samples with common law $M_{\mathbb{A}}$, where $\mathbb{A}$ solves the SOHB model \eqref{equationA} and $M_\mathbb{A}$ is defined by \eqref{eq:vonmises}. In \cite[Theorem 4.1]{degondfrouvellemerino17}, it has been shown that $c_4(\kappa)$ can actually be expressed as a function of a certain number $p$ of averaged quantities
\[c_4(\kappa) = F(\langle g_1\rangle_{M_{\mathbb{A}}},\ldots, \langle g_p\rangle_{M_{\mathbb{A}}}),\]
where $g_i : \mathrm{SO}_3({\mathbb R})\to \mathcal{M}_3({\mathbb R})$ and $F:\mathcal{M}_3({\mathbb R})^p\to{\mathbb R}$ are smooth functions. The IBM simulation thus defines an estimator $\hat{\kappa}$ of the concentration parameter such that 
\[c_4(\hat{\kappa}) = F(\hat{g}_1,\ldots,\hat{g}_p),\]
where $\hat{g}_i$ is the average of $g_i$ obtained by replacing $M_{\mathbb{A}}$ by the empirical measure of the $N$ body-orientations of the particles. We can then measure the bias by taking the expectation of the Taylor expansion of the previous expression around the point $(\langle g_1\rangle_{M_\mathbb{A}},\ldots, \langle g_p\rangle_{M_\mathbb{A}})$ : 
\[c_4(\hat{\kappa}) = c_4(\kappa) + \delta\mathbf{\hat{g}}\cdot \nabla F + (\delta\mathbf{\hat{g}})^\mathrm{T} (\mathrm{Hess}\,F) \delta\mathbf{\hat{g}} + R,\]
where $\delta\mathbf{\hat{g}} = (\hat{g}_1,\ldots,\hat{g}_p)^\mathrm{T} -  (\langle g_1\rangle_{M_\mathbb{A}},\ldots, \langle g_p\rangle_{M_\mathbb{A}})^\mathrm{T}$ and $R$ is a remainder. The gradient $\nabla$ and Hessian $\mathrm{Hess}$ are defined within the Euclidean framework given by \eqref{eq:innerproduct}. By the chaos hypothesis ${\mathbb E}[\delta\mathbf{\hat{g}}] = 0$ and by the central limit theorem, the term of order two behaves as $N^{-1}$. Since SO$_3({\mathbb R})$ is compact, higher order moments of $\delta\mathbf{\hat{g}}$ can be controlled by a classical argument based on Hoeffding's inequality \cite[Lemma 5.5 and Theorem 5.29]{vershynin2012introduction}. This ensures that ${\mathbb E}[R]$ is $\mathcal{O}(N^{-2})$. We therefore obtain a biased estimator: 
\[{\mathbb E}[c_4(\hat{\kappa})] = c_4(\kappa) + \frac{a}{N}+\mathcal{O}(N^{-2}),\]
where $a \in {\mathbb R}$ depends on the derivatives of the considered functions and on the variance of the estimator \eqref{eq:flux} where the particles are replaced by independent identically distributed samples  with law $M_{\mathbb{A}}$. The fact that $a>0$ can be empirically verified on Fig. \ref{fig:slopeN} but has not been proved yet. For each $N$, the fluctuations around the average (biased) value can be monitored by computing the standard deviation of the 10 independent simulations. Fig. \ref{fig:stddevN} shows this standard deviation as a function of $N$ in a log-log-scale (blue dots). Although fluctuations remain significant with only 10 simulations per data point, by a standard linear regression (solid orange line) we obtain that the size of the standard deviation behaves as $N^{-\beta}$ with $\beta\simeq 0.54$.  which is close to the value $\beta = 1/2$ which we expect from an application of the central limit theorem.

\section{Rare events}\label{sec:rareevents}

Although the scenario described in Section \ref{sectionnumericalexperiments} of the main text is the most common one, the IBM sometimes leads to different, slightly more complex scenarios which are described in the present section. Now, the IBM is initialized by drawing $N$ positions independently uniformly in the cubic domain 
$\mathcal{D} = [0,L]\times[0,L]\times[0,L]$ with periodic boundary conditions and $N$ body-orientations independently from the von Mises distribution $M_{\mathbb{A}(0,\mathbf{x})}$ where $\mathbb{A}(0,\mathbf{x})$ is given by \eqref{perptwist} with $\xi=2\pi/L$ (winding number equal to~$1$).


\subsection{From milling orbit to helical wave}
\label{sec:millingtohelical}

\begin{figure}[htp!]
\centering
\subfloat[]{\includegraphics[height= 4.5cm]{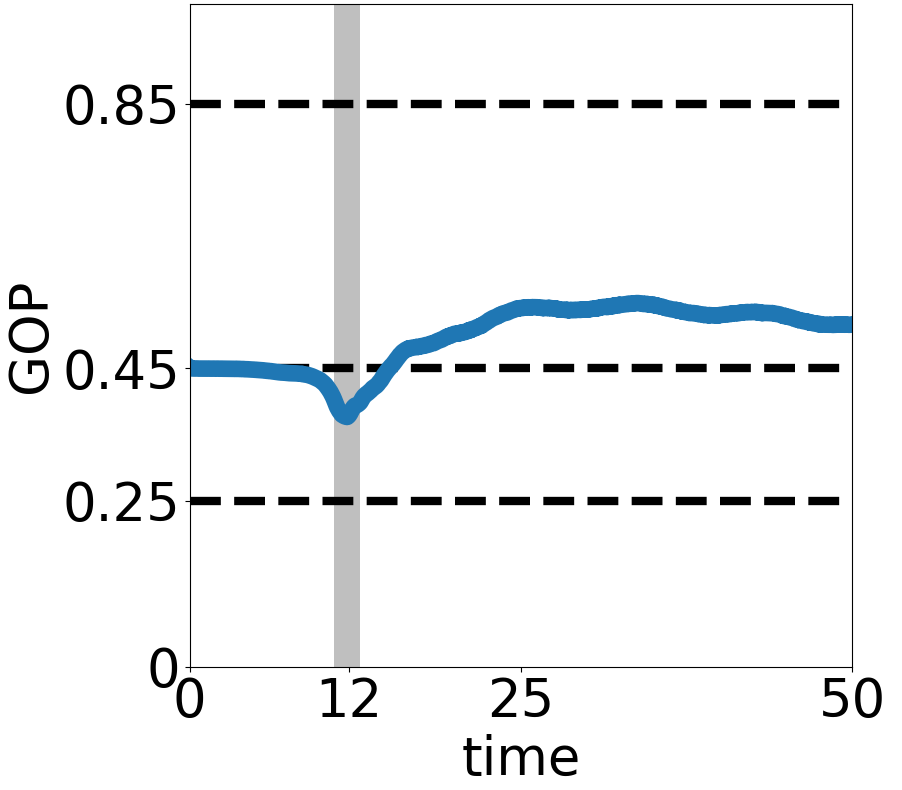}\label{subfig:milling_to_helical_op}}
\subfloat[]{\includegraphics[height= 4.5cm]{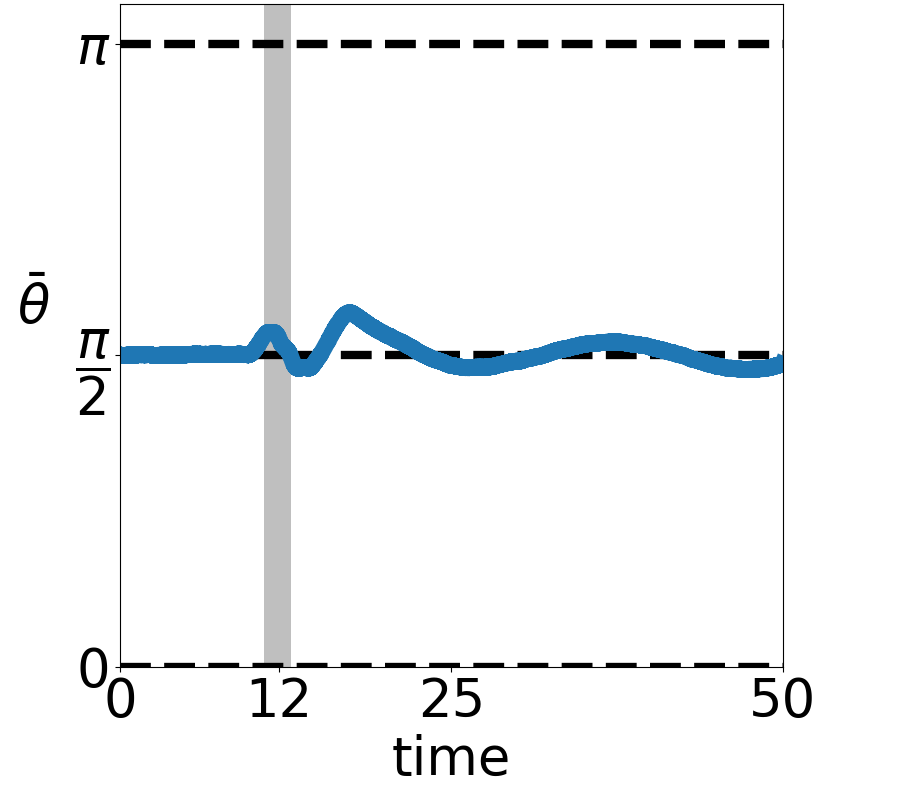}\label{subfig:milling_to_helical_theta}}
\subfloat[]{\includegraphics[height= 4.5cm]{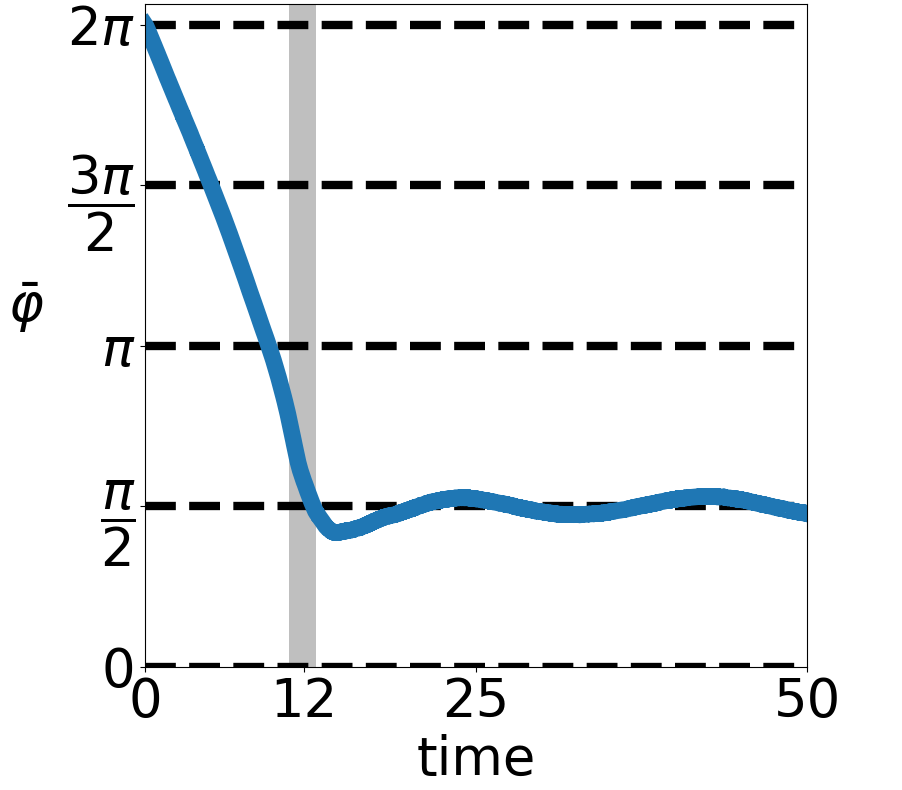}\label{subfig:milling_to_helical_phi}}

\subfloat[]{\includegraphics[height= 4.5cm]{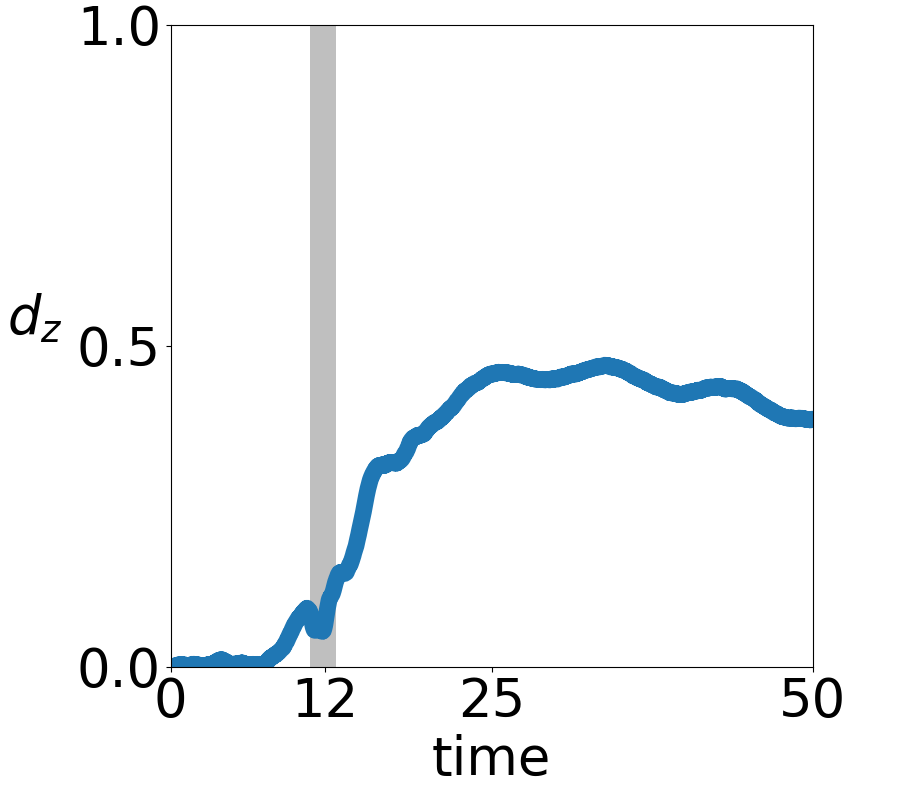}\label{subfig:milling_to_helical_d_z}}
\subfloat[]{\includegraphics[height= 4.5cm]{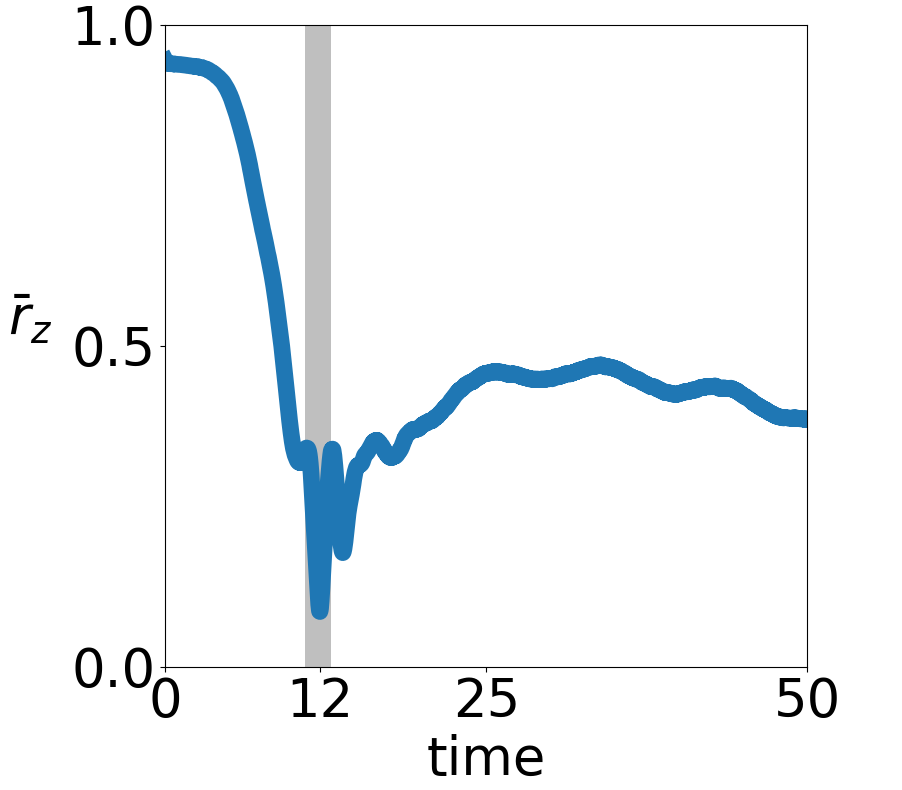}\label{subfig:milling_to_helical_r_z}}
\subfloat[]{\includegraphics[height= 4.5cm]{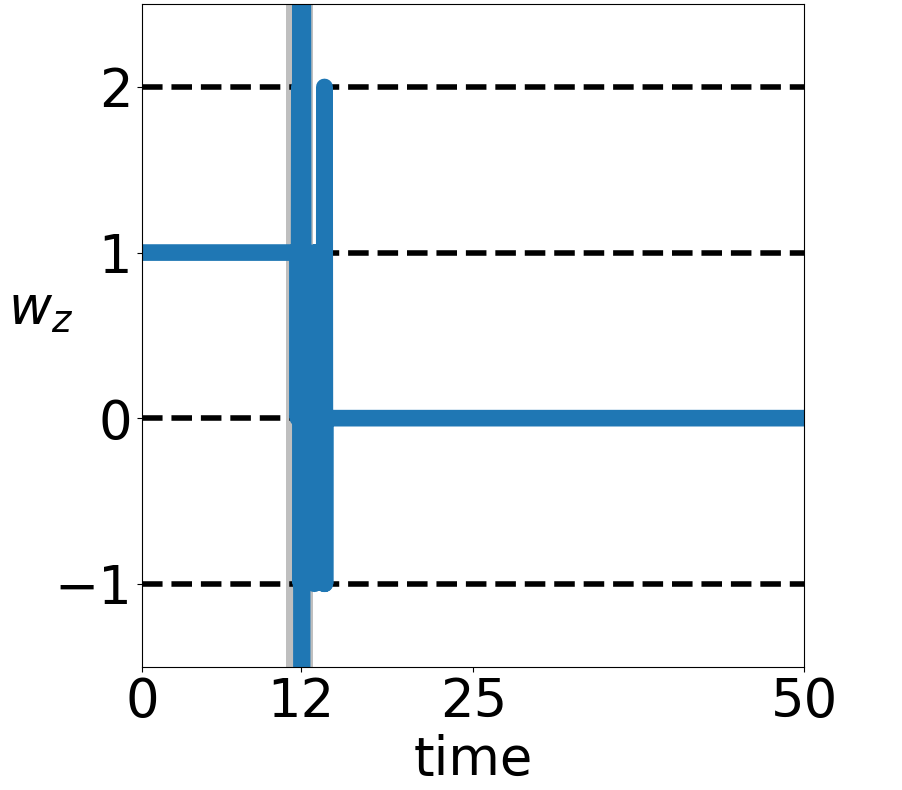}\label{subfig:milling_to_helical_wn_z}}

\subfloat[]{\includegraphics[height= 4.5cm]{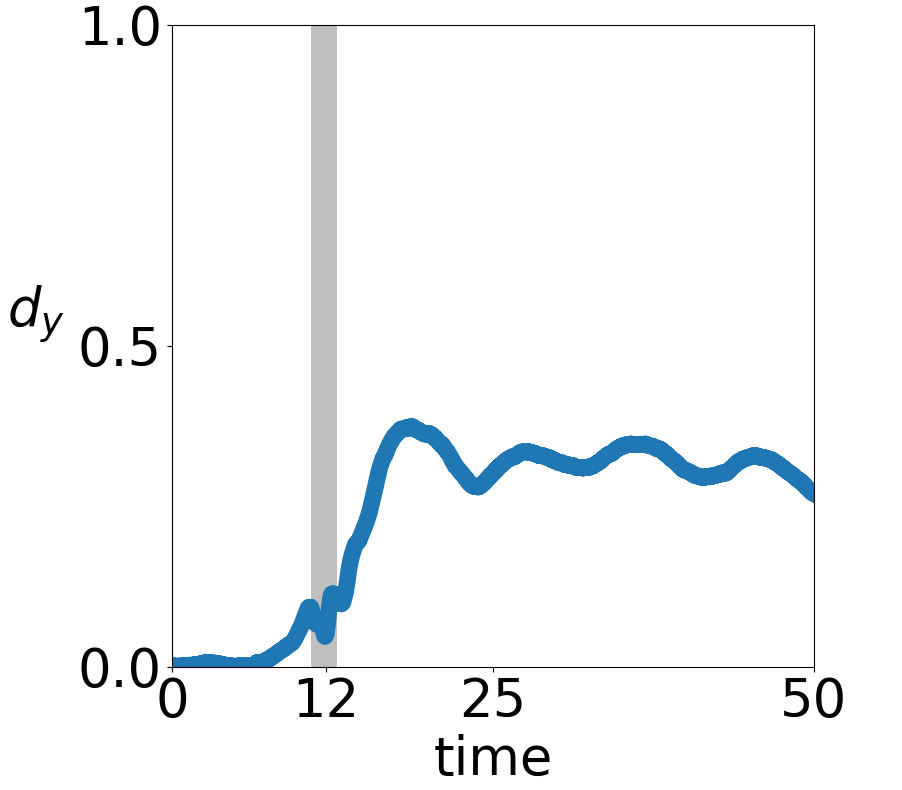}\label{subfig:milling_to_helical_d_y}}
\subfloat[]{\includegraphics[height= 4.5cm]{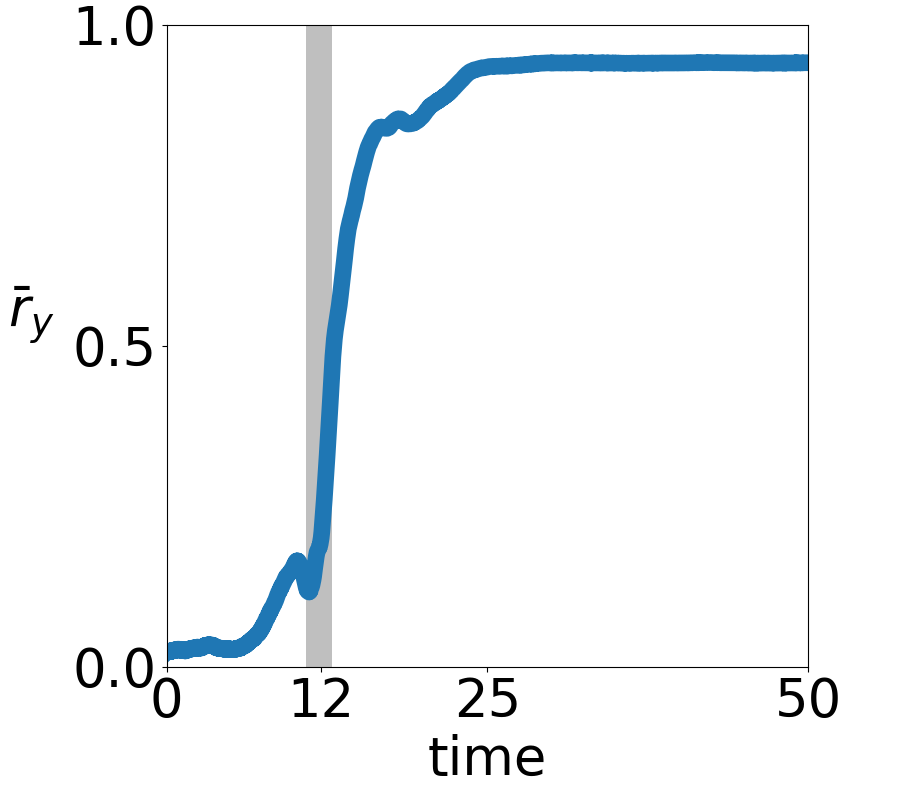}\label{subfig:milling_to_helical_r_y}}
\subfloat[]{\includegraphics[height= 4.5cm]{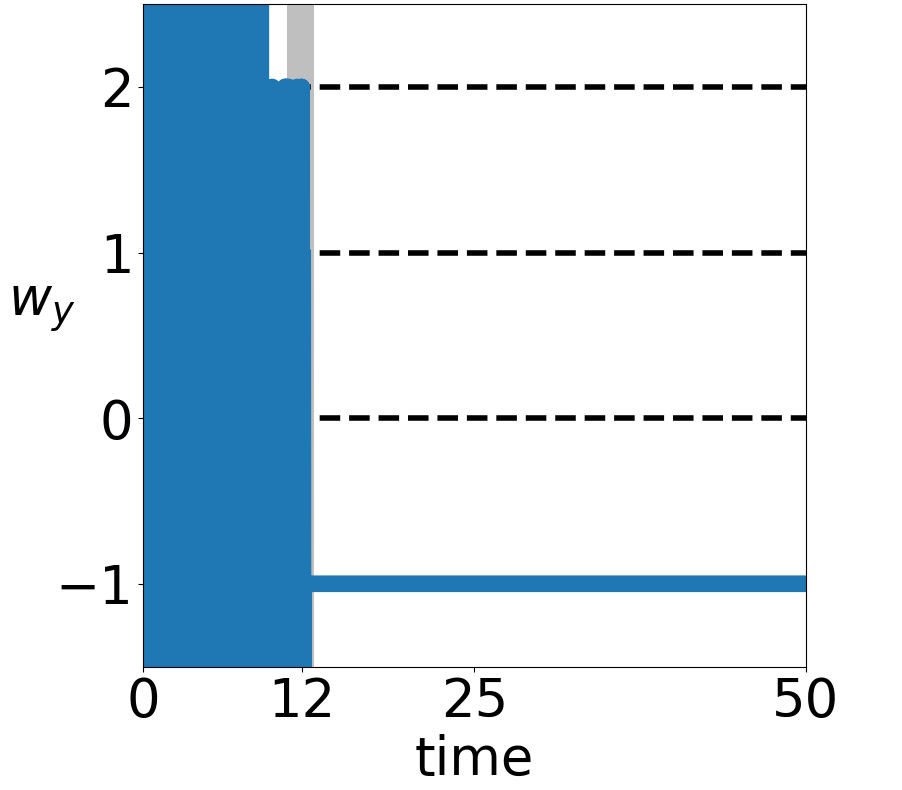}\label{subfig:milling_to_helical_wn_y}}
\caption{Transition from a MO to a HW: example of a solution of the IBM for an initial condition sampled from \eqref{eq:inimill_nt} in the rare case where it leads to a HW. The following indicators are plotted as functions of time:  
(a) GOP
(b) Pitch $\bar \theta$ of $\bar \Omega$. 
(c) Yaw $\bar \varphi$ of  $\bar \Omega$. 
(d) Distance of center of mass of RPZ curve to the origin $d_z$. 
(e) Mean distance of RPZ curve to the origin $\bar r_z$. 
(f) Winding number of RPZ curve $w_z$. 
(g) Distance of center of mass of RPX curve to the origin $d_x$. 
(h) Mean distance of RPX curve to the origin $\bar r_x$. 
(i) Winding number of RPX curve $w_x$. 
Gray shaded zones highlight a small region around the time of minimal GOP. 
Parameters: $N=1.5\cdot10^6$, $R=0.025$, $L=1$, $D=0.1$, $\nu=40$, $c_0=1$.
See caption of Fig. \ref{fig:topo_protec_graphs_mills} for further indications. See also Videos \ref{vid:milling_to_helical_particles} to \ref{vid:milling_to_helical_RPX} in Section \ref{appendix:listvideos}.}
\label{fig:milling_to_helical_graphs}
\end{figure}

Here, we report on the occurrence of transitions from a MO to a HW. Among twenty independent simulations, this transition occurred only once (the other cases being a transition from a MO to a FS). We run the IBM and record the time-evolution of a set of indicators as shown in Fig. \ref{fig:milling_to_helical_graphs} (see also supplementing videos~\ref{vid:milling_to_helical_particles} to \ref{vid:milling_to_helical_RPX} in Section \ref{appendix:listvideos}). 

As shown in Fig. \ref{subfig:milling_to_helical_op}, the GOP does not converge towards GOP$_2$ characterizing the FS, but towards an intermediate value between GOP$_1$ (which characterizes MO or HW) and GOP$_2$. As explained in Section \ref{appendix:generalisedorderlevels}, such values of the GOP can be attained by a generalized helical wave solution (as can be observed in Video \ref{vid:milling_to_helical_particles}). The pitch $\bar \theta$ (Fig. \ref{subfig:milling_to_helical_theta}) and yaw $\bar \varphi$ (Fig.~\ref{subfig:milling_to_helical_phi}) behave like in the milling-to-flocking transition (see Figs. \ref{subfig:milling_to_flocking_theta} and~\ref{subfig:milling_to_flocking_phi}) except for small-amplitude, slow-frequency oscillations appearing after the topological transition time. This may be due to some competition between two attractors, the FS and the HW, which being alternately stronger and weaker, generate this oscillatory behavior. Note that a transition to a HW cannot occur when the global direction of motion at the transition time is not one of the principal axes of the square domain since a HW along another direction is not compatible with the periodic boundary conditions (see Section \ref{sec:millingtoflockviahelical}). This is confirmed by the final values of $\bar \varphi$ and $\bar \theta$ (both equal to $\pi/2$) which correspond to a global direction of motion oriented along the $y$-axis (in what follows, in reference to \eqref{eq:inihelic_nt} and to avoid confusion, we will still call that direction, the $x$ direction). 

The second and third lines of figures in Fig. \ref{fig:milling_to_helical_graphs} show the triplets of topological indicators $(d_z, \bar r_z, w_z)$ and $(d_x, \bar r_x, w_x)$ which materialize the MO and HW structures respectively. The mean distance of the RPZ-curve to the origin $\bar r_z$ (Figs. \ref{subfig:milling_to_helical_r_z}) decreases, revealing an increase of the disorder. Simultaneously, the distance of its center of mass to the origin $d_z$ increases (Figs. \ref{subfig:milling_to_helical_d_z}) showing a transition trend to a FS. The winding number $w_z$ (Fig. \ref{subfig:milling_to_helical_wn_z}) jumps from $1$ to $0$ at the time of maximal disorder. However, $d_z$ and $\bar r_z$ do not reach zero, showing that complete disorder across $z$ is not reached. Since the final state of the system is a generalized helical wave state (see Section \ref{appendix:generalisedorderlevels}), we do not necessarily expect that complete disorder will be reached along the $z$-direction. In the mean time, $\bar r_x$ starts from $0$ (complete disorder) and increases up to a value close to unity, showing the build-up of a HW. The quantity $d_x$ increases during some time but eventually decreases to $0$ (not shown in the figure) as it should for a HW. Finally, the winding number $w_x$ is undefined in the initial stage, as it should for complete disorder, but builds up to $1$ at the time where the winding number $w_z$ drops to $0$. There is a transfer of non-trivial topology from an MO structure to a HW structure.

\subsection{From milling to flocking via a helical wave state}
\label{sec:millingtoflockviahelical}

\begin{figure}[ht!]
\centering
\subfloat[]{\includegraphics[height= 4.5cm]{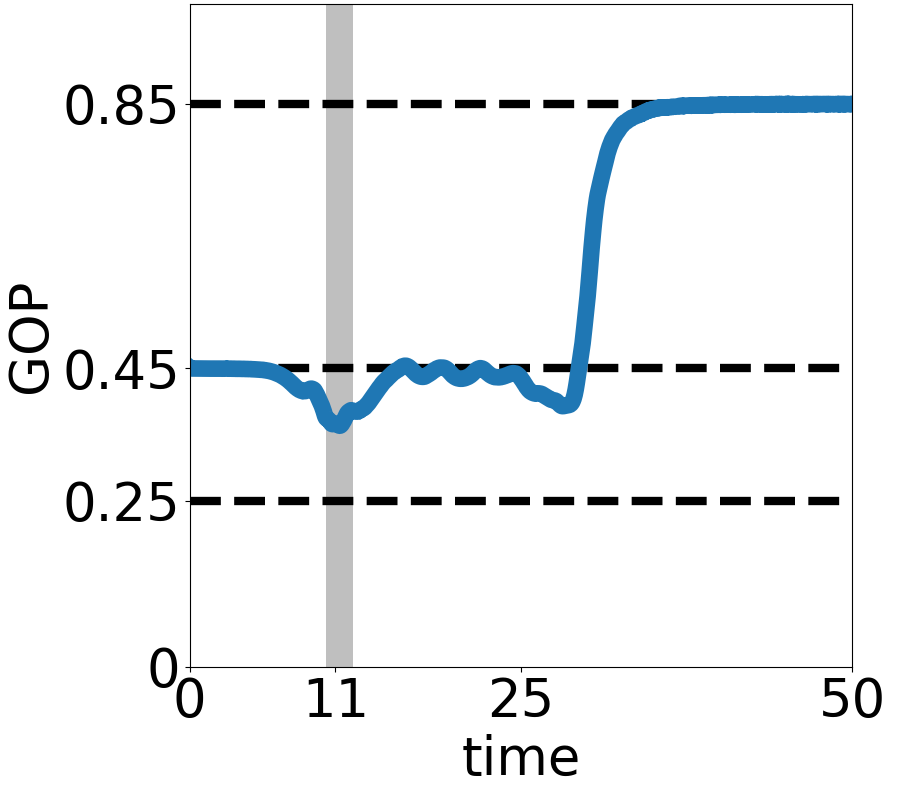}\label{subfig:milling_to_helical_then_flock_op}}
\subfloat[]{\includegraphics[height= 4.5cm]{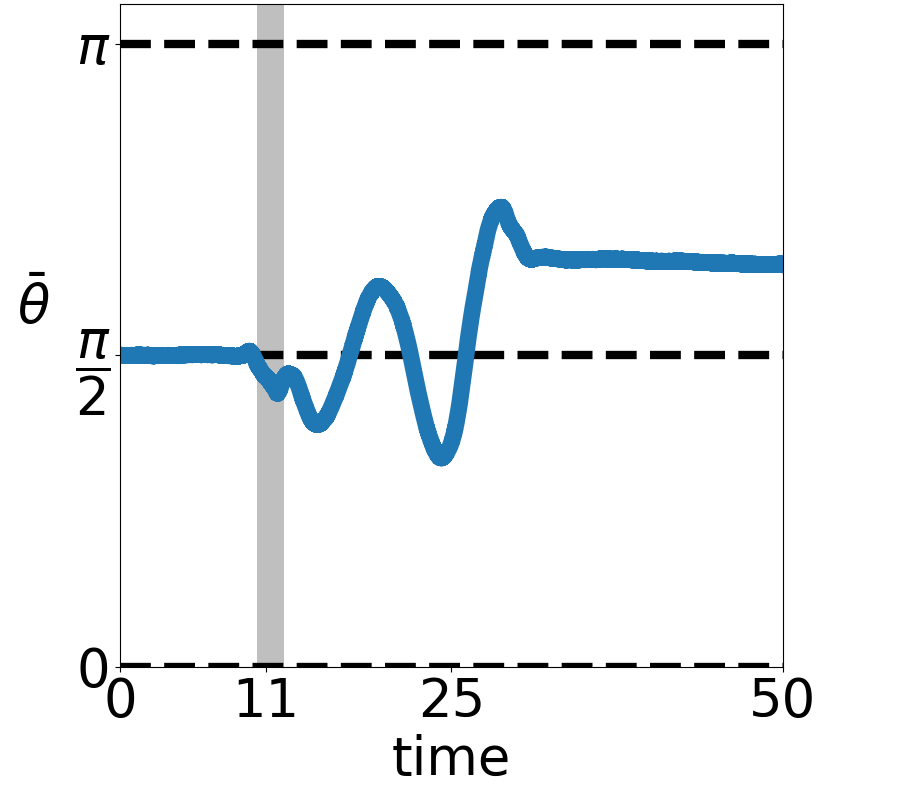}\label{subfig:milling_to_helical_then_flock_theta}}
\subfloat[]{\includegraphics[height= 4.5cm]{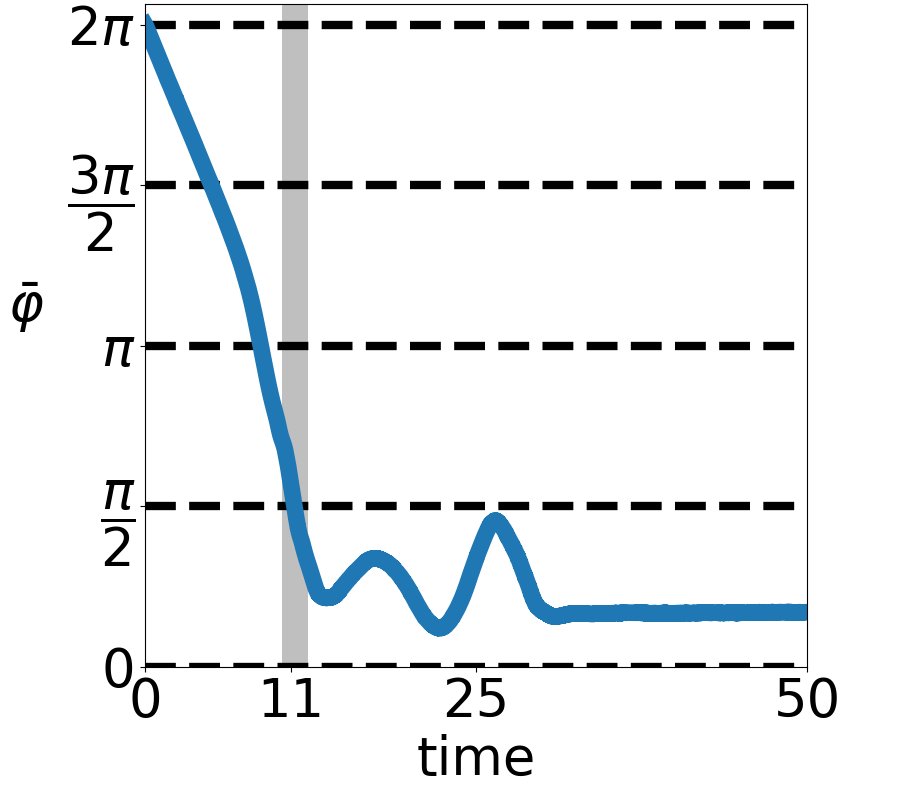}\label{subfig:milling_to_helical_then_flock_phi}}

\subfloat[]{\includegraphics[height= 4.5cm]{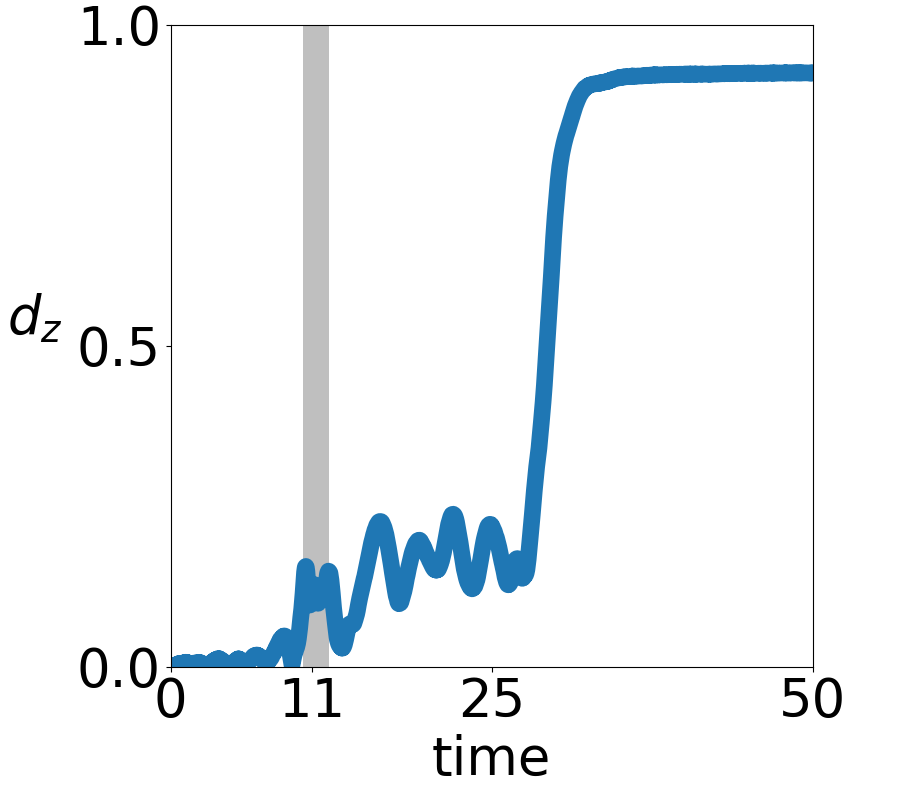}\label{subfig:milling_to_flock_via_helical_d_z}}
\subfloat[]{\includegraphics[height= 4.5cm]{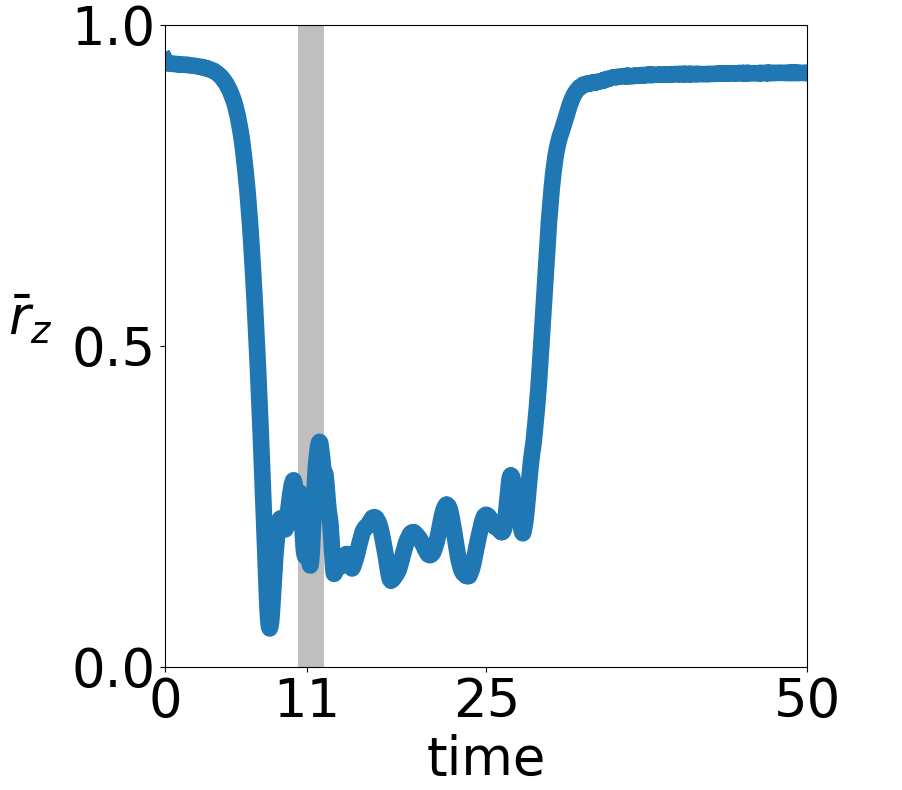}\label{subfig:milling_to_flock_via_helical_r_z}}
\subfloat[]{\includegraphics[height= 4.5cm]{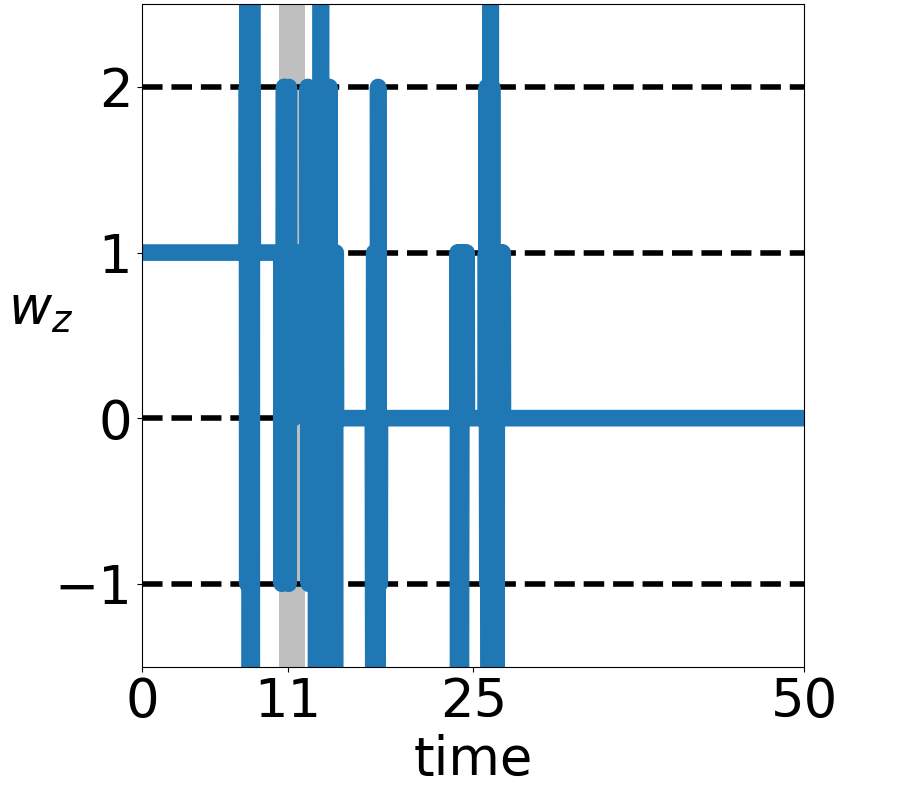}\label{subfig:milling_to_flock_via_helical_wn_z}}
\caption{Transition from a MO to a FS via an unstable HW: example of a solution of the IBM for an initial condition sampled from \eqref{eq:inimill_nt} in the rare case where it leads to a FS through a transient HW. The following indicators are plotted as functions of time:  
(a) GOP
(b) Pitch $\bar \theta$ of $\bar \Omega$. 
(c) Yaw $\bar \varphi$ of  $\bar \Omega$. 
(d) Distance of center of mass of RPZ curve to the origin $d_z$. 
(e) Mean distance of RPZ curve to the origin $\bar r_z$. 
(f) Winding number of RPZ curve $w_z$. 
Gray shaded zones highlight a small region around the time of minimal GOP. 
Parameters: $N=1.5\cdot10^6$, $R=0.025$, $L=1$, $D=0.1$, $\nu=40$, $c_0=1$.
See caption of Fig. \ref{fig:topo_protec_graphs_mills} for further indications. See also Videos \ref{vid:milling_to_helical_then_flock_particles} and \ref{vid:milling_to_helical_then_flock_RPZ} in Section \ref{appendix:listvideos}. 
}
\label{fig:milling_to_helical_then_flock_graphs}
\end{figure}

In some rare cases an intermediate unstable HW can be observed. Note that due to the periodic setting, an HW cannot be stable for most of the the global directions of motion. Although stable or unstable HW typically appear in one over twenty of our simulations, it should be kept in mind that the occurrence frequency also depends on the geometry of the domain and that this phenomena may be more frequent for other simulation settings. The procedure is the same as in the previous section. Fig. \ref{fig:milling_to_helical_then_flock_graphs} shows the results (see also supplementing videos \ref{vid:milling_to_helical_then_flock_particles} and \ref{vid:milling_to_helical_then_flock_RPZ} in Section \ref{appendix:listvideos}). 

The transition stage between the MO and FS is significantly longer than in the previous situations. During that phase, the GOP (Fig. \ref{subfig:milling_to_helical_then_flock_op}) oscillates between the value $\Psi_1$ characterizing the MO and lower values, i.e. lower order. Likewise, there are significant variations of the pitch $\bar \theta$ (Fig. \ref{subfig:milling_to_helical_then_flock_theta}) and yaw $\bar \varphi$ (Fig. \ref{subfig:milling_to_helical_then_flock_phi}). As in the previous section, this could be explained by antagonist effects of different attractors (the MO and HW) and subsequent oscillations of the system between them. Video \ref{vid:milling_to_helical_then_flock_particles} reveals large scale band structures similar to a HW except that the global direction of motion is not one of the principal axes of the square domain. As, in most cases, this cannot be compatible with the periodic boundary conditions, such state cannot persist in time. The relatively long-time persistence of this stage could be explained in the present case by the fact that the global direction of motion seems to oscillate around the direction given by $\mathbf{e}_1+\mathbf{e}_2$ (i.e. $\varphi = \pi/4$ and $\theta=\pi/2$) which is theoretically compatible with the periodic boundary conditions, provided the wave length $\xi$ is changed from $2 \pi / L$ to $\sqrt{2} \pi / L$. This state does not seem to be stable as shown by the large oscillations of $\bar \varphi$ and $\bar \theta$. 
The topological indicators $(d_z, \bar r_z, w_z)$ shown in the second line of figures of Fig. \ref{fig:milling_to_helical_then_flock_graphs} also display large oscillations. The quantity $\bar r_z$ drops, and at the same time, $d_z$ remains small, while the winding number $w_z$ has strong oscillations, indicating a state of large disorder across $z$, which is consistent with the fact that the temporary HW order is organized in a different direction. However, we see that $w_z$ has a calmer period between two series of oscillations. This calmer period corresponds to the interval of time during which the temporary HW order prevails. Eventually the triplet converges to the value $(1,1,0)$ characterizing the FS.

\bibliographystyle{abbrv}
\bibliography{biblio_topoprotec}

\end{document}